# FERROELECTROMAGNETS. FIFTY YEARS AFTER DISCOVERY

**I. E. Chupis**

Department of Theoretical Physics, B.I.Verkin Institute for Low Temperature Physics and Egineering,

47 Lenin Ave., Kharkiv 61103 Ukraine

E-mail: iechupis@mail.ru

In memory of G.A.Smolenskii

**Contents**

# 1. Introduction

## 1.1 History of the discovery

Electric and magnetic phenomena are known for a long time and their use is quite common. The existence of spontaneous ordering of electric dipoles (ferroelectricity) and spins (magnetism) in solids are of significant scientific interest. Ferroelectrics and magnets separately are widely employed in modern technology.

*But is simultaneous existence of ferroelectricity and magnetism possible in a crystal?*

Such possibility was experimentally proved about fifty years ago. A new class of materials was named ferroelectromagnets**.**

**Ferroelectromagnets** are compounds in which magnetic and ferroelectric (or antiferroelectric) spontaneous orderings exist simultaneously. The discovery of ferroelectromagnets was preceded by an intensive and fruitful development of the physics of magnetic phenomena and ferroelectricity. The search for new ferroelectrics led to the discovery of the materials with a perovskite structure and a significant iron ion concentration by a group of Leningrad physicists [1, 2]. Smolenskii and Ioffe in 1958 pointed on the principal possibility of the coexistence of ferroelectricity and magnetism in a perovskite compounds [3]. In perovskite-type compounds $ABO_3$ (see Fig.1), the angles in the (cation $B$)-oxygen-(cation $B$) chains are close to 180º . Hence, the magnetic ions in octahedral $B$-positions can become ordered due to an indirect exchange interaction involving oxygen ions. According to the model theories put forth in those years, the ferroelectric (FE) ordering in perovskite lattice appears mainly due to the displacements of A and B ions. The ordering is facilitated by the presence of $A$ ions with stereochemically active unshared pair of $6s$-electrons ($Pb^{2+}, Bi^{3+}, Tl^+$ ), and the transition metal ions

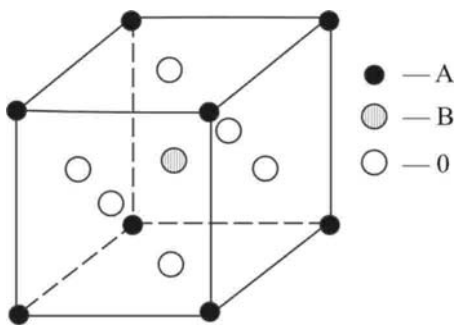

Fig. 1. The ideal unit cell of perovskite $ABO_3$. The B ion is at the centre of the cube, while the oxygen ions are at the centres of the faces.

—A
—B
—0

($Ti^{4+}, Zr^{4+}, Nb^{5+}, Ta^{5+}, W^{6+}, Mo^{6+}$) in the $B$-sublattice, which have a noble gas shell after the removal of $s$ and $d$- electrons [4]. These ions, however, do not have a magnetic moment. In order to satisfy the necessary conditions for the coexistence of ferroelectricity and magnetism, FE active ions as well as magnetic ions were introduced into the octahedral positions. Thus in 1961 the first ferroelectromagnets combining FE and antiferromagnetic (AF) properties were synthesized. Firstly, polycrystalline solid solutions $(1-x)Pb(Fe_{2/3}W_{1/3})O_3 - xPb(Mg_{1/2}W_{1/2})O_3$ were received [5]. Then in single crystals $Pb(Fe_{1/2}Nb_{1/2})O_3$ [6] and $BiFeO_3$ [4, 7- 9] were first observed FE and AF ordering. Later the presence of FE and AF orders was revealed in hexagonal rare-earth manganites (Bertaut *et al.* [10], 1963). The coexistence of ferroelectricity and antiferromagnetism with a weak ferromagnetic moment (AWFM) was observed in a boracite family (Schmid *et al.* [11], 1966). An intensive experimental and theoretical studying of ferroelectromagnets during the next two decades yielded many interesting results (see the reviews [12], [13]). About fifty ferroelectomagnetic compounds and dozens solid solutions were revealed. Besides the perovskite-type compounds, hexagonal rare-earth manganites, rare- earth molybdates, boracites, orthorhombic compounds $BaMF_4$, ferroelectromagnets with other crystallographic structures were discovered. The FE ferrite $Fe_3O_4$ with an inverted spinel structure and the oxide FE-AF $Li(Fe_{1/2}Ta_{1/2})O_2F$ with a pseudoilmenite structure are examples of such compounds. These ferroelectromagnets displayed the FE or antiferroelectric ordering and ferromagnetic (FM) or mainly AF and AWFM magnetic ordering. AF helical-type ordering accompanied by the emergence of FE properties was revealed in the triclinic $Cr_2BeO_4$ .Most experimental studies were devoted to synthesis and definition of the type of electric and magnetic order.

*1.2 Magnetoelectric effects*

The coexistence of electric and magnetic orders in ferroelectromagnets opens (besides the employment of FE and magnetic properties of crystal separately) new possibilities of the cross-controlling of electric (magnetic) parameters by magnetic (electric) field. These new possibilities appear due the coupling of magnetic and electric subsystems in matter which named magnetoelectric (ME) coupling.

The coupling of electric and magnetic fields in *dynamic* phenomena is well known and is contained in Maxwell's equations. The possibility of a *static* ME effects was also discussed in the 19-th century. Curie pointed out the possibility of a static ME phenomena in crystals on the basis of symmetry considerations [14]. But only about sixty years later the interest in the static

ME effects was revived. Landau and Lifshits indicated the possibility of the existence of an equilibrium electric polarization in magnetically ordered crystals, proportional to the magnetic field strength, as well as an equilibrium magnetization proportional to the electric field strength (the linear ME effect (LMEE)) [15].The LMEE was studied by Dzyaloshinskii [16] from the point of view of magnetic symmetry. Dzyaloshinskii showed the possibility of the LMEE in the antiferromagnet $Cr_2O_3$ [16] where this effect was soon discovered by Astrov [17].

The ME effects in external fields are described in the Landau theory by writing the expansion of the free energy of a single-phase crystal, i.e.,

$$-F(\vec{E},\vec{H}) = -F_0 + P_i^S E_i + M_i^S H_i + \frac{1}{2}\ \chi_{ij}^E E_i E_j + \frac{1}{2}\ \chi_{ij}^M H_i H_j + \alpha_{ij} E_i H_j \qquad (1.1)$$

$$+ \frac{\beta_{ijk}}{2} E_i H_j H_k + \frac{\gamma_{ijk}}{2} H_i E_j E_k + ...$$

with $\vec{E}$ and $\vec{H}$ as the electric field and the magnetic field, respectively; $\vec{P}^S$ and $\vec{M}^S$ are the spontaneous polarization and magnetization; $\hat{\chi}^E$ and $\hat{\chi}^M$ are the tensors of a dielectric and magnetic susceptibilities. From the expression (1.1) one obtains the electric polarization

$$P_i(\vec{E},\vec{H}) = -\frac{\partial F}{\partial E_i}\ = P_i^S + \chi_{ij}^E E_j + \alpha_{ij} H_j + \frac{\beta_{ijk}}{2} H_j H_k + \gamma_{jik} H_j E_k + ... \qquad (1.2)$$

and the magnetization

$$M_i(\vec{E},\vec{H}) = -\frac{\partial F}{\partial H_i} = M_i^S + \chi_{ij}^M H_j + \alpha_{ji} E_j + \frac{\gamma_{ijk}}{2} E_j E_k + \beta_{jik} E_j H_k + ... \qquad (1.3)$$

The response of the system to the action of an external electric and magnetic field is described by the generalized susceptibility tensor

$$\hat{X} = \begin{pmatrix} \hat{\chi}^E & \hat{\chi}^{EM} \\ \hat{\chi}^{ME} & \hat{\chi}^M \end{pmatrix}$$

where $\chi_{ik}^E, \chi_{ik}^M$ and $\chi_{ik}^{EM} = \chi_{ki}^{ME}$ are the second-rank tensors of dielectric, magnetic, and ME susceptibilities respectively:

$$\chi_{ik}^{E} = \frac{\partial P_i}{\partial E_k} \quad , \qquad \chi_{ik}^{EM} = \frac{\partial P_i}{\partial H_k} \quad ,$$

$$\chi_{ik}^{ME} = \frac{\partial M_i}{\partial E_k} \quad , \qquad \chi_{ik}^{M} = \frac{\partial M_i}{\partial H_k} \quad .$$

(1.4)

The constant $\alpha$ in (1.1) describes the LMEE:

$$\alpha_{ik} = \chi_{ik}^{EM} = \chi_{ki}^{ME} = \left( \frac{\partial P_i}{\partial H_k} \right)_{H=0} = \left( \frac{\partial M_k}{\partial E_i} \right)_{E=0} \quad . \tag{1.5}$$

Pay attention to the definition of LMEE (5) where the constant $\alpha$ is not zero at $H = E = 0$. This definition results from the expansion (1.1) where a weak field is supposed. LMEE is possible in magnetically ordered crystals of certain symmetry. Magnets showing a LMEE are commonly called **magnetoelectrics.** The vast majority of researches on the ME effect is devoted to the LMEE. The terms in the free energy (1.1) with the coefficients $\beta$ and $\gamma$ describe nonlinear (higher- order) ME effects. Nonlinear ME effects can occur in any materials, even if they are not magnets.

The ferroelectromagnet is not certainly a magnetoelectric. It differs from other materials as, besides the ME effects induced by external fields, it also exhibits *spontaneous ME effects* due to strong electric and magnetic internal fields. This may give rise to qualitatively new ME effects which do not have an analogue in other crystals. Unlike in the magnetoelectrics with weak induced ME effects, in ferroelectromagnets strong spontaneous effects manifest themselves near the phase-transition temperatures. The strongest value of ME effect was predicted in the ferroelectromagnet (see below).

The ME effects may be described by using the free energy in terms of electric and magnetic moments $\widetilde{F}(\vec{P}, \vec{M})$ where

$$\widetilde{F}(\vec{P}, \vec{M}) = F(\vec{E}, \vec{H}) + \vec{M}\vec{H} + \vec{P}\vec{E} \quad . \tag{1.6}$$

For equilibrium state the differentiation of $\widetilde{F}$ leads to the equations

$$\frac{\partial \widetilde{F}}{\partial \vec{P}} = \vec{E}, \qquad \frac{\partial \widetilde{F}}{\partial \vec{M}} = \vec{H} \quad . \tag{1.7}$$

The repeated differentiation (1.7) on the components of electric and magnetic fields gives the expressions for the susceptibilities (1.4) [18]. In a simple isotropic case the susceptibilities are the following:

$$\chi^E = AD, \quad \chi^M = BD, \quad \chi^{EM} = \chi^{ME} = -CD, \tag{1.8}$$

$$A = \partial^2 \widetilde{F}/\partial M^2, \quad B = \partial^2 \widetilde{F}/\partial P^2, \quad C = \partial^2 \widetilde{F}/\partial M \partial P, \quad D = (AB - C^2)^{-1}. \tag{1.9}$$

The condition of the stability $D > 0$ means that [18]

$$(\chi^{ME})^2 < \chi^E \chi^M \quad . \tag{1.10}$$

This inequality determines the upper bound for the absolute magnitude of the ME susceptibility. It can be seen that the larges value of the ME susceptibility, and hence of ME effects, should be expected in crystal having large values of $\chi^E$ and $\chi^M$, i.e., in ferroelectromagnets. Since the susceptibilities are the strongest near the phase transition temperatures, the strongest ME effects are expected in ferroelectromagnets with close electric and magnetic temperatures of the orderings [12].

### 1.3 *Evolution, terminology, and the subjects of the review*

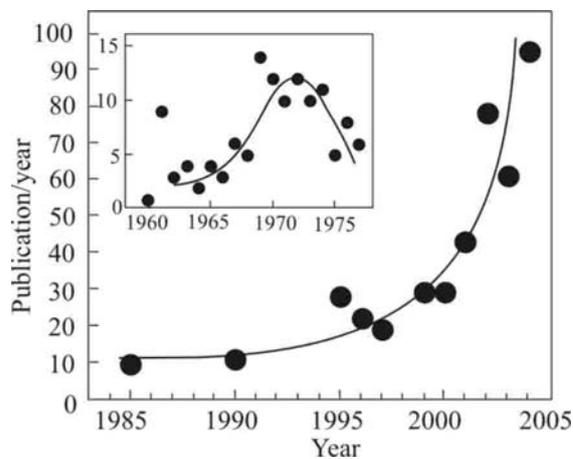

**Fig. 2**. Publications per year with "magnetoelectric" as a key word [19] according to the *Web Science* [20].

After the discovery of the LMEE and the ferroelectromagnets the intensive investigation of the ME effects in different compounds began. The maximum of this activity was observed during seventeen years when the book by O'Dell on electrodynamics of ME media appeared [21], and the first international conference on ME interaction phenomena in crystals (MEIPIC) started [22] (Fig. 2). Then the first manifestations of the strong ME effects in the ferroelectromagnets were also observed [11, 23]. The time period 1961- 1981 of the high interest in ME phenomena is called here as "The first ME renascence". But after two decades the research activities somewhat decreased. A large interval between the first and the second conferences MEIPIC indicates this decline [24]. "The second ME renascence" started at the beginning of the 21-st century. After the third MEIPIC in 1997 [25] a quick growth of the interest in ME phenomena is observed. This growth became almost exponentially large from the discovery of "colossal ME effects" in hexagonal $YMnO_3$ (2002, [271]) and in the spiral ferroelectromagnet $TbMnO_3$ (2003, ,[26]). The number of publications after 2005 is more than a hundred per year. Such colossal activity in the investigations of ME phenomena gives the hope that practical using of ferroelectromagnets in modern technology may be soon realized. Unfortunately, the misnomer is happenning in some papers. The term "magnetoelectric" is sometimes used for the compounds without the LMEE. The presence of a linear dependence of electric polarization $P$ of $H$ in a strong magnetic field is erroneously considered as the manifestation of LMEE (see (1.5)).

The elastic subsystem of a crystal may also posses a spontaneous ordering. Ferroelectrics, ferromagnets and ferroelastics belong to the so-called ferroics [27]. Recently the *ferrotoroidics* were added to the group of ferroics. The ferrotoroidic state is parametrized by a spontaneous toroidal moment. The toroidal dipole moment $T$ is the lowest moment of the third independent electromagnetic multipole family in addition to electrical and magnetic moments. The expression for the toroidal moment has the form [28]

$$\vec{T} = (10c)^{-1} \int (\vec{r}(\vec{r}\vec{j}) - 2r^2 \vec{j}) d\vec{r} \quad .$$

$(1.11)$

Here c is the light velocity, $\vec{j}$ is a current density. If $\vec{j}$ is a spin current, $\vec{j} = rot\vec{S}$, then $\vec{T}$ is a magnetic toroidal moment. A vortex spin structure in magnets may induce $\vec{T}$. The spin vortices like planar quadratic or hexagonal head-to-tail arrangements of spins carry a toroidal moment. The spontaneous toroidal moment exists in magnetoelectrics with nonzero antisymmetric components of the ME tensor $\alpha_{ik}$, and $\vec{T}$ is the vector dual to its antisymmetric part, $T_i^0 \sim \varepsilon_{ijk}\alpha_{jk}$. The simplest example of a system with the spontaneous toroidal moment is a two-sublattice AF, for which spontaneous electric polarization and magnetization are absent. Few ferroelectromagnets are simulteneously toroidics (known example is a nickel-iodine boracite $Ni_3B_7O_{13}I$ ). As the toroidal moment is a higher order moment in comparison with the electric and magnetic ones, the role of a toroidal moment is significant especially in crystals without spontaneous electric and magnetic orderings. The studying of ferrotoroidics began not long ago [29], and the subject of the present review is restricted by ferroelectromagnets only.

In 1994, Schmid [30] started calling ferroelectromagnets *"multiferroics"*. In multiferroic, according to the definition, two or more of ferroic properties (ferroelectric, ferromagnetic, ferroelastic or ferrotoroidic) are joined in the same phase. The term "multiferroic" used for ferroelectromagnet is hardly relevant as it requires additional information: one has to indicate which of the ordering subsystems coexists in every case. In some last publications the term "magnetoelectric multiferroic" was also used.

The primary term "ferroelectromagnet" is more informative and it is used in this paper although the term "multiferroic" is rather popular last time. Moreover, almost all publications about multiferroics are devoted to the studying of the electric and magnetic properties.

Last time the ME science develops extraordinarily quickly. Not only single-phase systems but also laminates and composites are investigated. The summary of the studying of the ME phenomena in composites is the subject, for example, of the monographs [31, 19]. The revival in the studying of the LMEE effect has been lighted in the recent review [19].

In this book mainly the single-phase ferroelectomagnets are considered because they are convenient for studying owing to their simpler structure in comparison with composite systems. The book is organized as follows. Firstly, a brief review of the main ME effects received from the first renascence up to the second renascence is presented. An impressive revival of ME studying of ferroelectromagnets at the beginning of the 21-st century is observed more detailed. In the conclusion some aspects of possible employments of ferroelectromagnets in techniques and engineering as well as the perspectives of the MEE sciences are discussed.

## II  The first magnetoelectric renascence (1961 – 1981)

### 2.1 *The nature of ME coupling*

First two decades the experimental data were accumulated, the versions of the nature of ME interaction and the possibilities of new interesting ME effects were discussed. A ferroelectromagnet differs from other materials in which, besides the ME effects induced by external fields,  the spontaneous ME effects due to strong electric and magnetic internal fields are also exhibited. This may give rise to the qualitatively new ME effects which do not have an analogue in other crystals. Ferroelectromagnets have an important advantage over magnetoelectrics and other crystals as the ME effects in these materials are the strongest.

In order to characterize the ME interactions, it is convenient to divide them into two parts, viz. the magnetoisotropic (exchange) and anisotropic parts.

The *exchange isotropic ME energy* describes the change in the exchange interaction due to electric polarization, as well as the effect of the spin-exchange field on the magnitude of the crystal polarization. As a rule, the unit cell of a ferroelectromagnet is quite complicated. The magnetic cations are separated by nonmagnetic ions, and the exchange interaction is indirect. During a FE transition, the lattice ions are displaced, thus changing the equilibrium distance between the magnetic ions and hence the exchange integral. Besides, the change in the electric field due to the Stark effect upon an FE transition in the crystal perturbs the wave functions of the magnetic electrons as well as the electrons of the intermediate atoms participating in the indirect exchange. Consequently, the overlapping of the electron wave functions changes, thus changing the value of the transport integral of magnetic electrons, as well as the strength of the indirect exchange interaction. Conversely, the change in the polarization due to magnetic ordering can be considered as a result of the Zeeman splitting of levels due to the exchange field. In turn, this splitting accompanied by the displacement of ions is a result of the electron-phonon interaction.

The *anisotropic ME energy* characterizes the change in the energy of magnetic anisotropy due to electric polarization and the effect of the anisotropic magnetic spin field on the magnitude of polarization. As a result of the FE ordering, the internal electric field causes a Stark splitting of the electron levels of the magnetic ions, changing the spin-orbit and spin-spin dipolar interactions, and hence the energy of the single- ion magnetic anisotropy as well. On the other hand, the magnetic anisotropy field changes the electron states of ions due to the Zeeman effect. Consequently, the electric dipole moment also changes.

Besides, the indirect interaction between electric and magnetic subsystems occurs through an *elastic* subsystem. The presence of an elastic subsystem in the crystal leads to the emergence of *electrostriction* which accompanies the FE ordering and changes the crystal size. Thus, in turn, changes the magnetic state due to *magnetosriction*, and vice versa.

Theoretical investigations of the ME interaction mechanisms were mainly qualitative that time. A phenomenological Landau's theory uses the free energy density which can be represented as a power series in the projections of the electric dipole moment $\vec{P}(\vec{r})$ and the magnetic moment $\vec{M}_s(\vec{r})$ ($s$ is the number of the magnetic sublattice). In general case, the ME part of the free energy may be written in the form:

$$F_{ME} = -\left( \Gamma_{ss'}^{j} P^{j} + \frac{1}{2} \gamma_{ss'}^{jj'} P^{j} P^{j'} \right) \vec{M}_s \vec{M}_{s'} - \left( \Lambda_{ss'}^{jll'} P^{j} + \frac{1}{2} \lambda_{ss'}^{jj'll'} P^{j} P^{j'} \right) M_s^l M_{s'}^{l'} + \dots$$

(2.1)

In this formula, $j$, $j'$, $l$, $l'$ are the indices of the projections. The first two terms in (2.1) are isotropic exchange ME energy, the last two terms describe an anisotropic ME coupling. All terms in (2.1) are invariants with respect to the crystal symmetry of nonordering state (paraphrase). Thus, the presence of specific terms in the expansion (2.1) depends on the symmetry of paraphrase of the crystal. Notice that for the description of multisublattice magnets, besides the crystal symmetry, it is necessary to take into account the transpositions of magnetic ions under the operations of the crystal symmetry [32]. Apparently it is connected with the concrete surroundings of the magnetic ions in the elementary cell.

The energy of the exchange ME interaction in ferroelectromagnet with a paraelectric phase without a central symmetry (the first term in (2.1) with the coefficient $\Gamma$) was obtained in [33]. The calculations of the exchange ME energy with the coefficient $\gamma$ in (2.1) [34, 35, 12] gave the value $\gamma \approx 10^{-7}$ dyne$^{-1}$ cm$^2$ for the parameter of the exchange ME interaction. For $P^2 \approx 10^9$ CGS units and $M^2 \approx 10^5 Gs^2$, the exchange ME energy is equal to $F_{ME}^{ex} \approx 10^7 \ erg \cdot cm^{-3}$.

The anisotropic ME terms in (2.1) with $s = s'$ describe the change of a single-ion magnetic anisotropy due to the presence of electric polarization. The second anisotropic term in (2.1) appears in the forth approximation of the perturbation theory: in the second order with respect to the spin-orbit interaction and the second order with respect to the Stark splitting [36, 37]. If the magnet has few magnetic sublattices, then the coupling of the spin-orbit energy with the energies of indirect exchange and crystal field leads to the anisotropic magnetic energy. The corresponding anisotropic ME energy is described by the terms with $s \neq s'$ in the free energy (2.1).

The contribution of the electromagnetic elastic interaction to the ME energy [36] can be estimated by putting $F_{ME}^{el} \sim Cu_E u_M$, where $C$ is the elastic constant, $u_E$ is the electrostriction, and $u_M$ is the magnetostriction. Assuming $C \approx 10^{12}$ ergcm$^{-3}$, $u_E \approx 10^{-3} - 10^{-2}$, and $u_M \approx 10^{-5} - 10^{-4}$, we get $F_{ME}^{el} \approx 10^4 - 10^6 \ erg \cdot cm^{-3}$. This quantity may assume values of the order of $10^7 \ erg \cdot cm^{-3}$ for compounds with rare-earth ions, where magnetostriction may attain the values $u_M \approx 10^{-3}$. In other words, the electromagnetic elastic energy may play a significant role in the materials with large values of electrostriction and magnetostriction.

The exchange ME interaction has electrostatic nature, while the anisotropic ME energy which appears from spin-orbital and magnetic dipole couplings is relativistic. Hence, in crystals where the exchange energy prevails over the anisotropic magnetic energy, the exchange ME energy considerably exceeds the anisotropic ME energy, $F_{ME}^{ex} >> F_{ME}^{an}$. This condition is implemented, for example, in the magnetoelectric $Cr_2O_3$ [38].

*2.2  Main experimental results. First colossal ME effects*

The first ferroelectromagnets were synthesized in the Soviet Union and the theory of these materials was also developed there. It is not surprising, therefore, that the largest number of publications appeared from the USSR. Swiss, French, and Japanese scientists made also a significant contribution to the synthesis and investigation of ferroelectromagnets.

The investigation of the nature and magnitude of the ME coupling presumes the existence of sufficiently perfect single crystal with a low electric conductivity. Only a few such ferroelectromagnets were known that time. Besides, the investigation of LMEE requires that the samples should not only be a single crystal, but also have preferably a single domain character, since the ME susceptibility of LMEE in 180º domains has the opposite sign. In spite of these difficulties, two colossal ME effects more were observed that time.

The ME effect is usually called "colossal" when the change of electric (magnetic) property due to magnetic (electric) influence is of order of ten or more per cent. It is possible, for example, in the ferroelectromagnets with improper FE transition where electric polarization is induced by magnetic ordering (see Sec.IV). The value of the spontaneous polarization in such ferroelectromagnets is usually small, by two or three orders smaller than in a good classical FE like $BaTiO_3$. In the improper FE of magnetic origin the temperature of FE ($T_c$) and magnetic ($T_m$) ordering coincide.

From the said above it follows that the presence of a weak FE subsystem in ferroelectromagnet with improper FE ordering of magnetic origin is favorable to the observation of colossal ME effects. The magnetic origin of electric polarization leads to the significant response of it on the change of magnetic state. Therefore, it is not surprising that firstly the colossal ME effects were observed in the nickel-iodine boracite where $T_c = T_m$.

**Ni-I boracite** ( $Ni_3B_7O_{13}I$ ) possesses FE and AWFM properties and belongs to the group of boracites with the general formula $M_3B_7O_{13}X$, where $M$ is a bivalent metal ion $M = Cr, Mn, Co, Fe, Cu,$ or $Ni$, and $X = Cl, Br$ or $I$. The ferroelectromagnetic boracites were synthesized and subsequently investigated extensively by Swiss physicists. In the $Ni - I$ boracite the FE and AF orderings with weak ferromagnetic moment arise below 64K. The vector $\vec{P}$ is directed along the [001] axis, while the spontaneous magnetic moment is in a plane perpendicular to $\vec{P}$. If the direction of the magnetic moment is taken as the y-axis, the magnetic point group for such a state is given by $m'm2'$.

**The first colossal ME effect** was discovered in 1966. Asher *et al.* [11] observed a clearly marked coupling between the directions of spontaneous polarization and magnetization in the $Ni - I$ boracite: a rotation of $\vec{P}$ through 180° is accompanied by a rotation of $\vec{M}_S$ through 90°, and vice versa. This phenomenon was explained from the point of view of the group theory [39], as well as thermodynamically [40]. Using the simplest model of two magnetic sublattices, it can be shown that the thermodynamic potential of the $Ni - I$ boracite contains the term $P_z (L_x^2 - L_y^2)$ ($\vec{L} = \vec{M}_1 - \vec{M}_2$ is AF vector) which gives rise to a polarization $P_z \sim (L_x^2 - L_y^2)$ upon an AF transition. This term also indicates that a rotation of $\vec{M}_S$ from $y$-axis to the $x$-axis, corresponding to the $L_x \to L_y$ transformation, reverses the sign of polarization. In the state $m'm2'$ under consideration, the nonzero components of ME susceptibility of LMEE are given by $\alpha_{yz}^{EM} = \alpha_{zy}^{ME}$ and $\alpha_{zy}^{EM} = \alpha_{yz}^{ME}$. In other words, the following relations hold:

$$p_z = \alpha_{zy}^{EM} H_y, \quad m_y = \alpha_{yz}^{ME} E_z \ ,$$
$$p_y = \alpha_{yz}^{EM} H_z, \quad m_z = \alpha_{zy}^{ME} E_y \ . \tag{2.2}$$

In these formulas, $\vec{p}(\vec{m})$ is the change in the electric (or magnetic) moment induced by a magnetic (electric) field. This change is much smaller than the spontaneous moment and is described by a term of the form $\vec{P}[\vec{L} \times \vec{M}]$ in the thermodynamic potential. If the magnetic field is directed along the $y$-axis ($y \parallel \vec{M}_S$), a polarization $p_z \sim L_x M_y \sim L_x H_y$ is induced, i.e. the ME susceptibility of the linear effect is given by $\alpha_{zy}^{EM} = \alpha \sim L_x$. The $p_z$ vs. $H_y$ dependence obtained experimentally in [11] during the remagnetization of the crystal in the direction $y \parallel \vec{M}_S$ has a butterfly-shaped ME hysteresis loop (Fig. 3). The initial state (segment 1) for $\vec{H} \parallel -y$ is a single domain state with $\vec{H} \parallel \vec{M}$ and $\alpha > 0$. A decrease and reorientation of the field, which reverses the direction of the magnetization without altering the direction of $L_x$, leave $\alpha$ unchanged $\alpha \sim L_x$ (segments 2 and 3). In a field which reverses the direction of $L_x$ ($H_c \approx 6kOe$), $\alpha$ changes the sign abruptly, and segment 4 corresponds to the values of $\alpha$ in stronger fields. A repeated reversal of magnetization is accompanied by a change in the ME signal along the segments $5 \rightarrow 6 \rightarrow 7 \rightarrow 8$. The value of $\alpha$ at 15K is $3.8 \times 10^{-4}$. The spontaneous electric polarization of the $Ni - I$ boracite is small, $P_S \cong 7.6 \times 10^{-2} \mu C / cm^2$ (for example, $BaTiO_3$ has $P_S \cong 26 \mu C / cm^2$).

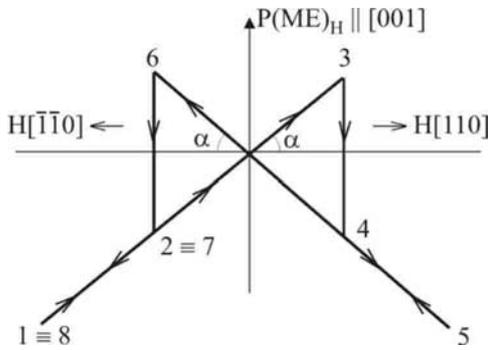

**Fig. 3**. ME hysteresis loop for Ni − I boracite at 46 K [11]. The magnetic field is directed along the ±[110] direction (in pseudocubic indices).

The second colossal ME effect in external magnetic field was discovered also in the $Ni - I$ boracite in 1978. Baturov $et$ $al.$ [23] found a strong dependence of the dielectric permittivity on the magnitude and direction of the magnetic field. The largest variation of $\varepsilon_{zz}(H)$ was observed in the vicinity of $T_c = T_m$, where a 90º rotation of the magnetic field $H = 12.5$ $kOe$ led to a 30% change in the value of $\varepsilon_{zz}$ as compared to its value at $H = 0$ (Fig.4).

Besides the mentioned (colossal) ME effects the evidences of a weak ME coupling were received. The LMEE was studied in other boracites [41 -43], in the FE – ferrimagnet $Fe_3O_4$ [44]. In the $Ni - Cl$ boracite, a peak in the ME susceptibility was revealed near 9K during a transition to a weak ferromagnetic state [45]. The magnetization induced by an alternating electric field was observed in the $Pb(Fe_{1/2}Nb_{1/2})O_3$ [46], $PbMn_2O_4$ [47], and $Fe - Br, Co - Cl, Co - I$, $Fe - Cl$ boracites [48, 49].

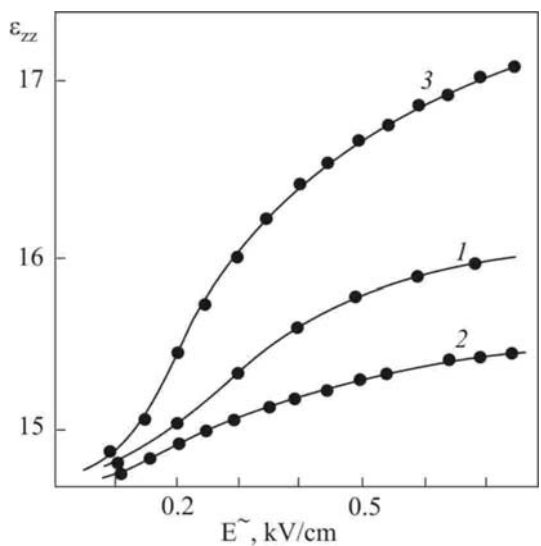

**Fig. 4**. Dependence of the permittivity of Ni–I boracite on the amplitude of the alternating electric field $E$ at 62 K in a magnetic field H= 12.5 kOe [23]: 1 – H=0; 2 - $\vec{H} \parallel \vec{M}_s$ ; 3 – $\vec{H} \perp \vec{M}_s$ .

A break in the temperature dependence of dielectric permittivity during magnetic ordering was at first observed in the $BiFeO_3$ [50, 51]. Further such a break was detected in the $Cu - I$ and $Cu - Br$ boracites [52], $BaMnF_4$ [53, 54], as well as in certain ferroelectromagnetic solid solutions.

In the AF- pyroelectric $BaMnF_4$ in the temperature dependence of the dielectric permittivity $\varepsilon$ (T) there was observed the break at the AF ordering temperature $T_m \cong 30K$ [53, 54]. Fox $et$ $al.$ [55] attributed this break to a weak ferromagnetism and an anisotropic ME energy $F_{anis} = \lambda_{ik} L_i M_k P^2$. However, it follows from the general theoretical concepts that a break in $\varepsilon(T_m)$ appears even in the isotropic case and does not impose any additional restrictions on the crystal symmetry. In any ferroelectric antiferromagnet there exists isotropic exchange ME energy of the form

$$F_{ex}^{ME} = \gamma P^2 L^2 . \qquad (2.3)$$

This energy considerably exceeds the relativistic anisotropic ME energy $F_{anis}$. The exchange ME energy causes a change in the dielectric permittivity

$$\Delta \varepsilon \sim \gamma L^2 \sim \gamma M_0^2 \sim \gamma (T_m - T)^{2\beta} \qquad ((T \le T_m) , \qquad (2.4)$$

where $M_0$ is the magnetic moment of the sublattice. The theoretically calculated value of $\Delta \varepsilon$ for $BaMnF_4$ at $T = 0$ with the help of the expression for the ME exchange interaction parameter $\gamma$ in terms of microscopic constants [35] led to the values which are in agreement with the experimentally observed results for $\Delta \varepsilon$ (Fig.5). The value of the magnetic critical index $\beta$, estimated from the $\varepsilon(T)$ dependence shown in Fig.5, is found to be of order of 0.35. Neutron diffraction measurements [56] yield the value $\beta \cong 0.32$.

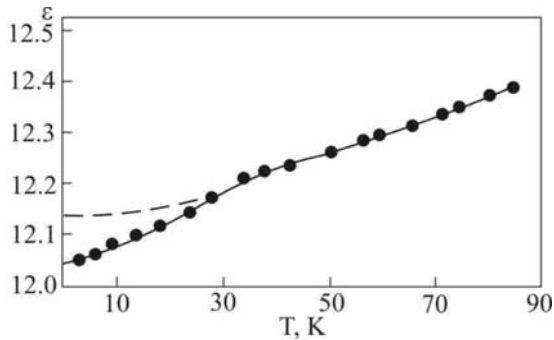

**Fig. 5**. Permitivity along the pyroelectric axis as a function of temperature in BaMnF₄ [54].

In fact, a break in the temperature dependence of the dielectric susceptibility (2.3), (2.4) is the result of the renormalization of the FE energy by the exchange ME energy. **In optical phenomena this ME effect was at first observed in the ferroelectromagnet** $BaCoF_4$ [57]. In optical spectrum the line with the frequency $\omega = 331 cm^{-1}$ at $T = 4.2$ K sharply decreases its frequency with the increasing of temperature and shows a break at the Neel temperature $T_N = 69.6$K (Fig 6). Due to the renormalization of electric polarization by isotropic exchange ME energy (2.3) the change of optical frequency at $T \le T_N$ is [58]

$$\frac{\Delta \omega}{\omega} = \frac{\gamma}{2} \chi^E L^2, \quad L^2 \sim (T_N - T)^{2\beta} . \qquad (2.5)$$

The estimation [58] of the value of the frequency break in $BaCoF_4$ showed a good agreement with the experimental data. The value of the critical index $\beta \approx 0.1$ is a typical for a layered magnet as $BaCoF_4$ is.

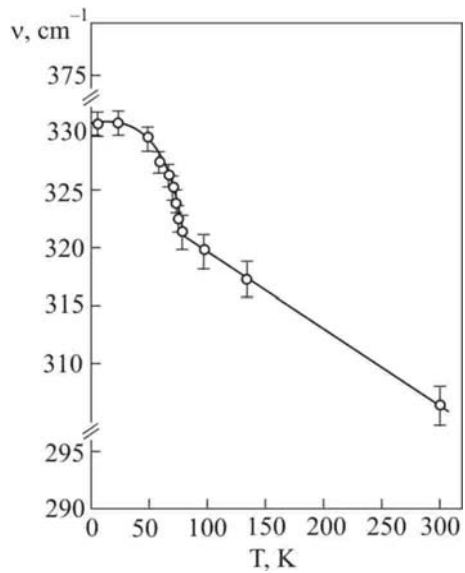

**Fig. 6.** *Adapted from* [57]. The temperature dependence of the spectral line in a combinative light diffusion spectrum of BaCoF₄ .

Among other ME measurements in ferroelectromagnets, the visual observation of narrow-band ferromagnetic domains inside the FE domains in the $Ni - Cl$1 and $Ni - Br$ boracites [64] and the observation of the maximum in the electric signal during a reversal of magnetic sublattices (spin-flop) in $BiFeO_3$ [65] are worth mentioning. The latter effect is in agreement with the theoretical concepts on the jump in electric polarization as a result of a spin flop [66].

*2.3 Theoretical predictions*

The discovery of ferroelectromagnets opened the perspectives of various interesting ME effects and inspired the appearance of many theoretical investigations of ME properties and their possible applications in sciences and technology.

### 2.3.1. *Spontaneous ME effects*

The theoretical investigations of the equilibrium properties of ferroelectromagnets have been carried out within the framework of the Landau theory of second-order phase transitions by using the expansion of the free energy in powers of the FE and magnetic moments. The classes and special groups allowing for ferroelectromagnetic structures with different mutual orientations of $\vec{P}$ and $\vec{M}$ were given in [59-63]. The cases when the crystal symmetry permits exchange ME invariants which are linear in polarization were indicated in [68].

The studying of the second-order thermal phase transitions in a ferroelectromagnet within the framework of the Landau theory [36, 18, 67] led to the following results: (1) the "twinning law" is satisfied for $T = T_c < T_m$, just as in the case of an ordinary FE transition; (2) the temperature dependence of the magnetic (FE) susceptibility shows a break at the temperature of the FE (magnetic) transition; and (3) the magnetic (FE) susceptibility undergoes a jump at the temperature of the FE (magnetic) transition.

The break in the temperature dependence of the susceptibility is a first-order effect in the ME interaction, while the jump in the susceptibility is a second-order effect. Owing to a break in the temperature dependence $\chi^E(T)$ at $T = T_m$ the dielectric susceptibility below $T_m$ will differ from the value obtained by extrapolation of the temperature dependence $\chi^E(T)$ from the paramagnetic phase to the region $T < T_m$. This difference, which we denote $\Delta\chi^E$, is proportional to the square of the appearing magnetic order parameter $M$ :

$$\Delta\chi^E \propto M^2(T) \qquad (\text{T} \le T_m < T_c).\tag{2.6}$$

Similarly,

$$\Delta\chi^M \propto P^2(T) \qquad (T \le T_c < T_m).\tag{2.7}$$

The sign of the $\Delta\chi$ in (2.6) and (2.7) depends on the sign of the ME interaction constant and can be either positive or negative.

As mentioned above, the break is a first- order effect with respect to the ME interaction. Hence, we should primary expect an experimental confirmation of this effect rather than the weaker susceptibility jump which is of the second order with respect to the interaction. Indeed, the ferroelectromagnets show a break rather than a jump in the dielectric permittivity at the magnetic transition temperature (see Sec.2.2). Investigations of the temperature dependence of $\Delta\chi^E(\Delta\chi^M)$ at $T \le T_m(T_c\}$ can give information about the critical index of the emergent order parameter.

The predicted temperature dependence of ME susceptibility near phase transition temperatures is $\chi^{ME} \propto (T_c - T)^{-1/2}$ in the vicinity of $T_c$ $(T_c < T_m)$ [18, 67], and $\chi^{ME} \propto (T_m - T)^{k-1}$ for $T_c > T_m$ [68]. The index $k$ is the exponent in the temperature dependence of the magnetization $\vec{m} = \sum_s \vec{M}_s$ (s is the number of the magnetic sublattice) and $m \propto (T_m - T)^k$. The exponent $k$ is equal to $1/2$ for a ferromagnetic transition or transition to a weakly ferromagnetic phase whose magnetization is due to the second-order invariants in the free energy. In this case, the ME susceptibilities has an anomalously high value in the vicinity of the second phase transition temperature $T_m$ : $\chi^{ME} \propto (T_m - T)^{-1/2}$. If, however, the small ferromagnetic moment is due to the fourth-order invariants, we have $k =3/2$, and $\chi^{ME} \propto (T_m - T)^{1/2}$.

Generally speaking, the electric and the magnetic transition temperatures do not coincide, $T_c \ne T_m$. But if the FE transition is improper and induced by the magnetic transition, these two temperatures coincide. In this case, the ME energy contains the terms that are even in the magnetic moments and odd in the components of a polarization vector. This means that $\vec{P}$ may become of the order of $\vec{M}^2$ at the temperature below $T_m$ (improper FE transition). Such a transition takes place in the nickel-iodine boracite.

Thermodynamical considerations of ferroelectromagnets of that time were restricted by homogeneous magnetic and electric states. The terms with space derivatives in the expansion of ME free energy (2.1) were absent. The induction of the FE ordering by a magnetic subsystem was expected only in the crystals without the centre of the symmetry (the terms with a first degree of $P$ in (2.1)).

The possibility of an improper FE transition in magnets with helical-type magnetic ordering was at first formulated in [12]. This possibility exists in magnet (and in the case of centrosymmetrical paramagnetic phase), if its ME energy contains terms of the type [12]

$$P_i M_S^l (\partial M_{S'}^{l'} / \partial x_k).$$ (2.8)

Here $i, k, l, l'$ indicate the projections, and $s, s'$ are the numbers of the magnetic sublattices.

It was pointed to the spiral magnet $Cr_2BeO_{44}$ with centrally-symmetric $D_{2h}^{16}$ paraelectric phase where spontaneous polarization appeared simultaneously with AF helical magnetic ordering at $T_c = T_m = 28\,\mathrm{K}$ [69].

The ME energy of the type (2.8) was later [70] called inhomogeneous ME energy and used for the description of ME effects in the magnetic domain walls. The case of $Cr_2BeO_4$ was in detail considered in 1985 [71]. Thus, $Cr_2BeO_4$ is the first discovered ferroelectromagnet where the electric polarization is induced by spiral magnetic structure.

### 2.3.2 Phase transitions in external fields

One of the manifestations of the ME interaction is a shift in the magnetic (electric) transition temperature due to the electric (magnetic) field. In ferroelectromagnets with $T_m < T_c$ the shift in the magnetic transition temperature $\Delta T_m$ is proportional to the electric field strength [36], i.e., $\Delta T_m \propto E$, irrespective of the crystal symmetry. The shift in the temperature of the FE ordering in ferroelectromagnets with $T_c < T_m$ is proportional to the first degree of magnetic field, i.e. $\Delta T_c \propto H$, if the crystal symmetry allows a linear ME effect (in other words, if ferroelectromagnet is simultaneously a magnetoelectric). In the ferroelectromagnet of any symmetry, the shift $\Delta T_c$ is proportional to the square of the magnetic field $H$. Thus, if we neglect the paraprocesses, the exchange ME interaction in FE-AF leads to the following change in the FE transition temperature in a magnetic field:

$$\Delta T_c = T_c - T_c^0 = \gamma H^2 C \pi^{-1} (2\Delta_0)^{-2}. \tag{2.9}$$

Here $T_c^0$ is the FE transition temperature in the absence of a magnetic field, $\Delta_0$ is the exchange constant, and $C$ is the Curie-Weiss constant.

The ME interaction may change the order of a phase transition, or induce additional phase transitions in the magnetic and FE subsystems. Phenomenological considerations of possible ME effects in uniaxial FE-ferromagnet and FE-antiferromagnet in the magnetic field for any values of the parameters in the free energy were fulfilled in [72 -74 ]. Numerous ME phase diagrams (16 for FE- ferromagnet and 40 for FE-AF) showed the possibility of many interesting ME effects. For instance, the possibilities of the switching on and the switching off of electric polarization by magnetic field (Fig.7) exist. In FE- ferromagnet at $T < T_m, T_c$ when $\vec{M}_s^2$ is a constant, only a weak anisotropic relativistic ME energy couples electric and magnetic subsystems of crystal. In contrast to this in the magnets with several magnetic sublattices (AF, ferrites) a strong exchange ME interaction connects spins and electric polarization at low temperatures. Below we formulate some results of the works [72-74] for uniaxial FE-AF with two magnetic sublattices. The free energy of the crystal is following:

$$F = \Delta(\vec{M}_1 \vec{M}_2) - \vec{H}(\vec{M}_1 + \vec{M}_2) - \frac{1}{2}\beta(M_{1z}^2 + M_{2z}^2) - \beta_1 M_{1z} M_{2z} -$$
$$- \frac{1}{2}\kappa P^2 + \frac{1}{4}\delta P^4 - \frac{1}{2}\gamma P^2 (\vec{M}_1 \vec{M}_2). \tag{2.10}$$

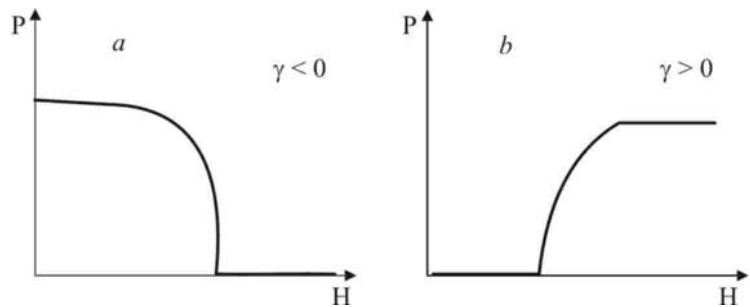

**Fig. 7**. Possible dependences of electric polarization on magnetic field in ferroelectromagnet.

In (2.10) we took into account only the strongest exchange ME energy (the last term). It is important that this energy is a scalar and, therefore, exists in the crystal of any symmetry. The magnetic field $H$ and electric polarization $P$ are directed along the easy axis $z$, the exchange constant $\Delta$ and the constant $\delta$ are positive. Other restrictions on the parameters in (2.10) for the received in [66, 72-76] ME phase states are absent.

The value and character of the ME effects essentially depend on the sign exchange ME constant $\gamma$. If $\gamma < 0$, and $\beta - \beta_1 < 0$, $2\gamma^2 M_0^2 \delta^{-1} < \Delta$, $\kappa < -\gamma M_0^2$ ($M_0$ is the magnetic moment of the sublattice), then the magnetic field may switch off the electric polarization (Fig.7 $a$). The condition $\kappa < -\gamma M_0^2$ means not so large values of a spontaneous polarization $P_s^2 = \kappa \delta^{-1}$. Thus, it is the condition $P_s^2 < \gamma M_0^2 \delta^{-1}$. For the perovskite compounds we can put $\delta \sim 10^{-11} cm^2 dyne^{-1}$ [77]. Taking the values $\gamma \sim 10^{-7} cm^2 dyne^{-1}$, $M_0 \sim 10^2 Gs$ we receive the condition $P_s < 10^4$ CGS units. This condition is not strong and may be realized, for instance, in the improper FE and near the FE transition temperature.

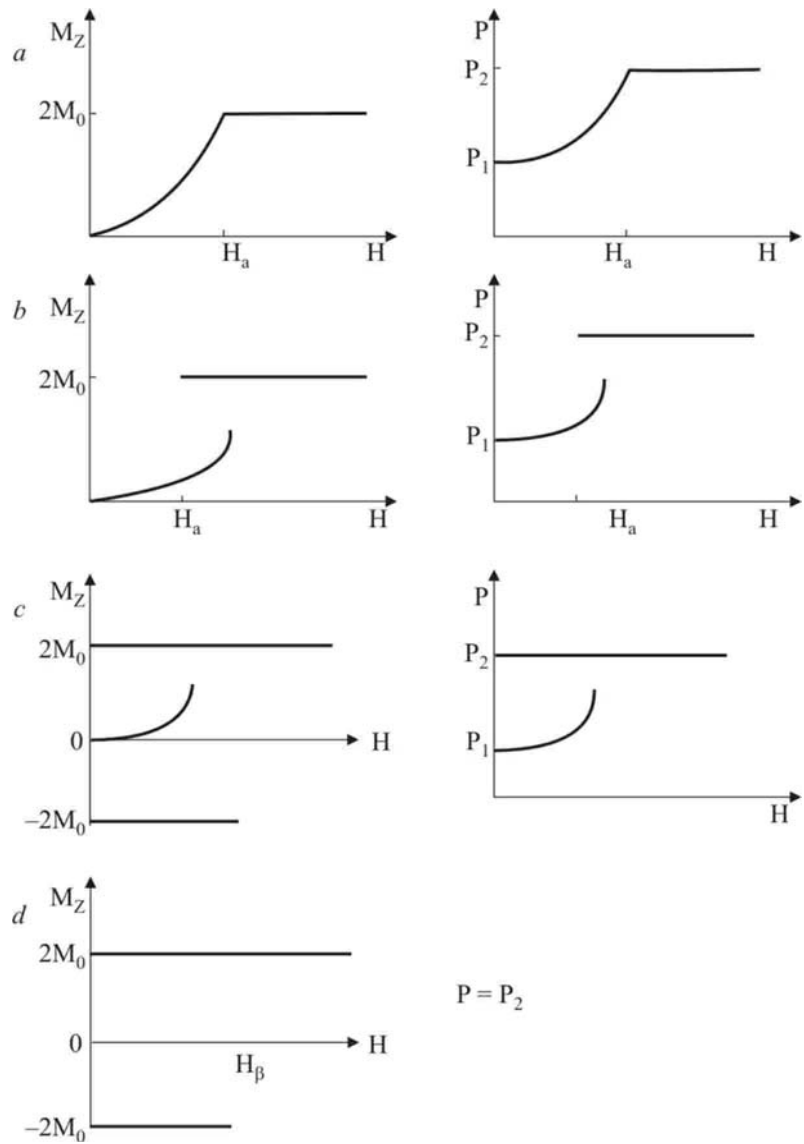

**Fig. 8**. Dependences of the electric polarization P and magnetization $M_z = M_{1z} + M_{2z}$ of uniaxial FE-AF on magnetic field $H_z$ in the case $\gamma > 0$, $3\gamma^2 M_0^2 \delta^{-1} < \Delta$, $\beta - \beta_1 < 0$ for different values of the parameter $\kappa$.

The case $\gamma > 0$ is much more rich in the ME effects [76]. For example, in the case $\gamma > 0, \beta - \beta_1 < 0, \ 3\gamma^2 M_0^2 \delta^{-1} < \Delta, \quad \beta + \beta_1 > 0$, the evolution of the states $a \to d$ with the increasing the FE parameter $\kappa$ is shown in Fig.8. Here $H_a$ is a spin-flip field, $P_{1,2}^2 = (\kappa \mp \gamma M_0^2)\delta^{-1}$, $H_\beta = (\beta + \beta_1)M_0$ is a field of magnetic anisotropy. The $a-$ state exists if $\gamma M_0^2 \leq \kappa \leq \delta(2\Delta - \beta - \beta_1)\gamma^{-1} - 5\gamma M_0^2$. It is seen that the continuous increasing $\kappa$ leads to the change of the second-order spin-flip transition at $H = H_a$ by the first order ($b-$state) to the additional phase transition of the first kind in magnetic field in FE subsystem and the appearance of a new phase ($b-$ and $c-$states). As a result, the AF state is replaced by the ferromagnetic (FM) $d-$state if $\kappa \geq \gamma M_0^2 + \delta(2\Delta - \beta - \beta_1)\gamma^{-1}$, i.e., if $\kappa \geq 10^{-3}$. If we use the previous values of the constants $\gamma, \delta, M_0$ and consider $\Delta \sim 10^2$, then the temperature transition $AF \to FM$ due to the exchange ME coupling is possible if electric polarization is sufficiently large, $P \sim 10^4 - 10^5 \ CGS$ units (of order of polarization in $BaTiO_3$).

If $\gamma > 0, \beta - \beta_1 > 0, \beta > 0$, the spontaneous AF vector is directed along the $z-$axis and it is parallel to the direction of the magnetic field. In this case the first-order phase transition (spin-flop) takes place at $H = H_{sf}$ in the magnetic subsystem (Fig.9). As it is seen in Fig.9, the electric polarization also has a jump at a spin-flop transition [66]. The increasing of the value $\kappa$ decreases the value $H_{sf}$, and for $\kappa \sim \Delta\delta\gamma^{-1}$ the spin flop disappears. The transition from AF to FM state in the magnetic field occurs as the first-kind transition of a metamagnetic type without the intermediate spin-flop phase.

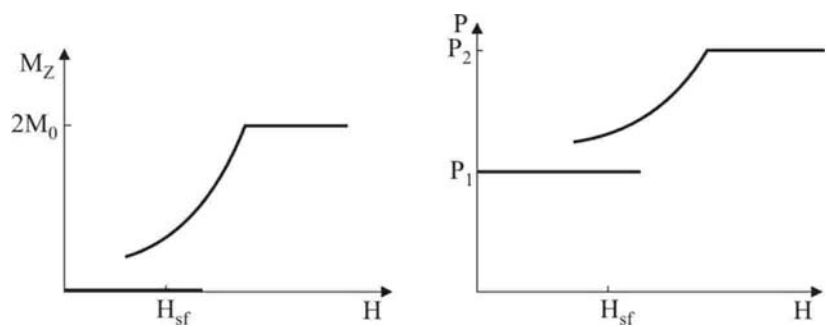

**Fig. 9**. ME states of uniaxial FE-AF for $\gamma > 0, \ 3\gamma^2 M_0^2 \delta^{-1} < \Delta, \ \beta > 0, \ \beta - \beta_1 > 0$.

Besides the pointed ME effects in the FE-AF , the theory also predicts other ME effects, for example, the possibilities of an "easy magnetic axis – easy magnetic plane" transition due to the influence of the electric polarization; the appearance a tetracritical point as well as a critical point of the second-order transition in the field-temperature phase diagram [76].

### 2.3.3 High-frequency properties

At low temperatures ( $T << T_c, T_m$ ), small oscillations of electric and magnetic moments relative to equilibrium positions propagate in the form of spin waves and polarization waves, which are coupled by the ME interaction. The quanta of these coupled oscillations were called ferroelectromagnons [77]. Ferroelectromagnons exist in a ferroelectromagnetic phase of the crystal with the coexisting static electric and magnetic moments. As a result, ferroelectromagnons are sensitive to the electric as well as to the magnetic fields.

The spin waves are coupled very weakly with the polarization waves in FE –FM. Here, the ME interaction takes place through the relativistic interactions, and the FM resonant frequency is much smaller than the electro-dipole optical frequency [78 – 80].

In the ferroelectromagnets with two (or more) magnetic sublattices, the coupling between the spin waves and the polarization waves can be stronger due to significantly stronger ME exchange interaction [81]. The analysis showed that in FE-AF the ME coupling of spin and polarization excitations exists only in the case of a noncollinear orientation of spins in the ground state. In uniaxial FE-AF without weak ferromagnetism, the noncollinear orientation of spins may be caused by an external magnetic field which controls the coupling of waves. In this case the relative spin frequency variation due to ME interaction is given by $(\Delta \omega / \omega) \sim (H / H_0)^2$, $H_0$ being the exchange field.

A ferroelectric antiferromagnet with a weak ferromagnetism (AWFM) is subjected to the Dzyaloshinskii relativistic ME interaction, for example, with energy in the form $\Lambda [\vec{M}_1 \times \vec{M}_2] \cdot \vec{P}$ [82] or $\lambda_i P^2 [\vec{M}_1 \times \vec{M}_2]_i$. This interaction couples the polarization oscillations with the upper spin branch even in the absence of the magnetic field [83].

The strongest coupling between the spin and polarization waves should be expected in the FE-ferrimagnet [84, 85]. Here, the polarization oscillations are connected by the exchange interaction with the lower as well as upper branches of the spin waves. The magnitude of this coupling is larger than in FE-AF by a factor of $\Delta / \beta \approx 10^2$. This strengthening of the coupling is due to the fact that the interacting branches become closer and lie in the same IR region. This means that a resonant interaction between the FE wave and the upper spin wave is possible, for example, when they approach one another upon a decrease in FE mode near the FE transition temperature.

The spin and polarization waves in a ferroelectromagnet interact not only with each other, but also with other elementary excitations in a crystal. Owing to the electro- and magnetoelastic interactions, a coupling arises between the ferroelectromagnons and acoustic phonons [86, 87]. The former also interact with the electromagnetic waves due to electric and magnetic dipole interactions [86, 88].

The presence of electric and magnetic coupling subsystems in a ferroelectromagnet means the possibility of ME resonances in alternating external fields, i.e. a resonant absorption of magnetic energy at the FE frequency and a resonant absorption of electric energy at the spin wave frequency [77, 80, 81, 84, 85, 89]. The maximum FE resonances are expected in strong exchange fields or when the spin frequency $\omega_s$ and electro-dipole frequency $\omega_p$ are close enough for resonance. In the FE- ferrimagnet with $T_m > T_c$ (for example, $Fe_3O_4$) and $\omega_p > \omega_s$, the FE frequency can approach the spin frequency, for instance, in the vicinity of the Curie temperature $T_c$, where the frequency of the soft FE mode decreases considerably [84, 85]. Near the resonance the amount of the electric field energy absorbed by the spin waves is of the same order as energy absorbed by the FE oscillations.

Besides the resonance effects, various nonlinear processes were theoretically investigated for ferroelectromagnets in alternating external fields. The possibility of the ME frequency doubling was indicated, viz. the onset of polarization (magnetization} oscillations at twice the frequency of the magnetic (electric) field [33, 90]. Various methods of investigating the spectrum of elementary excitations of a ferroelectromagnet by creating instability in it with the help of external fields were considered. The parametric excitation of spin waves in a FE-AF by an alternating electric field seems to be the most effective approach [91, 92]. Owing to the ME exchange interaction [92], the threshold field for exciting spin waves in the FE-AF is small and can be easily attained experimentally. The excitation of spin waves of the frequency $\omega = \omega_{1s} + \omega_{2s}$ occurs in the condition of a nonresonance pumping of the polarization waves by electric field with the frequency $\omega << \omega_p$. The threshold field $e_{th}$ is defined by the expression

$$e_{th} = \frac{2\eta_s}{\gamma} \frac{\delta P_0}{\Delta} \left( \frac{\omega}{gM_0} \right)^2, \qquad (2.11)$$

where $g$ is the gyromagnetic ratio, and $\eta_s$ is the relative decrement of damping of spin waves. Assuming $(\omega/gM_0) \cong 10$, $.\eta_s \cong 10^{-2}$, $\gamma \cong 10^{-7} dyne^{-1}cm^2$, $\delta \cong 10^{-11} dyne^{-1}cm^2$, $P_0 \cong 10^4 - 10^5 CGS$ units, and the exchange constant $\Delta \cong 10^2$, we get $e_{th} \cong 10$ V/cm. This value is two or three orders lower than the threshold field induced by the Dzyaloshinskii ME interaction [91]. The values of the threshold excitation fields in FE-FM [93] and FE ferrimagnet are also several orders higher than those obtained from formula (2.11).

The frequency of the alternating electric field in the case of parametric excitation (2.11) is of order of spin frequencies. Under certain conditions, an electric field with a much higher frequency can also excite spin waves, although much stronger excitation fields are required for this purpose. A rapidly oscillating electric field which varies slowly in the space can excite spin waves in a FE-AF at values of the field exceeding the threshold value (2.11) by several orders of magnitude [94, 95].

Ferroelectromagnons can also be excited by an electric current in semiconductors [82] or by an electron beam passing through a dielectric in a narrow cylindrical channel [96]. In contrary to the similar way of the excitation of spin waves in magnets [97] , this method of excitation in ferroelectromagnets is electrostatic, and the rate of growth of the waves does not contain the small factor $(v/c)^{2/3}$, where $v$ and $c$ are the velocity of spin wave and the light, respectively. Both spin and FE oscillations are excited. However, this method cannot be used to excite the soft FE mode which is characterized by the absence of an electrostatic field.

### 2.3.4  *Possible applications*

The above review showed that the ME researches in the "first renaissance" period only began and were mainly qualitative in nature. The experimental results in the studying of ME effects in ferroelectromagnets were not numerous, theoretical investigations prevailed. However the experimental proof was obtained for the coupling of polarization and magnetization and for the possibility of controlling the magnetization by an electric field or the electric polarization by a magnetic field. A significant ME interaction was also observed in heterophase systems [98]. Some perspectives of scientific and practical uses of ferroelectromagnets were already visible.

The dependence of the ME susceptibility on the orientation of the moments in the domains made it possible to exercise a ME control over the spin- and FE states. An investigation of the temperature and field dependence of the ME susceptibility may provide information on the orientation (spin-flop, spin-flip) and thermal phase transitions. The magnetic critical index can be determined from the temperature dependence of the dielectric permittivity. Similarly, the FE index can be determined from the temperature dependence of the magnetic susceptibility.

In order to use ferroelectromagnets for practical purposes, it is desirable to have compounds with low losses, low electric conductivity, a considerable magnetic moment and ME susceptibility, as well as sufficiently high values of the electric and magnetic transition temperatures (above room temperature). There were no ferroelectromagnets satisfying these requirements at that time.

The devices based on ferroelectromagnets can be divided into three types: (1) devices using the FE or magnetic properties separately, (2) devices which simultaneously employ the FE and magnetic properties, but without any ME interaction, and (3) devices whose action is based on ME effects. The first type of devices is quite common.

The review by Wood and Austin [99] contains a detailed table of possible applications of ME crystals as well as the characteristics of 15 different devices whose operation range varies from audio to optical frequencies and which include modulators, phase inverters, switches, rectifiers, stabilizers, etc.

The application of the ME interaction (devices of the third type), for example, switching or modulation of electric polarization by a magnetic field, makes it possible to obtain a magnetically operated optical device in the visible and infrared (IR) spectral regions. The operation of this device is based on the variation of the linear birefringence by changing the polarization under the action of the magnetic field. The magnetic field required to switch the polarization may not be very high and is just a few kilooersteds for $Ni-I$ boracite. Usual electric type of $\vec{P}$-switching would have required strong electric fields of the order of several kilovolts per centimeter. Samples in the form of thin plates are required to produce such fields, while in the magnetic method of polarization switching, massive samples can be used.

The predicted ease with which the spin waves can be parametrically excited by an alternating electric field in FE-AF makes it possible to use them for creating UHF spin wave oscillators in spin-wave electronics. FE ferrimagnets also seem to be effective for inducing spin waves in the IR region with the help of an electric field. Nonlinear ME interaction, which is manifested especially strongly near the transition temperature, can be used to produce a nonlinear ME optical device for, say, frequency doubling.

Ferroelectromagnets can also be used as thin-film wave-guides in integral optics and fibre communication technology.

The preferable using for measurements and practical applications of ferroelectromagnets with close temperatures of magnetic ($T_m$) and FE ($T_c$) transition was predicted [12]. The value of the ME susceptibility is proportional to the magnetic susceptibility as well as the dielectric susceptibility which are strongest at their phase transition temperatures.

### III. Between the first and second renascences (1981 - 2001)

*3.1 Preface*

That time in the ME investigations can be characterized as a period of accumulation, expansion and deepening of ME knowledge. Investigations of ME effects presume the existence of sufficiently good single crystals with low electric conductivity. At the beginning of 90's there had been known about 50 ferroelectromagnets (see Table 1 in [12]), and the synthesis of new ferroelectromagnets went on vigorously. Most ME measurements were made on the ferroelectromagnets where electric $T_c$ and magnetic $T_m$ ordering temperatures were not the same. The observed ME effects were weak, no new "colossal" ME effects were discovered.

As before, the LMEE was a preferable subject of the ME researches, although the nonlinear ME effects were also measured. The studying of the LMEE required a real domain structure to be taken into account, since the ME susceptibility of the LMEE in 180° domains had the opposite sign. A preferential orientation of the moments in crystals was created by applying an electric and magnetic fields near the phase transition temperatures. Naturally, the interest to the studying of the domain structures of ferroelectromagnets had increased. The ME investigations required the measurement techniques development.

A large interval between the first conference MEIPIC-1 (1973) and the second MEIPIC-2 (1993) says about a decline in the ME sciences, which was mainly stipulated by the weakness of the observed ME effects. However, the ME investigations continued and after the third conference MEIPIC-3 (1997) the activity in the ME researches appreciably increased (see Fig. 2). The development of a microscopic theory of ME phenomena promoted the better understanding of the nature of ME coupling. The discovery of a novel method for studying AF domains by nonlinear spectroscopy was the most considerable contribution to the ME measurement techniques. This discovery became the prerequisite for the future revival of ME sciences.

*3.2 ME measurements*

*3.2.1 Spontaneous ME effects*

An important aspect of experimental investigations is a verification of the previous results and their more precise definition. The measurement of the temperature dependence of dielectric permeability along a pyroelectric axis $\varepsilon_a$ in FE-AF $BaMnF_4$ [100] did not show the presence of weak ferromagnetism. The temperature dependence of $\varepsilon_a$ below $T_N$ was in a good agreement with the temperature dependence of the correlation function $\langle \vec{S}_j \vec{S}_{j+1} \rangle$ that confirmed the exchange nature [35] of the observed break in the $\varepsilon_a(T_N)$ in the contrast with the interpretation [55]. A small maximum in the dielectric constant $\varepsilon_a(T)$ at $T < T_N$ was also observed [100]. The magnitude and the temperature of this anomaly depended on the external magnetic field. The authors supposed that the observed maximum was a result of a jump of $\varepsilon_a(T)$ (the second order effect with respect to ME interaction) predicted by the thermodynamic theory [12].

A similar anomaly (break and small maximum) in the temperature dependence of dielectric permeability below the Neel temperature was detected in the FE-AF hexagonal manganite $YMnO_3$ [101].

The evidence of simultaneous appearance of electric and magnetic ordering in $EuMn_2O_5$ became especially important experimental result of [102]. The $EuMn_2O_5$ belongs to the family of the rare-earth manganates $RMn_2O_5$ ( $R = Nd - Lu, Y$ or $Bi$ ) with orthorhombic perovskite structure (space group $Pbam$ ). Previous neutron diffraction studies revealed an incommensurate AF ordering of $Mn^{3+}$ and $Mn^{4+}$ moments below $T_N$ of about 40 K in $RMn_2O_5$ . Sanina *et al.*

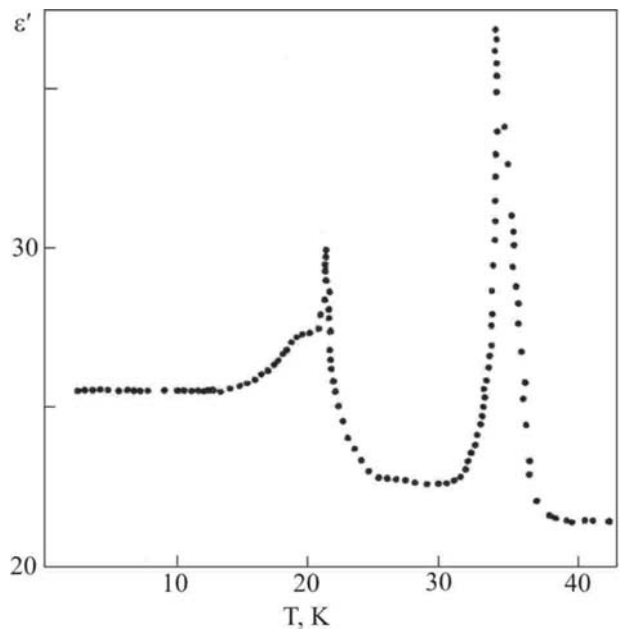

[102] were the first to observe an anomaly in the temperature dependence of dielectric permeability of $EuMn_2O_5$ at AF ordering (Fig.10). It meant a simultaneous appearance of magnetic and electric ordering in the crystal. The correlation of magnetic and dielectric properties in the $EuMn_2O_5$ observed in [103] confirmed this conclusion. The second anomaly in the dielectric constant in Fig.10 at $T$ =22 K was explained by a possible heterogeneity of magnetic structure [102].

**Fig. 10.** *Adapted from* [102]. Real part of the dielectric permittivity $\varepsilon'$ as a function of temperature in $EuMn_2O_5$ .

Systematic study of the $RMn_2O_5$ started when the first ferroelectromagnet

$EuMn_2O_5$ from these rare-earth manganites was discovered. The FE ordering was also observed in the $YMn_2O_5$ [104], $YbMn_2O_5$ [105], $ErMn_2O_5$ [106] and in almost all $RMn_2O_5$ [107,108]. The value of electric polarization in this **new family of ferroelectromagnets**, $P_s \sim 10nC/cm^2$, is smaller than in the $Ni-I$ boracite. The temperature dependence $P_s(T)$ may be not monotonic. For example, in $EuMn_2O_5$ the electric polarization changes the sign at low temperature [109]. Although the spontaneous polarization in $RMn_2O_5$ is very small, the colossal ME effects will be further discovered in this class of crystals.

**New ferroelectromagnets**, solid solutions $Bi_{1-x}La_xFeO_3$ with FE and AF ordering ($0.06 \leq x \leq 0.24$) [110] and FE ferrimagnetic spinel $Co_{3-x}Mn_xO_4$ ($x =1.25$) [111] were synthesized and studied.

Most ferroelectromagnets known that time had the AF magnetic ordering. There were only a few ferroelectromagnets with FM or ferrimagnetic (FIM) properties (see Table 1 in [12]). More significant ME effects were expected in FM and FIM magnets since they have considerably larger value of magnetic susceptibility than AF has. The scientists sought the ways to receive ferroelectromagnets with FM and FIM ordering. FE and FIM properties were discovered in the $LuFe_2O_4$ [112]. This **new ferroeletromagnet** has ordering at high temperatures, $T_c =320$ K, $T_m$ =220 K, that is important for possible future uses. Charge and spin ordering process in the mixed-valence system $LuFe_2O_4$ was experimentally and theoretically studied [113]. A theoretical model was based on the assumption of the localization of a charge-density wave at the $Fe$ -sites and the interchange interaction between the neighboring layers.

### 3.2.2 Induced ME effects

**Static ME effect** in a constant magnetic field was studied in the FE-AF boracite $Cr_3B_7O_{13}Cl$ [114]. It was revealed that the ME effect in this compound is characterized essentially by the nonlinear quadratic term in (1.2) with the coefficient $\beta_{333} = 1.5 \times 10^{-18} s/A$ at 4.2 K. A linear component of the ME effect was found to exist with the coefficient $\alpha_{33}$ of the order of $3.6 \times 10^{-5}$, rather small compared with those of other boracites. The value of the electric polarization in the $Cr-Cl$ boracite $P_s \cong 2.5 \mu Ccm^{-2}$. The results of the measurements permit to suppose the existence of a very weak spontaneous magnetization parallel to the twofold polar axis.

Rare-earth manganese oxides $RMn_2O_5$ revealed many temperature and field phase transitions in ferroelectromagnetic phase [115]. A clear LMEE with the coefficient $\alpha \sim 10^{-3} - 10^{-2}$ was detected only in $TbMn_2O_5$, in the compounds with $R = Y, Eu, Ho$ and $Gd$ the values of $\alpha$ were smaller than $5 \times 10^{-6}$. The coefficients of the nonlinear quadratic ME effect $\beta \sim 10^{-15} s / A$ for $TbMn_2O_5$, and $\beta \sim 10^{-17} - 10^{-16} s / A$ for other manganese oxides. Large values of the coefficients $\beta$ in the $RMn_2O_5$ compared with those in the boracites testify to a stronger ME coupling in this new class of ferroelectromagnets.

The phase transitions driven by a magnetic field were observed in the $RMn_2O_5$ ( $R = Er, Yb, Tm$ ). Anomaly dependence of a dielectric constant of the $ErMn_2O_5$ on the magnetic field at low temperature was interpretated as a structure transition [116].

The magnetic fields used in [115, 116] were not so large. A stronger magnetic field led to the spin-flop transition from one incommensurate structure to another one in the $BiFeO_3$ [117]. In a strong magnetic field $H_c$ of order of exchange one (hundreds $kOe$ ) in the $EuMn_2O_5$ [118] and the $BiFeO_3$ [119], the first-kind phase transitions from an incommensurate to a homogeneous magnetic structure at $H = H_c$ were observed. In a homogeneous phase the linear dependence of electric polarization on magnetic field was noticed at $H > H_c$. It was declared by the authors as a linear ME effect. However, such an interpretation is in contradiction with the definition of the LMEE as the linear dependence of $P(H)$ in a weak field (1.5).

Among ferroelectromagnets, a particular place is taken by the magnetite $Fe_3O_4$ with a high ferrimagnetic ordering temperature ( $T_m \sim 500K$ ), where the LMEE and nonlinear ME effects were earlier observed below the Verwey temperature $T_v \sim 120K$ [120]. The FE properties of $Fe_3O_4$ were apparently established by the observation of FE hysteresis loop below $T_v$ [121]. The value of a spontaneous polarization in $Fe_3O_4$ was determined as $P_s \sim 1 \mu C cm^{-2}$, which is only ten times smaller than the electric polarization in a classical FE $BaTiO_3$. The $Fe_3O_4$ being the first magnet known to the mankind also happened to be ferroelectromagnet!

**Optical studying** of magnets and ferroelectromagnets is one of the most sensitive and exact instrumental methods. It has a ME nature: the alternating electric field of electromagnetic wave interacts with the spins of the magnet changing their state. Firstly the evidences of the ME interaction were found in the linear optics. So, the birefringence of the FE-AF $BiFeO_3$ was measured at room temperature in the rhombohedral phase [122]. The Neel temperature was found for the first time optically, and a critical exponent was derived near $T_N$.

*The first photoinduced ferroelectromagnet* was received by a strong optical pumping of the AF $EuCrO_3$ at $T \leq 80K < T_N$ [123]. The AF ordering in this crystal is created by the spins of chromium. Optical pumping changed the state of the rare-earth ions of $Eu^{3+}(^7F_0 \rightarrow ^7F_1)$. As a result, a homogeneous ordering of spins of $Eu^{3+}$ and electric dipoles appeared.

Optical studying of ferroelectromagnets has been showing a significant progress since 1993 due to the improvement of experimental techniques and the use of lasers.

The static and dynamic ME effects depend essentially on the disposition and orientation of spins in the equilibrium state, i.e. on the space magnetic symmetry. Diffraction experiments with the using of x-rays, electrons, and neutrons are powerful tools for studying the symmetry and structure of magnets. However, diffraction measurements do not allow to determine completely the magnetic structure in magnets with complex noncollinear arrangements of spins. For example, to study the neutron, it is necessary to have a large single-crystalline sample in a magnetic single-domain state. Such criteria are difficult to meet, and powder samples are often used instead. These principal or technical restrictions of the neutron or x-ray diffraction experiments were overcome. The novel method for studying a space magnetic symmetry in noncentrosymmetrical crystals was developed by using nonlinear spectroscopy [124]. The nonlinear optical method of the second harmonic generation (SHG) was applied to study the AF domains in magnetoelectric $Cr_2O_3$, piezomagnetic $\alpha - Fe_2O_3$ and FE-AF $YMnO_3$. The determination of a magnetic structure from SHG was based on the relation between the Furies components of the induced nonlinear polarization $P_i(2\omega)$ and the electric field components $E_j(\omega), E_k(\omega)$ of the light:

$$P_i(2\omega) = (\chi_{ijk}^{(i)} + \chi_{ijk}^{(c)}) E_j(\omega) E_k(\omega), \qquad (3.1)$$

where $\chi_{ijk}^{(i)}$ and $\chi_{ijk}^{(c)}$ are the crystallographic and the magnetic SHG susceptibility tensor components, respectively. The $\chi_{ijk}^{(i)}$ is a polar time-invariant ($i$-type) tensor of rank 3 allowed only in noncentrosymmetrical media. The $\chi_{ijk}^{(c)}$ is a time-noninvariant tensor which originates from the breaking of the time-reversal symmetry by the spin ordering and contains an ME term with the coefficients $\gamma_{ijk}$ in (1.2) and (1.3). The possibility of the ME doubling of the frequency was pointed earlier in [33, 90]. The tensor components $\chi_{ijk}^{(c)}$ and $\chi_{ijk}^{(i)}$ are defined by crystallographic and magnetic space groups, respectively [125]. The intensity of the second harmonic signal $I(2\omega) \propto |P_i(2\omega)|^2$ depends on magnetic symmetry through the tensor $\chi_{ijk}^{(c)}$ in (3.1). Thus, the possibility to distinguish different magnetic states appears. The noncollinear spin structure of the hexagonal manganites $RMnO_3$ ($R = Sc, Y, Ho, Er, Tm, Yb, Lu$) was studied by this novel optical method [126]. It is interesting to adduce some of these results.

The manganites $RMnO_3$ have crystallographic space group $P6_3cm$ and high FE ordering

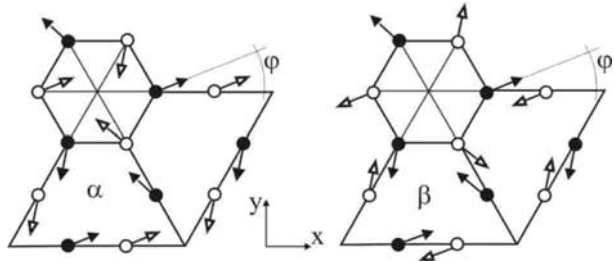

temperatures $T_c >> T_N$. Below the Neel temperature the triangular ordering of the $Mn^{3+}$ spins occurs. There are six possible magnetic states with different magnetic symmetry. Two of them in Fig.11 are distinguished by parallel ($\alpha$) or antiparallel

**Fig. 11**. Planar triangular magnetic structures of hexagonal $RMnO_3$. A projection of the Mn spins at $z = 0$ (closed arrows) and $z = c/2$ (open arrows) on the $xy$ plane of the magnetic unit cell shows parallel ($\alpha$ model) or antiparallel orientation ($\beta$ model) of neighboring spins on one line. The axes perpendicular to the glide planes $c$ (mirror planes $m$) are the $x(y)$ axes. The sixfold axis is the $z$ axis [126].

($\beta$) orientation of corresponding spins at $z = 0$ and $z = c/2$ in the unit cell. Depending on the angle $\varphi$ between the spin and the $x$-axis being 0º, 90º, or between $\alpha$ and $\beta$ models, one gets six different magnetic states. Fig.12 shows the intensity of the second harmonic (SH) spectra for the light incident along the optical axis ($\vec{k} \parallel z$) in the cases $\vec{P}(2\omega) \parallel \vec{E}(\omega) \parallel x$ ($\chi_{xxx}^{(c)} \neq 0$) and $\vec{P}(2\omega) \parallel \vec{E}(\omega) \parallel y$ ($\chi_{yyy}^{(c)} \neq 0$). At the Neel temperature $T_N = 72$ K, the SH signal vanishes confirming its purely magnetic origin. Below $T_N$ and at $T > T_R = 41$ K, $\chi_{xxx}^{(c)} \neq 0$, $\chi_{yyy}^{(c)} = 0$, that means the space magnetic symmetry $P\,6_{\underline{3}}\,c\,m$. Below the reorientation temperature $T_R$, on the contrary, $\chi_{xxx}^{(c)} = 0$, $\chi_{yyy}^{(c)} \neq 0$, and the space magnetic symmetry group is $P\,6_{\underline{3}}\,\underline{c}\,\underline{m}$. The change of magnetic structure goes along with a 90º spin rotation and a change of the AF domain structure.

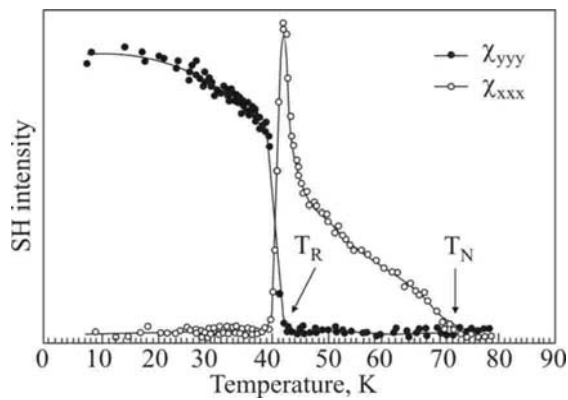

Thus, high spectral and spatial resolution of the SHG method made it possiblle to study complex magnetic structures which were indistinguishable when the neutron and other diffraction techniques were used.

**Fig. 12**. *Adapted from* [126]. Temperature dependence of the $\chi_{xxx}^{(c)}$ (open circles) and the $\chi_{yyy}^{(c)}$ components (closed circles) of SH tensor in the $HoMnO_3$.

*3.3 Theoretical investigations*

*3. 3.1 Domains. Inhomogeneous ME effects*

It is well known that a real FE and a magnet in the ordering state have domain structures. In the ferroelectromagnet two types of the domains (electric and magnetic) exist simultaneously. If the ME interaction is absent or negligibly weak, the FE and magnetic domains are independent.

The term "ferroic" was introduced by Aizu [27] to describe the known at that time three types of twinned crystals, in which one or more twin components could be switched to other equivalent states by using suitable external driving "forces". These "forces" are the electric and magnetic fields and the mechanical stress for the domains which differs in the orientation of spontaneous polarization $\vec{P}_s$, the magnetization $\vec{M}_s$ and the spontaneous deformation $\varepsilon_s$. In Aizu's nomenclature the *fully feroic* means that the external field can realize switching between all possible domain states; the *partially ferroic* means the possibility of switching between some, but not all possible domain states.

The term "multiferroic" was introduced by Schmid [30] for the materials in which two or more primary ferroic properties occur simultaneously in the same phase, and in which the magnetic point group has been reliably established by ME, optical, dielectric, magnetic and related studies on single crystals and single domains. Domain aspects, structural and symmetry requirements for the coexistence of ferroelectricity, ferromagnetism and ferroelasticity in crystals were analyzed in [30, 132].

The influence of ME interaction on the domain states in the ferroelectric ferromagnet (FE-FM) due to a weak anisotropic ME coupling (term with the coefficient $\lambda$ in (2.1)) was analyzed in [127], where there was shown the possibility of the bound state of the FE and FM domain walls, when they oscillate relative to one another with a frequency

$$\omega_0 \cong \frac{d}{l_0}\left[\frac{3}{2}\lambda M_0^2 f\right]^{1/2}.$$

Here a phenomenological constant $f = Z^2 / mv_c$, where Z, $m$ and $v_c$ are the charge, mass and volume of the elementary cell, respectively; $d$ and $l_0$ are the widths of the FE and FM domain walls, respectively. For the values $f \sim 10^{26} c^{-2}, \lambda \sim 10^{-9} dyne^{-1}cm^2, M_0 \sim 10^2 Gs$, $d / l_0 \sim 10^{-2}$, we have $\omega_0 \sim 10^8 - 10^9 c^{-1}$. This bound state of the FE and FM domains may be destructed by the electric field $E > E_c \propto (\lambda M_0^2 P_0^2) dl_0^{-1} \sim 10^5 V / m$.

In a single domain of FE-FM, the coupling between magnons and optical phonons is weak because of the large difference in spectral regions. However, experimental data show that the frequency of the FE domain wall (FEDW) excitations lies in the spin wave (HF) region of spectrum. The analysis [128] of exciton spectrum in a single magnetic domain with one FEDW showed that FEDW induces the spin state localized near the bottom of the magnon zone. Local excitations of the FEDW induce the running spin waves. The generation of spin waves by the oscillations of FEDW in an alternating electric field was considered. The estimation of the coefficient of losses K of electric energy due to the generation of spin waves in FE-FM led to the value $K \sim 10^{-7} - 10^{-9} \ cm^{-1}$ in the case of nonresonant generation. The possibility of the resonant ME generation appeared when the influence of the magnetic subsystem on the dynamics of FEDW was taken into account [129]. In this case the resonant generation of spin waves by electric field is possible on the frequency

$$\omega' \propto \lambda P_0 M_0 f^{1/2} (d/l_0)^{3/2} \sim 10^{10} (d/l_0)^{3/2}.$$

The situation when FE-FM contained both FE and FM domain walls was also considered. In this case the resonant ME generation is also possible on the single frequency which is proportional to the ME constant $\lambda$ .

The excitation of electric polarization  in FE-FM by oscillations of magnetic domain wall as a whole in alternating magnetic field was studied theoretically, too [130]. As the frequencies of the oscillations of magnetic domain walls are essentially smaller than the FE frequencies, the resonance excitation of running waves of electric polarization by magnetic field through the excitation of the FM domain walls is impossible. Electric polarization excitations are localized on the magnetic domain wall.

According to theoretical views, the magnetic (electric) domain wall is a localized magnetic (electric) nonlinear excitation. The low-energy localized excitations of both the magnetic and the electric subsystems were studied in hexagonal perovskites $RMnO_3$ ( $R$ is a rare-earth ion or $Y$ ) with a triangular AF structure [131]. The expressions for the inhomogeneous magnetization and electric polarization in the region of localized excitations were obtained. It was noticed that all parameters of the domain walls may be controlled by external electric field directed along the hexagonal axis of the crystal. This effect will be demonstrated later (see Sec. IV).

The induction of electric polarization by magnetic heterogeneity, due to the ME energy (2.8) and the terms with the second derivative $P_i M^{l} \partial^2 M^{l'} / \partial x_k \partial x_j$ , was called the *ingomogeneous ME effect* [70]. The existence of electric polarization in magnetic domain walls was considered. It was pointed that the ingomogeneous ME effect (IME) may have not only a relativistic and exchange-relativistic nature but also an exchange nature if the crystal symmetry allows the ME term $P_i \vec{M} \partial^2 \vec{M} / \partial x_k^2$.

The analysis of the appearance of electric polarization due to a helical magnetic ordering in $Cr_2BeO_4$ was done in [71]. Static electric polarization in this rhombic AF is induced by the exchange-relativistic ME energy

$$P_x \left[ d_1 \left( G_{1z} \frac{dG_{1x}}{dz} - G_{1x} \frac{dG_{1z}}{dz} \right) + d_2 \left( G_{2z} \frac{dG_{2x}}{dz} - G_{2x} \frac{dG_{2z}}{dz} \right) \right] , \qquad (3.2)$$

where $\vec{G}_1$ and $\vec{G}_2$ are the AF vectors. The electric polarization is perpendicular to the modulation vector $q_z$ and lies in the plane ($x, z$) of the cycloid. According to the experiment [69], the absolute value of electric polarization is of order $10^{-6} - 10^{-4}$ to that of conventional FE. The enhancement of electric polarization in magnetic domain wall in the FE ordering in ferroelectromagnets with $T_c$ less than $T_m$ and not close to it was predicted in [133].

The FE and LMEE of exchange nature near the surface of the magnet of any symmetry were studied. These effects appear due to the change in value of magnetic moment near the surface and are described by an exchange ME energy of the scalar form $\vec{P}(\partial \vec{M}^2 / \partial \vec{n})$, where $\vec{n}$ is a normal to the surface. The effects were studied in the FM and ferroelectromagnetic films and in the two-layered contact system of the FE and FM [134].

### 3.3.2 The nature of ME interactions

In searching for significant ME effects the main attention was paid to the studying of the LMEE. The largest value of the LMEE was supposed for the case of exchange nature of this effect.

The **exchange nature of the LMEE** in $Ni - I$ boracite was shown in [40] as well as later in [135, 136] there was given a detailed description of this compound whith 24 magnetic ions $Ni^{2+}$. The exchange ME energy which is responsible for the LMEE is in the form [136]

$$F_{ME}^{ex} = \Gamma \sum_i \vec{L} \vec{M} P_i . \qquad (3.3)$$

Here $P_i$ is a component of the electric polarization, $i = x, y, z ..$

The LMEE and piezomagnetic effects of exchange nature were predicted in hexagonal perovskites $RMnO_3$ with noncollinear triangular AF structure [137]. The symmetry of these compounds permits the **exchange ME energy**

$$F_{ME}^{ex} = D\vec{C}(\vec{B}_1 E_x + \vec{B}_2 E_y),$$ (3.4)

where $\vec{C}, \vec{B}$ are the AF vectors, $\vec{E}$ is an electric field.

The existence of the exchange ME energies (3.3) and (3.4) in these compounds means the possibility of the direct (one-magnon) absorption of electromagnetic field by exchange spin branches [138]. This resonant method is very convenient for studying the exchange magnons and ME constants.

According to the phenomenological theory, for FE- AWFM boracites [135] the temperature dependences of the components $\alpha_{23}$ and $\alpha_{32}$ of the LMEE in orthorhombic boracites should be identical, $\alpha \propto (T_m - T)^{1/2}$, near $T_m$. However, the experimental data for $Ni - Cl$ and $Ni - Br$ boracites indicated a significant difference in the temperature dependences: when $\alpha_{23} \propto (T_m - T)^{1/2}$, the component $\alpha_{32}$ showed a sharp peak near $T_m$ [41]. A similar peak occurs for $\alpha_{32}$ in $Co - I$ [139] and $Co - Br$ [140] boracites. The analysis, which takes into account the multiple-sublattice magnetic structure of orthorhombic boracites, shows that the observed peaks in the component $\alpha_{32}$ are attributable to the contributions of AF vectors in the LMEE through the terms $\partial L / \partial E$ describing the AF response on electric field [141]. In multiple-sublattice magnets it is necessary to take into account the ME susceptibility (1.4), (1.5) as well as the AF response on electric field [141].

It is usually assumed that the magnetic moment is of spin nature because of the "freezing" of orbital moment of $d -$ ions. However, many ferroelectromagnets contain rare-earth ions with nonzero orbital moment which together with the spin one contributes to the total magnetic moment. The contribution from the orbital moment into the ME susceptibility may be large. For instance, in the FE-AF $Tb_2(MoO_4)_3$, the manifestation of the ME interaction exceeding by two orders of the ME interaction in spin magnets with zero orbital moment was observed [142]. Not only a considerable magnitude of the ME effects, but also qualitatively new effects may be expected in the ferroelectromagnets with "no frozen" orbital moment.

**The influence of orbital moment** on the spectrum of ferroelectromagnets can be noticed, as the operator of orbital moment does not commute (in contrast to the spin operator) with the electric polarization operator. The presence of nonzero orbital moment results in the appearance of a low FE mode and the high-frequency magnetic modes, the change of excitation character: the precession of the electric polarization vector and the oscillation of the magnetic moment. The peculiarities of the spectrum would manifest themselves in the appearance of additional resonance frequencies under the absorption of magnetic and electric fields at the transition into a ferroelectromagnetic phase [143].

The ordering of orbital moments was suggested to explain the ferromagnetic ordering in the ferroelectromagnet $BiMnO_3$ [144]. According to this model, in $BiMnO_3$ the ordering of the $d_{z^2}$ and $d_{x^2-y^2}$ orbital states (orbitals) of $Mn$ ions in the $ac$-planes takes place. The orbitals in the adjacent $ac$-planes are directed in such way to overlap a half-full and empty $e_g$-orbitals along the chains $Mn-O-Mn$. Thus, the condition for the ferromagnetic exchange interaction in the chain is realized for all $Mn$ ions.

The $BiMnO_3$ was the only known (at that time) single-phase ferroelectromagnet which had ferromagnetic order (see Table 1 in [12]). Meantime the FM is characterized by the largest magnetic susceptibility between magnets that is important for the increasing of the value of ME effects (1.10). The understanding of the reasons of these unique phenomena requires that the electronic states of the crystal should be taken into account. The urgent necessity of microscopic considerations appeared.

First-principles investigation of magnetic and electric properties of $BiMnO_3$ was accomplished by using the local spin density approximation (LSDA) [145]. The $BiMnO_3$ was chosen because of its unique FM property and relatively simple structure. The origin of the differences between bismuth manganite and other perovskite manganites was determined by calculating total energies and band structures. The results indicated that a covalent bonding between the bismuth cations and oxygen anions in $BiMnO_3$ introduced additional interactions compared with the rare-earth manganites, in which the rare earth - oxygen interaction was essentially purely ionic. These additional orbital interactions in turn stabilized different magnetic and structural phases.

It is known (see Sec1.1) that the presence in the B-positions of perovskites of metal ions with a noble gas shell after the removal of $s-$ and $d-$electrons promotes the FE displacements ("$d^0$" rule). The first-principle calculations [146] confirmed this rule and showed that $d$-electrons of the transition metal reduced the tendency for the off-center FE distortion. But the "$d^0$" rule can be broken by the distortion of non-Jahn-Teller type which creates an asymmetric potential in spite of the $d$ occupation of the magnetic cation. In the $BiMnO_3$ the asymmetry is created by the covalent bonding $Bi-O$ [146]. The work [146] was a start to the fundamental microscopic investigations of the ME interaction. Further the LSDA method of studying of electronic zone structure of ferroelectromagnets was often used.

In the conventional theory the FE are insulators. The FE transition is a particular case of a structural transition which is accompanied by the appearance of electric polarization. The positive ions are displaced relative to the negative ions leading to a spontaneous polarization. Electric polarization consists of ionic and electronic parts. The ionic part is generated by the lattice distortion changing the ionic distances. Electronic polarization is conditioned by the electron displacement relatively to the nuclear and results in the distortion of the electron cover of ion.

**Electron participation in the electric polarization** of perovskites is significant. The contribution of electrons to electric polarization increases with the decreasing of the bond between the electron and the nuclear and with the appearance of novel degrees of freedom. Thus, in the $PbTiO_3$ there is the hybridization between the $Pb$ $6s$ and O $2p$ electrons in contrast to the $BaTiO_3$ with the ionic $Ba-O$ interaction. It results in the larger value of electrical polarization in the $PbTiO_3$ than in the $BaTiO_3$.

Expected is an active participation of the electrons in electrical polarization of magnetic compounds with the mixed valency and at the phase transition metal- insulator.

The theory of electronic ferroelectricity was also presented in [147]. In this theory the FE phase transition is driven by a change in the electronic structure. The electronic FE is possible in the insulating phase of the Falikov-Kimbal model, in which the Coulomb attraction between the localized $f$ - holes and itinerant $d$ - electrons leads to a built-in coherence between the $d$ and $f$ states. The predicted values of polarization in this model were around $10\,\mu C/cm^2$, similar to those found in the perovskite FE.

### 3.4 Potential ferroelectromagnets. Perspectives

Many *potential ferroelectromagnets,* especially of the perovskite type, were synthesized at that period of time [148]. The synthesis of the solid solutions based on the first ferroelectromagnets led to the discovery of compounds with high (above room temperature) temperatures of electric and magnetic orderings in the systems $BiFeO_3-LaMnO_3$, $Pb(Fe_{2/3}W_{1/3})O_3-PbTiO_3$ and $Pb(Fe_{2/3}W_{1/3})O_3-BiFeO_3$ [149]. For example, the compound $xBiFeO_3-(1-x)LaMnO_3$ is the FE-AF with $T_N=448\,K$, $T_c>1100\,K$ for $x=0.8$. The FE-AF compound $xPb(Fe_{2/3}W_{1/3})O_3-(1-x)BiFeO_3$ with $x=0.5$ has $T_N=448\,K$, $T_c\approx800\,K$. But all these compounds were produced only in a polycrystalline form and required additional investigations. The results of the ME measurements in ceramic spaces have only a preliminary character. The verification of the ME properties in a single crystal may essentially correct a previous result. So, the ME investigation of a single crystal $Pb(Fe_{2/3}W_{1/3})O_3$ did not reveal a spontaneous electric polarization [150] declared before [5]. A weak FM was also found in AF $Mn-Cl$ and $Mn-Br$ boracites [151].

Magnetic states of some single-phase magnets, in which the FE ordering was discovered later, had been studied. Magnetic transitions between the incommensurate and commensurate phases were revealed by neutron diffraction in the $MnWO_4$ [152]. In orthorhombic perovskites $Pr\,NiO_3$ and $NdNiO_3$ with mixed valency of nickel, the metal-insulator transition was accompanied by an unusual coexistence of the AF and FM orderings [153]. In the $RNiO_3$ compounds, the change in crystal symmetry at metal-insulator transition was first observed in $YNiO_3$ [154]. In the FM piezoelectric $Ga_{2-x}Fe_xO_3$, a linear magnetostriction [155] and a toroidal ordering [156] were discovered.

Usually magnetic ions were introduced in the FE matrix to synthesize ferroelectromagnets [12]. The other possibility is an introduction of the FE ions in the magnetic material [157]. For this purpose the ferric spinels are the most perspective magnets because of their large magnetization and high magnetic ordering temperature (above room temperature). The ferric spinels have a symmetry group without space inversion $(F\bar{4}3m)$, a low conductivity, small dielectric losses $(tg\delta)$, and a high dielectric constant $\varepsilon$. The introduction of the FE ions in the ferrites matrix increases $\varepsilon$ significantly. For example, in the $Ni-Zn$ ferrite the value $\varepsilon \sim 10^4$ was reached [158].

The use of a multiphase structure (layered systems, composites, nanostructures) may significantly increase the value of the ME coupling (see [19]) and create a strong spectral nonreciprocity [159, 160]. It was shown that in the two-layered system metal-dielectric and, consequently also in ferroelectromagnet, strong nonreciprocal polaritons running in the dielectric along the contact surface may exist. It is possible in the presence of a constant electric field directed normally to the contact surface or in a constant magnetic field oriented in the contact plane, and also in the crossed electric and magnetic fields [161]. This possibility appears due to the dynamical ME energy

$$F_{ME}^d = \frac{v_c}{mc}\vec{P}\left[\vec{\Pi} \times \vec{H}\right] \quad , \qquad (3.5)$$

Where $\vec{\Pi} = mv_c^{-1}\vec{v}$ is the momentum density, $m$ is the mass of the charge, $v_c$ is the elementary cell volume, and $c$ is the light velocity. The electric polarization $\vec{P}$ consists of ionic as well as electronic contribution. The dynamical ME energy (3.5) is the energy of an interaction between the electric polarization with the effective electric field $\vec{E}_{ef} = c^{-1}[\vec{H} \times \vec{v}]$ produced by the motion of charge $e$ with velocity $\vec{v}$ in the magnetic field. This energy is a scalar, and therefore it is present in any crystal.

The ME energy (3.5) creates a strong nonreciprocity for electromagnetic waves of infrared and optical ranges of the spectrum running in dielectric (phonon polaritons) at the contact with a metal. The number of polariton modes is different for the waves running in the opposite directions. For the fixed direction of a constant field the polaritons with given frequency propagate only in one direction, i.e. the effect of rectification of electromagnetic waves takes place. The spectrum of the surface polaritons in isotropic dielectric at the boundary with a metal (($x, y$) is a contact plane) in the magnetic field $H_y$ of opposite directions is shown in Fig. 13. Here $\omega_{1,2}$ are phonon polariton modes in the presence of magnetic field, $\omega_1 = \omega_0 - \omega_H, \omega_2 = \omega_0 + \omega_H$, where $\omega_H = gH$, $g = |e|/mc$ ; $\Omega_1 \cong \omega_0$, $\Omega^2{}_2 \cong \omega_0^2 + 4\pi e^2/mv_c$. The frequency $\omega_0$ is an optical phonon mode in the absence of magnetic field. Magnetic field splits this mode into two modes, $\omega_1$ and $\omega_2$, in which electric polarization rotates around magnetic field in the opposite directions. The value of the frequency splitting $\omega_H$ is larger in the optical spectrum range where the electron contribution in polarization prevails. Then the gyromagnetic ration $g \sim 10^7\,CGS$ units, and in the magnetic field $H \sim 10\,T$ the value $\omega_H \sim 10^{12}\,c^{-1}$. The predicted strong nonreciprocity of the surface electromagnetic waves appears due to the nondiagonal components of dielectric susceptibility which are induced by the constant magnetic (electric) field of the pointed direction. As a result, the vortical toroidal excitations appear which push out an alternating field of electromagnetic wave to the surface of dielectric [162].

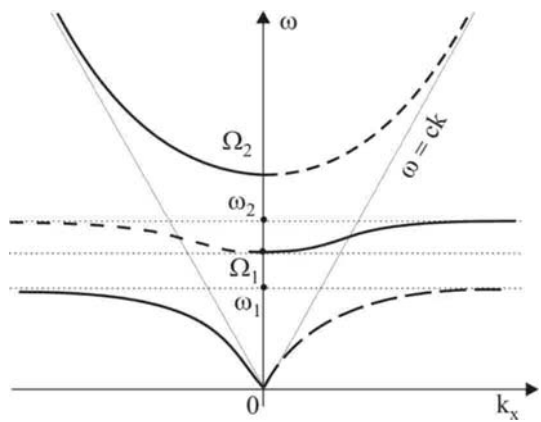

**Fig. 13**. Modes of surface polaritons in dielectric at the contact with a metal in the cases $\vec{H} \uparrow \uparrow Y$ (solid curves) and $\vec{H} \uparrow \downarrow Y$ (dashed curves).

In spite of optimistic theoretical predictions, the observed ME effects were weak. The distance between the possibilities and the use of ferroelectromagnets in techniques did not decrease during the forty years after the discovery of these compounds. But the discovery of the novel ferroelectromagnets and effective nonlinear optical methods of investigation made ready the conditions for the new revival of the ME science in the 21st century.

## IV.  The second magnetoelectric renascence (after 2000)

Intrinsic spontaneous electric and magnetic fields in the ferroelectromagnet form in it a coupled equilibrium ME structure (magnetic and electric domains). The value of this ME coupling is proportional to the magnitudes of electric and magnetic moments (2.1) and is larger in the case of a proper FE transition.

The ME interaction displays itself with the change of electric (magnetic) state. The ME response on the changes may be significant near the temperature and field phase transitions. In the ferroelectromagnet with a proper FE ordering and $T_c > T_m$, a dielectric susceptibility is large near $T_c$, but quickly decreases below $T_c$ in the ferroelectromagnetic state. In the case of $T_m > T_c$, the magnetic susceptibility is small in the ferroelectromagnetic phase. The FE and the magnetic susceptibilities have simultaneously their maximum values only at the improper FE transition induced by magnetic ordering. But the value of the electric polarization is very small at this FE transition.

However, at the beginning of the 21st century the gigantic ME effects in the ferroelectromagnets with the proper and improper FE orderings were discovered. The number of publications on the discovery and the prediction of colossal ME effects extremely increases. Owing to this, only the main ME results in the ferroelectromagnets are considered below. In Sec. 4, the gigantic ME effects in ferroelectromagnets with the improper FE transition of magnetic origin in the cases of homogeneous and modulated magnetic orderings are reviewed. Sec. 5 contains a description of the colossal ME effects in the ferroelectromagnets where the FE and magnetic ordering temperatures are not close. The outlook, conclusions and perspectives of the ME sciences are briefly discussed in Sec. 6.

## 4   Ferroelectromagnets with improper ferroelectric transition of magnetic origin

In the noncentrosymmetrical magnets with homogeneous magnetic structure an electric polarization can appear at the magnetic transition as a result of the ME interaction (linear in P terms in (2.1)). This improper FE transition of magnetic nature has been observed in the $Ni - I$ boracite [11]. Modulated magnetic structure, which will be further called "spiral structure", can induce the improper FE transition in the magnets of any symmetry. The **first example** of the ferroelectromagnet of this type is a centrosymmetrical FE-AF $Cr_2BeO_4$ [12]. The value of the electric polarization appeared at the improper FE transition is usually weak ( $P \leq 0.1 \mu C/cm^2$ ). The electric polarization being of magnetic origin is strongly dependent on the magnetic state. Not a large magnetic field can essentially change a small electric polarization. It was shown in the $Ni - I$ boracite. Recently, a colossal ME effect in the spiral magnet $TbMnO_3$ was obtained by Kimura *et al.* [26].

### 4.1 Magnetoelectric effects in spiral magnets with magnetic-induced ferroelectric ordering

### 4.1.1 Rare-earth orthorhombic perovskite manganites $RMnO_3$

The type of the AF structure and the Neel temperature in orthorhombic perovskites

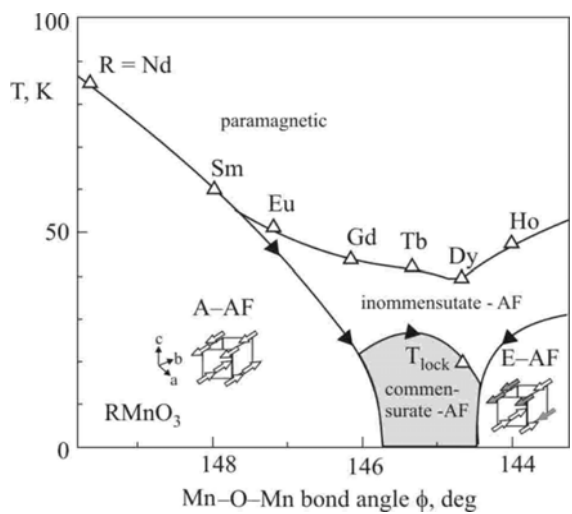

$RMnO_3$ ( $R$ is rare earth) depend on the $Mn - O - Mn$ bond angle $\phi$ (or ionic radius of $R$ ) (Fig.14, [163]). In Fig. 14 the $A$ -type of the AF order (left lower inset) corresponds to the FM order between the nearest-neighboring $Mn$ sites in the $(a,b)$ plane and the AF interaction between these planes. The E-AF- type (right lower inset) is a zigzag-type order. As displayed in Fig 14, a decrease of the angle $\phi$ changes the layered $A$ -type onto the $E$ -type in the $HoMnO_3$. The compounds with $R = Eu, Gd, Tb, Dy$ have an incommensurate AF structure and are potential ferroelectromagnets with the improper FE transition similar to the $Cr_2BeO_4$.

**Fig. 14.** *Adapted from* [ 163]. Magnetic phase diagram for $RMnO_3$ as a function of $Mn - O - Mn$ bond angle $\phi$. Open and closed triangles denote the Neel and lock-in transition temperatures, respectively.

### 4.1.1.1 Colossal magnetoelectric effects in $TbMnO_3$

The publication of Kimura *et al.* [26] with the revolutionary title "Magnetic control of ferroelectric polarization" removed the doubts of possible use of the ME effects in applications.

Electric polarization strongly connected with the incommensurate magnetic structure in the manganite $TbMnO_3$ was discovered. A magnetic field of a few Tesla could cause a 10% change in the dielectric constant and switch the direction of electric polarization. The last effect was called the "magnetic-field-induced electric polarization flop" [26].

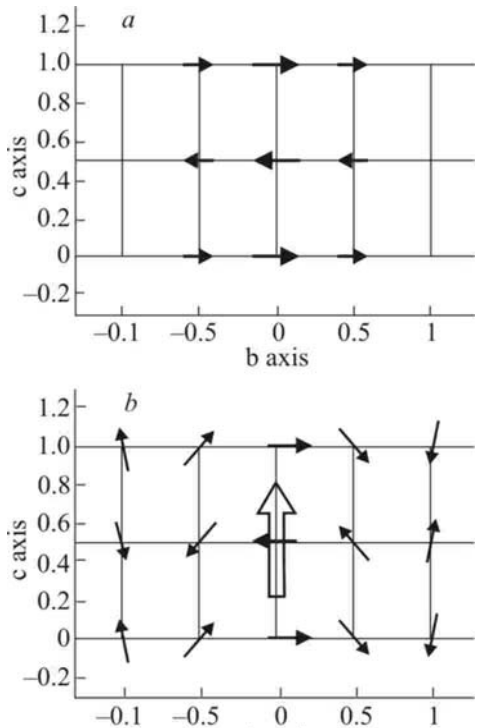

Fig. **15**. Schematic of the magnetic structure at (a) $T = 35\,\text{K}$ and (b) $T = 15\,\text{K}$, projected onto the $b - c$ plane. Filled arrows indicate direction and magnitude of Mn moments. The longitudinally modulated phase (a) respects inversion symmetry along the $c$ axis, but the spiral phase (b) violates it, allowing an electric polarization (unfilled arrow)[165].

The $TbMnO_3$ has the orthorhombically distorted perovskite structure (centrosymmetrical space group $Pbnm$) and the incommensurate sinusoidal AF structure of the $A$-type below $T_N \cong 42\,\text{K}$. Spins of the $Mn^{3+}$ and wave vector of the modulation $\vec{k}$ are directed along the $b$ axis [164]. The appearance of electric polarization along the $c$ axis below $T_c \approx 27\,\text{K}$ was firstly connected with the incommensurate – commensurate (lock-in) magnetic transition where the magnetic modulation wave vector $\vec{k}$ was locked at a constant value [26]. Later the existence of the additional $c$ component of the $Mn$ moments at $T \leq T_c$ was revealed by neutron diffraction [165]. The noncollinear incommensurate magnetic ordering breaks the inversion symmetry, and thus gives rise to the electric polarization along the axis $c$ (Fig. 15). The value of a spontaneous electric polarization increased at lowering temperature and was small, $P_c \sim 5 \times 10^{-2}\,\mu C / cm^2$. The first experiment was made in the magnetic field oriented along the $b$-axis in the range of magnetic fields (0 -9T) [26]. Then the temperature dependences of the electric polarization components along the $a, b$, and $c$-axes at various magnetic fields (up to 9 T) directed along the $a, b$, and $c$-axes were received [166]. No electric polarization along the $b$-axis ($P_b$) was induced by the application of $H$. In Fig. 16, the temperature profiles of the electric polarization $P_c$ in the magnetic fields are shown. A qualitative explanation of some properties of these ME diagrams may be given in the frame of phenomenological theory. Below such a consideration is proposed [167].

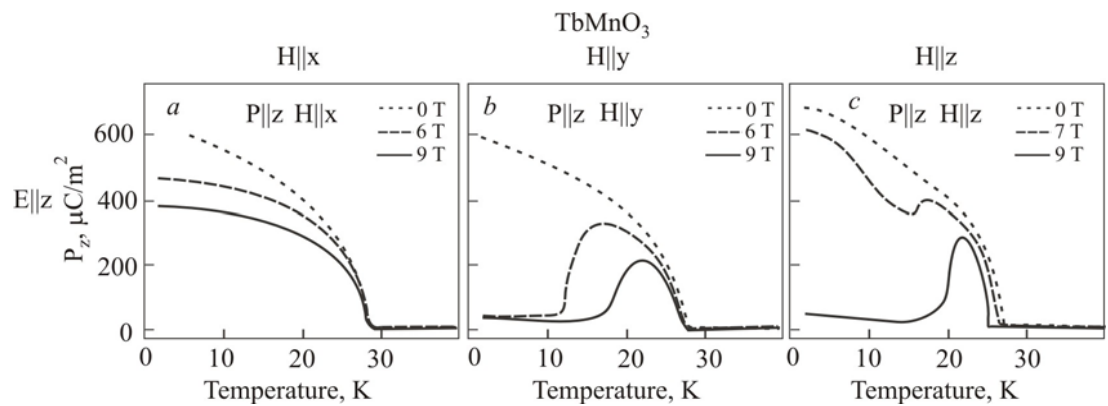

**Fig. 16**. *Adapted from* [166]. Temperature profiles of electric polarization along the $z$ axis at various magnetic fields (up to 9 T) along the $x, y,$ and $z$ axis for $TbMnO_3$.

An elementary cell of the $TbMnO_3$ has four $Mn^{3+}$ ions and four $Tb^{3+}$ ions. The AF sinusoidal ordering of the manganese spins with the incommensurate wave number $k \approx 0.295$ $b^*$ occurs at the Neel temperature $T_N \cong 42$ K. The spins and the wave vector $\vec{k}$ are directed along the $b$-axis ($A_y$ state [164]). The incommensurate ordering of the $Tb^{3+}$ spins with another wave vector (0, ~0.42, 1) takes place below $T' \approx 7$ K. The FE ordering appears at the temperature $T_c \approx 27$ K as a result of magnetic transition from the AF collinear ($A_y$) to the noncollinear ($A_y, A_z$) incommensurate state [165]. In phenomenological theory the electric polarization $P_c = P_z$ is induced by the ME energy

$$F_{ME}(P_z) = \nu P_z \left( A_z \frac{\partial A_y}{\partial y} - A_y \frac{\partial A_z}{\partial y} \right) . \tag{4.1}$$

Here and further the coordinates $a = x, b = y, c = z$.

At first we will consider the interval of the temperature $T' < T < T_c$ where the spins of $Tb$ ions are not ordered and it is possible to take into account only manganese spins. The ions $Mn^{3+}$ are in the positions [164]:

$Mn^{3+}$: 1(1/2, 0, 0), 2(1/2, 0, 1/2), 3(0, 1/2, 1/2), 4(0, ½, 0).

The designations of the magnetic state are the following:

$$\begin{aligned}
\vec{A} &= \vec{M}_1 - \vec{M}_2 - \vec{M}_3 + \vec{M}_4, \\
\vec{G} &= \vec{M}_1 - \vec{M}_2 + \vec{M}_3 - \vec{M}_4, \\
\vec{C} &= \vec{M}_1 + \vec{M}_2 - \vec{M}_3 - \vec{M}_4, \\
\vec{M} &= \sum_1^4 \vec{M}_n .
\end{aligned} \tag{4.2}$$

The transformations of the AF vectors $\vec{A}, \vec{G}, \vec{C}$, the magnetization $\vec{M}$ (4.2) and the electric polarization $\vec{P}$ under the actions of elements of the group symmetry *Pbmn* are shown in Table 1, where small symbols designate the magnetic vectors of terbium. Table 1 takes into account the transpositions of the numbering magnetic moments under the symmetry operations that can change the sign of the AF vectors [32].

| $r_i$ | $l$ | $2_{lx}$ | $2_{ly}$ | Components |
|-------|-----|----------|----------|------------|
| $r_1$ | +1 | +1 | +1 | $A_x, G_y, C_z, c_z$ |
| $r_2$ | +1 | +1 | −1 | $M_x, C_y, G_z, m_x, c_y$ |
| $r_3$ | +1 | −1 | +1 | $A_z, C_x, M_y, m_y, c_x$ |
| $r_4$ | +1 | −1 | −1 | $A_y, G_x, M_z, m_z$ |
| $r_5$ | −1 | +1 | +1 | $a_x, g_y$ |
| $r_6$ | −1 | −1 | −1 | $a_y, g_x, P_z$ |
| $r_7$ | −1 | −1 | +1 | $a_z, P_y$ |
| $r_8$ | −1 | +1 | −1 | $g_z, P_x$ |

Table 1. Irreducible representations of the group $P_{bnm}$

By the Landau theory of phase transition, all possible equilibrium phases of the system can be received from the free energy which is invariable under the elements of the symmetry of this crystal by minimizing this energy with respect to the variables. It follows from Table 1 that the Dzyaloshinskii-Moriya terms of the type $\vec{M} rot \vec{M}$ ($\vec{M}$ is a magnetic moment of the *Mn* ion) do not exist in the $TbMnO_3$. Thus, the modulated magnetic structure below $T_N$ has an exchange nature. It is important that the magnetic moments of *Mn* are even, while the electric polarization and the terbium moments are odd with respect to the inversion.

For the existence of the FE ordering, the energy must contain invariants with the first power of electric polarization. Since the symmetry group of the crystal is centrosymmetric, and the terbium subsystem is disordered, it follows from Table 1 that the electric polarization of magnetic origin at $T > T'$ can exist only in the modulated magnetic states. In the frame of the $A$-state, the ME invariants with the first degree of $P_i$ are the following:

$$P_x\left(A_x \frac{\partial A_y}{\partial y} - A_y \frac{\partial A_x}{\partial y}\right), \quad P_y \frac{\partial A^2}{\partial y}, \quad P_z\left(A_y \frac{\partial A_z}{\partial y} - A_z \frac{\partial A_y}{\partial y}\right). \tag{4.3}$$

It is seen from Table 1 and (4.3) that in the modulated magnetic phase the $P_y$ is also modulated, and its general value is zero in the agreement with the experiment [26]. The nonzero value of the electric polarization along the $x$-axis $P_x$ observed after the spin-flop in the field $H_y$ ("polarization spin-flop" [26]), one would be able to connect with the first term in (4.3). After this, the spin-flop $A_y = 0$. Turthermore, the $x$-component of the AF vector (if it exists at $H = 0$) is too weak in the absence of magnetic field (its value is smaller than 1% of $A_{y,z}$ according to the data in [165]). Neutron diffraction measurements [164, 168] revealed, besides $A_y$, a weaker $G_y$ peak and very weak $C_x$ and $C_z$ peaks. These additional peaks appear simultaneously with the $A_y$ peak and have the same wave vector. The most significant of these additional contributions into magnetic order below $T_N$ is the contribution from the $G_y$ state corresponding to the value AF component $G_y \sim 10^{-1} A_y$. It was supposed [167] that the behavior of the electric polarization $P_x$ in magnetic fields is a display of weaker $\vec{G}$ and $\vec{C}$ states in the AF configurations. Below we will show it taking into account two AF vectors $\vec{A}$ and $\vec{G}$, where $|A| >> |G|$. The Ginzburg-Landau functional of the $TbMnO_3$ may be written in the form

$$F = V^{-1} \int \{ \frac{1}{2}(a_1 \vec{A}^2 + a_2 \vec{G}^2) + \frac{1}{4}u(\vec{A}^4 + \vec{G}^4) + \frac{1}{2}\gamma[(\partial_y \vec{A})^2 + (\partial_y \vec{G})^2]$$

$$\frac{1}{2}\alpha[(\partial_y^2 \vec{A})^2 + (\partial_y^2 \vec{G})^2] + \frac{1}{2}wA_z^2 + \bar{\nu}A_x G_y - \vec{M}\vec{H} + \frac{1}{2}BM^2 + \frac{1}{2}\lambda_1(\vec{A}\vec{M})^2 + \frac{1}{2}\lambda_2(\vec{G}\vec{M})^2 + \quad (4.4)$$

$$\frac{1}{2}(\lambda'_1 \vec{A}^2 + \lambda'_2 \vec{G}^2)\vec{M}^2 + \frac{1}{2}\chi P^2 + \nu P_z(A_z \partial_y A_y - A_y \partial_y A_z) + \nu_1 P_x(G_y \partial_y A_y - A_y \partial_y G_y)\}dV .$$

Here, $\partial_y = \partial / \partial y$. Functional (4.4) contains only the main terms necessary for further analysis. The first four terms result from a homogeneous exchange interaction. The next two terms are inhomogeneous exchange energy of the FM interaction between the nearest neighbours (term with the coefficient $\gamma < 0$) and the AF interaction with the next neighbours (term with the coefficient $\alpha > 0$). The term with the coefficient $w$ is an anisotropic magnetic energy. Since the manganese spins are oriented along the y-axis in the collinear sinusoidal phase, the coefficient $w > 0$. Weak FM terms of the type $A_i M_k$ are omitted. The terms with the coefficients $\lambda, \lambda'$ describe an interaction of the AF vectors with the magnetic field owing to magnetization. The last two terms are the ME energy inducing electric polarization along the z- and x-axis, respectively.

The equilibrium AF states can be found in the form of harmonic rows. It is possible to take into account only the first harmonic since the third one is significantly small [168]. One can set in the noncollinear phase ($(T' < T < T_c)$

$$A_y = A_1 \cos ky, \quad A_z = A_2 \sin ky, \quad G_y = G_1 \sin ky. \tag{4.5}$$

Substituting (4.5) into the functional (4.4) and minimizing (4.4) with respect to $A_1, A_2, G_1, P_x, P_z, \vec{M}$ and the wave number $k$, the following expressions before the spin-flop transitions are obtained:

$$A_1^2 = (2u)^{-1}(L_1 - \Lambda_i^{(1)}h_i^2), \quad L_1 = 2(a_c - a_1) + w > 0, \quad h_i = H_i B^{-1}, \quad i = x, y, z$$

$$A_2^2 = (2u)^{-1}(A_0 - \Lambda_i^{(2)}h_i^2), \quad A_0 = 2(a_c - a_1) - 3w = 2\xi_1(T_c^0 - T) > 0, \quad k^2 = -\gamma / 2\alpha, \tag{4.6}$$

$$G_1^2 = 4(3u)^{-1}(G_0 - \Lambda_i^{(3)}h_i^2), \quad G_0 = a_c - a_2 = \xi_2(T_N - T), \quad a_c - a_{1,2} = \xi_{1,2}(T_N - T), \quad a_c = \alpha k^4.$$

$$M_i = H_i(B + \lambda_1 A_i^2 + \lambda_i' A^2)^{-1} \cong h_i, \quad \xi_{1,2} > 0 \quad.$$

Since each of the considered magnetic fields is much smaller than the exchange field, we have $B >> \lambda A^2$.

In the expressions (4.6), the coefficients $\Lambda_x^{(1)} = \Lambda_x^{(2)} = 2\lambda_1', \Lambda_z^{(3)} = \Lambda_z^{(3)} = \lambda_2'$, $\Lambda_y^{(1)} = \Lambda_z^{(2)} = 3\lambda_1 + 2\lambda_1', \quad \Lambda_z^{(1)} = \Lambda_y^{(2)} = 2\lambda_1' - \lambda_1, \quad \Lambda_y^{(3)} = \lambda_2 + \lambda_2'$. The electric polarization $P_z$ induced by the $A$-configuration is significantly larger than $P_x \propto A_1 G_1$, which arises only due to small G-distortion. The spontaneous value $|P_x| << |P_z|$; in the weak magnetic fields we can neglect $P_x$ because $G_1 << A_{1,2}$. The spontaneous polarization $P_z$ exists in the noncollinear magnetic state at $T < T_c^0 = T_N - 3w / 2\xi_1$, where $T_c^0$ is the FE transition temperature in the absence of magnetic field. The expressions for $P_z$ and $P_x$ are the following:

$$P_z = vkA_1A_2\chi^{-1}, \quad P_x = v_1kA_1G_1\chi^{-1} \quad. \tag{4.7}$$

Using (4.6), it is possible to interpret some observed temperature and field dependences of the electric polarization (4.7).

The electric polarization $P_z$ is described by the expressions

$$P_z = \nu k (2u\chi)^{-1}[(L_1 - 2\lambda_1' h^2)(A_0 - 2\lambda_1' h^2)]^{1/2}, h = h_x,$$
$$P_z = \nu k (2u\chi)^{-1}\{[L_1 - (3\lambda_1 + 2\lambda_1')h^2][A_0 + (\lambda_1 - 2\lambda_1')h^2]\}^{1/2}, h = h_y,$$
$$P_z = \nu k (2u\chi)^{-1}[(L_1 + (\lambda_1 - 2\lambda_1')h^2)(A_0 - (3\lambda + 2\lambda_1' h^2))]^{1/2}, h = h_z. \tag{4.8}$$

The electric polarization $P_z$ appears in the noncollinear AF phase when $A_2 \geq 0$. Therefore, the shift of the FE transition temperature in the magnetic fields of different orientations is the following:

$$T_c - T_c^0 = -\lambda_1' \xi_1^{-1} h_x^2; \quad (\lambda_1 - 2\lambda_1')(2\xi_1)^{-1} h_y^2; \quad -(3\lambda_1 + 2\lambda_1')(2\xi_1)^{-1} h_z^2. \tag{4.9}$$

The constant $\lambda_1 > 0$ since the magnetization decreases below $T_c$ [26]. Also $\lambda_1' > 0$ because the $P_z$ decreases in the magnetic field $H_x$ (see Fig. 16 a). Figs. 16 (a-c) show that the shift $T_c$ in the magnetic field is very small. It is noticeable and negative only in the magnetic field $H_z$ that is in a qualitative agreement with (4.9).

Spontaneous polarization $P_z \propto A_2 \propto \sqrt{A_0}$ (see (4.6)) and $P_z \propto \sqrt{T_c - T}$ near $T_c$. This dependence is observed in Fig. 16 a-c.

**In the magnetic field $\vec{H} \parallel x$** the electric polarization $P_z$ decreases with the increasing of the magnetic field (Fig.16 a). The electric polarization $P_x$

$$P_x = \nu_1 k (2u\chi)^{-1}[(L_1 - 2\lambda_1' h_x^2)(G_0 - \lambda_2' h_x^2)]^{1/2} \tag{4.10}$$

is very small in the magnetic fields up to 10 T. Measurements at higher fields $H > 10\,\text{T}$ [166] show a significant growth of $P_x$ up to $2 \times 10^{-2}\,\mu C / cm^2$ at $H \sim 14T$ [166]. In the frame of the supposed model, the increasing of $P_x$ (4.10) is possible if the constant $\lambda_2' < 0$. The electric polarization $P_x$ becomes noticeable due to smallness of $G_0 << L_1$ in the fields $|\lambda_2'| h_x^2 \geq G_0$. When the magnetic field $H_x$ grows, the $z$-component of the electric polarization decreases, while the $x$-component increases. According to the novel data obtained by using a high-resolution thermal expansion and magnetostriction measurements [169], the first order phase transition was observed between the states with $P_z \neq 0$ and $P_x \neq 0$ in the noncollinear AF phase, what contradicts to the previous report [166]. The observed first order phase transition may be explained by the presence of the $x$-component of AF vector $A_x$ induced by $G_y$ (term $\bar{\nu} A_x G_y$ in (4.4)) whose magnitude increases with the increasing of the magnetic field. A significant value of $A_x$ can provoke a spin-flop in the field $H_x$ at about 10 T. After the spin-flop the value $P_x \propto A_1 G_1$ grows in the magnetic field. Observation of nonzero polarization means that the modulation of the magnetic structure was preserved after the spin flop.

The observed growth of the electric polarization $P_z$ and $P_x$ with the decrease of temperature (Fig.16 a and Fig. 5 d in [166]) is confirmed by the formulae (4.6), (4.8), and (4.10).

**In the magnetic field** $\vec{H} \parallel y$ the electric polarization $P_z$ depends on the temperature not monotonically (Fig.16 b) and disappears with the increasing of magnetic field at the spin-flop transition. Since $P_z \propto A_y A_z$, after the spin-flop $A_y = 0$, and in the model under consideration, $P_z = 0$. The sharp decrease of the electric polarization at the spin-flop is seen in Fig.16 b. If after the spin- flop the AF vectors $A_y$ and $G_y$ are reoriented towards the $z$-axis, the following ME invariant is added to (4.4):

$$\nu_2 P_x (A_z \frac{\partial G_z}{\partial y} - G_z \frac{\partial A_z}{\partial y}),\qquad(4.11)$$

where

$$A_z = A_2 \sin ky, \ \ G_z = G_2 \cos ky,$$

$$A_2^2 = 4(3u)^{-1}(L_2 - \lambda_1' h_y^2), \ \ G_2^2 = 4(3u)^{-1}(G_0 - \lambda_2' h_y^2), \ \ L_2 = a_c - a_1 - w.\qquad(4.12)$$

$$P_x = 4\nu_2 k(3u\chi)^{-1}[(L_2 - \lambda_1' h_y^2)(G_0 - \lambda_2' h_y^2)]^{1/2}.$$

Since the constant t $\lambda_2' < 0$, the value of $P_x \propto A_2 G_2$ becomes noticeable after the spin-flop and grows with the increasing of magnetic field as in the previous case of $\vec{H} \parallel x$. The increase of $P_x$ after the spin-flop with the increasing of the magnetic field $H_y$ at first was observed by Kimura et al. [26] and called the "polarization flop". A process of "reorientation" of the electric polarization from the $z$-axis to $x$-axis ("polarization flop") is not similar to the flop in magnets in our model as it is not a rotation of the electric polarization vector. The electric polarization $P_x$ is nonzero initially and it increases gradually with the increasing of magnetic field, becoming noticeable at fields of the spin-flop field order owing to the presence of weak G states. The described behavior of the electric polarization is natural, since in an anisotropic crystal the rotations of electric polarization, unlike spins, are impeded, and it is easier for the electric polarization to change its modulus. The proposed here model of the behaviour of $P_x$ in the magnetic field is not the only one possibility to explain qualitatively the observed "polarization flop" in the TbMnO$_3$. These strong anomalies of the thermal expansion and magnetostriction at the reorientation of magnetic sublattices [169] verify that a complex process takes place in the lattice under magnetic spin flop.

**In the magnetic field** $\vec{H} \parallel z$, in the noncollinear magnetic state the value $A_z \neq 0$, and the electric polarization $P_z$ can be switched off by the magnetic field at the spin flop (Fig. 16 c). The negative sign of the derivatives $\partial P_z / \partial H_z$ and $\partial P_z / \partial T$ may be obtained from (4.6) and (4.7). After the spin-flop, the AF component $A_z = 0$, and the electric polarization $P_z = 0$. If after the spin-flop the modulated magnetic structure remains, then, as before, $P_x \propto A_1 G_1$, and its value must increase with the increasing of the magnetic field. However, the experiment [166] does not show any noticeable value of $P_x$ in the magnetic field $H_z$ [166]. Thus, after the spin flop the magnetic subsystem is not a modulated one, $P_x = P_z = 0$, and terbium manganite in the temperature range $T' < T < T_c$ is a paraelectric and a canted AF with nonzero $A_y, M_z$.

All previous considerations of the ME states in the $TbMnO_3$ concerned the temperature range $T' < T < T_c$ where the spins of terbium were disordered.

**At low temperature,** when $T < T' \approx 7K$, the spins of terbium ions have incommensurate (IC) ordering in the $(x, y)$ plane with the wave vector along the $y$-axis, $k_y = k \pm 0.42$ [164]. Later measurements showed a sinusoidal ordering of the terbium spins along the $x$-axis [165]. The phase transition from the IC into the commensurate (C) spins state in the magnetic fields $H_x$ and $H_y$ was observed in neutron diffraction [171, 172] and magnetostriction [169] investigations. A linear magnetoelastic coupling induced by magnetic field at the transition into the C state was revealed [172]. Field dependences of the electric polarizations $P_x$ and $P_z$ at $T = 2K$ can be seen in Fig.17. One can say here some considerations about a possible contribution of terbium spins into the observed linear ME coupling $P_x \propto H_{x,y}$ at $T < T'$ [172]. These remarks are based on the symmetric considerations.

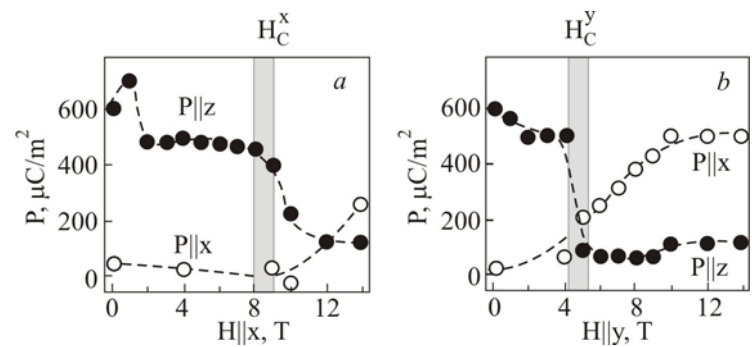

**Fig. 17**. *Adapted from* [172]. Electric polarization $P_x$ and $P_z$ of $TbMnO_3$ for $H_x$ and $H_y$ at T=2K .

**In the field** $H_x$**,** the symmetry of the *TbMnO$_3$* allows the invariant $P_x H_x g_y$ (see Table 1). One can suggest that the observed anomaly in the magnetic state of terbium at $H_x \sim 1$ T [165] is a result of the spin flop when the AF vector of terbium $g_x$ changes into $g_y$. This is the spin flop in the IC magnetic structure which exists up to $H_C^x \approx 9$ T (Fig 17(a)). In the IC state the average value of $P_x \sim H_x g_y$ is zero in consequence of the space modulation of $g_y$. The contribution to $P_x$ from the *Mn* subsystem is weak (see (4.7)). In contrary to $P_x$, the electric polarization $P_z \sim A_y A_z$ is not weak in the IC state. Magnetic phase transition of the IC into the C state takes place in the field $H_C^x$. In the C phase, the electric polarization $P_x \sim H_x g_y$ will be commensurable with the crystal structure [172]. Nonzero average value $P_x \sim H_x$ can be observed mainly in the single-domain crystal with the help of magnetic field. Apparently this case was realized in [26]. Contrary to the increase of $P_x$ with the increasing of magnetic field, the electric polarization $P_z$ decreases with the increasing of $H_x$ owing to the intrinsic reasons considered before and the monodomainization of the crystal by magnetic field. The $P_z$ appears due to magnetic inhomogenity, and in a single-domain magnetic state $P_z$ =0.

**In the field** $H_y$**,** the ME invariant $P_x H_y g_x$ exists. The transition IC $\rightarrow$ C takes place at $H_y \approx 4.5$ T (Fig.17$b$) as a result of the spin-flop transition. Then the spin flop $P_z = 0$ since $A_y = 0$. The value of $P_x \sim H_y g_x$, and it increases linearly with the increasing of the magnetic field up to $H \sim 10$ T.

**Electric control of spin helicity was** demonstrated in the *TbMnO$_3$* [173]. There is a number of ferroelectromagnets where electric polarization can be controlled by the magnetic field, but that time there were known only a few ferroelectromagnets whose magnetic structure could be changed by the electric field [11, 174, 175, 271]. A spin-polarized neutron scattering experiment [173] showed that the spin helicity, clockwise or counterclockwise, was controlled by the direction of the spontaneous polarization and hence by the polarity of the small electric field applied on cooling. In a noncollinear phase at $T < T_c$, the *TbMnO$_3$* has a spiral structure in the ($y, z$) plane and, hence, the vector chirality (vector which is perpendicular to the plane of spins) $\vec{C}$ is directed along the $x$-axis. Intensity of diffraction depends on the angle between the vectors $\vec{C}$ and the neutron spin $\vec{S}_n$ of neutron beam, $I_\uparrow \sim I_0(m_z - m_y)^2$ and $I_\downarrow \sim I_0(m_z + m_y)^2$, for $\vec{S}_n$ being parallel and antiparallel to the vector chirality $\vec{C}$, respectively. The intensity $I_\downarrow$ of the diffraction peak (4, +q, 1) was significantly larger than $I_\uparrow$ if before measuring the crystal was cooled in the electric field $E_z > 0$ ($E = 160 kV/m$), i.e. the single-domain FE state with spontaneous polarization $P_s > 0$ was created. In this case the helicity vector $C_x > 0$, and the counterclockwise spiral takes place (Fig. 18). If the crystal was cooled in the electric field $E < 0$, the intensity $I_\uparrow$ was larger than $I_\downarrow$, and the clockwise spiral ($P_s < 0, C_x < 0$) was realized (Fig. 18). If $m_z$ changes into $(-m_z)$, then $I_\uparrow$ changes into $I_\downarrow$ and vise versa. Notice that the changing of the sign of electric polarization $P_z$ to the opposite under the change of the sign of $M_z$ follows from the ME invariant (4.1). The reversal of poling electric field induces the reversal of magnitude relation between the intensities $I_\uparrow$ and $I_\downarrow$. It suggests that the spin helicity, clockwise or counterclockwise, can be controlled by not so large poling electric field. The reversal of spin helicity by the electric field at a temperature below $T_c$ requires a larger electric field ($E > 2 MV/m$) and could not be realized in this work [173].

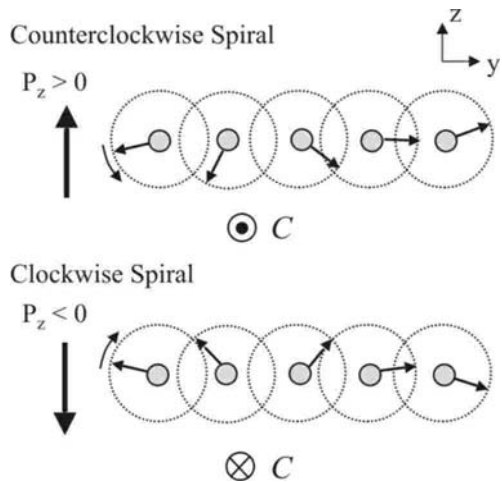

**Fig. 18**. *Adapted from* {173}. The relation between the spin rotatory direction (or helicity) and the direction of electric polarization in *TbMnO₃*.

Possible applications of ferroelectromagnets in optoelectronics revived an interest in the **serching for the ME excitations.** Elementary excitations in ferroelectromagnets (ferroelectromagnons) were predicted long ago [77] but at first were observed in the ferroelectromagnet $GdMn_2O_5$ [176] in 2003. In the IC magnetic state of *TbMnO₃*, the mixed ME mode excited by an alternating electric field was revealed in the terahertz spectral region [177]. Such excitations were also detected in the IC phase of *DyMnO₃* in the paraelectric state, therefore these excitations were called the "electromagnons". The frequency dependences of the imaginary part of the $x$-axis dielectric constant of *TbMnO₃* in the sinusoidal AF ($T = 30K$) and ferroelectromagnetic ($T = 20K$) phases revealed a broad maximum at the frequency $\nu_m \cong 20 cm^{-1}$ for all temperatures. No significant changes were observed when passing the FE phase boundary at $T = T_c \cong 28\,\text{K}$, although the damping of the mode decreased [177]. Magnetic field along the $z$-axis $B_z = 8\,\text{T}$ suppressed the ME excitations. In a large magnetic field, which is larger than the spin-flop field, the crystal is in a paraelectric and magnetic homogeneous states (see the previous considerations of the ME states in the $TbMnO_3$).

A significant change in the refraction index of $TbMnO_3$ at $T = T_c$ was predicted for the optical measurements of $\varepsilon_{zz}$ in an alternating electric field directed along the $z$-axis [178]. It was shown that a phason mode of the phase transition at $T_c$ is the Goldstone mode of electromagnons with the excitations of $A_z$ and $P_z$ (see below). Excitation spectrum and dielectric permittivity near the phase transition temperature would be calculated by using the Lagrange equations

$$\frac{d}{dt}\frac{\delta L}{\delta \dot{Q}_\alpha} = \frac{\partial L}{\partial Q_\alpha} - \frac{\partial}{\partial x_i}\frac{\partial L}{\partial(\partial Q_\alpha / \partial x_i)} + \frac{\partial^2}{\partial x_i^2}\frac{\partial L}{\partial(\partial^2 Q_\alpha / \partial x_i^2)} - \frac{\delta D}{\delta \dot{Q}_\alpha} \quad , \qquad (4.13)$$

where Lagrange's function $L = E_k - F$, $F$ is a potential energy (functional (4.4)), the density of kinetic energy is $E_k = \frac{1}{2}(\mu \dot{\vec{A}}^2 + \kappa \dot{\vec{P}}^2)$, the dissipative energy is $D = \frac{1}{2}(\eta_a \dot{\vec{A}}^2 + \eta_p \dot{\vec{P}}^2)$. To calculate the refraction index $n = qc/\omega$ (q is a vector of the electromagnetic wave with the frequency $\omega$), it is necessary to add the term $(-e_z P_z)$ to the potential $F$ (4.4) and take into account the connection between the $e_z$ and $P_z$ which follows from the Maxwell equations. The vectors $\vec{A}$ and $\vec{P}$ are the parameters of the considered phase transition. By solving equations (4.13) for the order parameters in the linear approximation with respect to small deviations $\vec{a}$ and $\vec{p}$ from their equilibrium values in a sinusoidal magnetic phase, one can find that the electromagnetic wave running along the $y$-axis excites only coupled $p_z$ and $a_z$ excitations. Restricting by the first harmonic, we put

$$p_z = p_0 \exp(iqy - i\omega t), \quad a_z = \exp(iqy - i\omega t)[a_1 e^{iky} + a_{-1} e^{-iky}) \, . \qquad (4.14)$$

The wave vector of the electromagnetic wave $q$ is significantly smaller than the modulation vector $k$. We are interested in the mode with q = 0, which is zero at the temperature $T_c$ of the FE transition (phason mode). The analysis of the coupling equations for the $p_0, a_1$ and $a_{-1}$ in the collinear phase ($T_c < T < T_N$) shows the presence of three modes of different symmetry. The AF mode with $\omega_3^2 = w/\mu$ and $a_1 = a_{-1}$ is not coupled with the electric polarization excitations. The AF excitations with $a_{-1} = -a_1$ are connected with the electric polarization excitations in the hybrid branches

$$\omega_{1,2}^2 = \frac{1}{2}\left[\omega_0^2 + \omega_p^2 \mp \sqrt{(\omega_p^2 - \omega_0^2)^2 + V}\right], \quad V = 8k^2\nu^2 A_1^2 / \kappa\mu, \quad (4.15)$$

$$\omega_0^2 = \mu^{-1}\left[w + \frac{2}{3}(a - a_c)\right], \qquad \omega_p^2 = \kappa^{-1}\chi.$$

The frequency $\omega_0$ is the AF frequency decreasing when $T \to T_c$, $\omega_p$ is the electro-dipole frequency, the $\omega_1$ and $\omega_2$ are modes of electromagnons with q = 0. The ME mixing occurs due to the interaction $V$ which appears in a modulated magnetic state with $k \neq 0, A_1 \neq 0$. The more exact value of the magnetic and electric transition temperature $T_c$ is the following [178]:

$$T_c = T_N - 3w/2\xi_1 + 6k^2\nu^2 w(u\chi\xi_1)^{-1}. \qquad (4.16)$$

From (4.15) and (4.16) one can see that the electromagnon mode $\omega_1 = 0$ at $T = T_c$. It means that the electromagnon mode $\omega_1$ is a phason mode of the considered ME transition when a magnetic transition from a collinear IC into a noncollinear IC state and the FE ordering derive simultaneously.

Usually the electric-dipole frequency $\omega_p$ is larger than the AF frequency $\omega_0$. Near the phase transition, when $\omega_1 \to 0$, one can consider that $\omega_2 \approx \omega_p >> \omega_1$ and take into account only a damping of the lower frequency branch $\omega_1$. Then the imaginary part of the dielectric permittivity $\varepsilon_2 = \mathrm{Im}\,\varepsilon_{zz}$ is the following:

$$\varepsilon_2 \approx \frac{\pi\kappa V}{\chi^2} \times \frac{\omega\Gamma_a}{\left[(\omega^2 - \omega_1^2)^2 + \omega^2\Gamma_a^2\right]}, \qquad \Gamma_a = \eta_a\mu^{-1}. \qquad (4.17)$$

From (4.17) it follows that the conditions for the observation of phason at the considered ME phase transition is the same as for the IC magnetic structure [179]. The damping (4.17) has maximum at the frequency

$$\omega_m^2 = \frac{\omega_1^2}{6}(2 - \Gamma^2 + \sqrt{16 - 4\Gamma^2 + \Gamma^4}) \ , \quad \Gamma = \Gamma_a / \omega_1 .$$

The damping curve has the Lorenz form when $\Gamma < \omega_1$, (Fig.19).

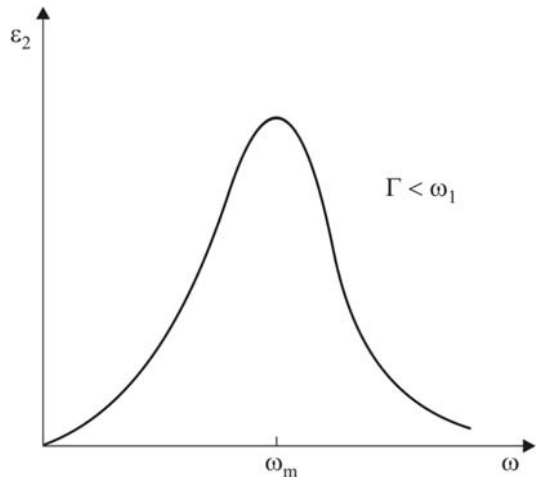

**Fig. 19.** Schematic frequency dependence of imaginary part $\varepsilon_2$ of a dielectric permittivity $\varepsilon_{zz}$ near $T_c$ in *TbMnO$_3$*.

Note that the excitations $(a_z, p_z)$ in the phason mode are perpendicular to the wave vectors $k_y$ and $q_y$, hence the phason mode $\omega_1$ is a soft TO mode.

Electromagnon spectrum in the FE and spiral magnetic phase of *TbMnO$_3$* at $T < T_c$ has been analyzed for different directions of the alternating electric field [180]. Besides the magnetic anisotropic term $wA_z^2$ in (4.4), another anisotropic energy $w'A_x^2$ was also taken into account owing to the orthorhombic symmetry of *TbMnO$_3$*. It was shown that in a linear approximation the magnons are coupled only with $(p_x, p_z)$ optical phonons in the waves running along the modulation vector $q = k_y$.

If the electric field is perpendicular to the spontaneous polarization ($e = e_x, p = p_x$), then the electromagnon frequencies are not soft. In this case the activation frequencies are determined by the magnetic anisotropic terms with the coefficients $w, w'$ and do not depend on temperature. This result is in a qualitative agreement with the experiment [177]. Two x-polarized electromagnon modes in [180] correspond apparently to the (1) and (2) modes observed by inelastic neutron scattering (Fig. 3d in [181]).

The electric field $e_z$ directed along a spontaneous polarization excites the electromagnons $(a_y, a_z, p_z)$ which are polarized in the spiral plane. Near the FE transition ($T < T_c$) one of the two modes is softened and in it the value $|a_y| << |a_z|$. At $T = T_c$, the frequency of this mode is zero, i.e. it is a phason mode in the spiral state. One may suppose that this mode is the sliding mode (3) ( Fig. 3 d in [181]).

Magnon-phonon modes coupled by the spin-orbit interaction in the helical FM were also analyzed in the theoretical investigation by Katsura et al.[182] (see 4.1.4).

*4.1.1.2 DyMnO$_3$ and other manganites*

The magnetic and the FE properties of $DyMnO_3$ are similar to those of $TbMnO_3$. The $DyMnO_3$ has the IC sinusoidal AF ordering below $T_N \sim 39$ K with the wave numbe r $k_y \approx 0.72$ and the electric polarization $P_z$ appeared below $T_c \sim 18$ K. The application of $H_x$ and $H_y$ causes the FE polarization vector to "switch" from $z$- to $x$-axis as it was observed in $TbMnO_3$. The ME effects in the magnetic fields $H_x$ and $H_y$ are similar to those in $TbMnO_3$. But the dielectric constant $\varepsilon_{xx}$ shows a gigantic change in the magnetic field $H_y \sim 5$ T near $T_c$ [183]. The value $\Delta\varepsilon / \varepsilon \sim 500\%$ is much larger than that in $TbMnO_3$. The behavior of the electric polarization along the $z$-axis is distinct from the case of $TbMnO_3$, where the spin-flop "switches off" the electric polarization in the canted AF phase. In the $DyMnO_3$, the $P_z$ does not vanish at high magnetic field up to 14 T. Possibly a larger magnetic field may induce the $A$-type AF phase and suppress the FE phase in $DyMnO_3$ [166].

**Solid solutions** $Tb_{1-x}Dy_xMnO_3$ also showed the transition from the collinear (sinusoidal) into noncollinear (spiral) magnetic structure with the FE ordering [184]. In the compounds $Eu_{1-x}Y_xMnO_3$ and $Gd_{1-x}Y_xMnO_3$ the magnetic transition into the spiral structure was accompanied by the FE ordering [185, 186]. Magnetic and dielectric investigations of the single crystal $Eu_{1-x}Ho_xMnO_3$ ($0 < x \le 0.5$) showed the appearance of the electric polarization $P_x$ at the temperature $T < T_c \approx 30$ K for $x \ge 0.2$ [187]. The value of the electric polarization was of the same order as in $TbMnO_3$. Decreasing of the temperature or increasing of the holmium content ($x$) led to the reorientation of the electric polarization from the $x$- to $z$-axis.

A strong ME coupling was discovered in the FE-AF $Eu_{0.75}Y_{0.25}MnO_3$ by studying of far-infrared spectrum [188]. Well-defined absorption peaks at 25 and 80 $cm^{-1}$ of the mixed magnon-phonon excitations – electromagnons were detected. A qualitative comparison of these experimental data with the theory [182] showed the discrepancy. The theory [182] predicts that electromagnons should be observed in the electric field $\vec{e}$ which is perpendicular to the electric polarization. However, in [188] electromagnons were detected only when the electric field was parallel to the spontaneous polarization $P_x$. This discrepancy was confirmed by the later measurements. The dielectric permittivity spectra of $Eu_{1-x}Y_xMnO_3$ in the composition range $0 \le x \le 0.5$ was studied in the terahertz frequency range [189]. Well-defined electromagnons were found for $x \ge 0.2$ below $T_c$ and only in the alternating electric field along the spontaneous polarization ($\vec{e} \parallel \vec{p} \parallel \vec{P}_s \parallel x$) that is in agreement with the results [188] and in contrary with the theoretical prediction [182].

### 4.1.2 Orthorhombic rare-earth manganates $RMn_2O_5$ (R=Pr – Lu, Y and Bi)

After the discovery of gigantic ME effects in the ferroelectromagnet $TbMnO_3$ a review of the similar colossal ME coupling started in the compounds $RMn_2O_5$, where a strong correlation of magnetic and electric properties was before detected [102]. Extraordinary changes

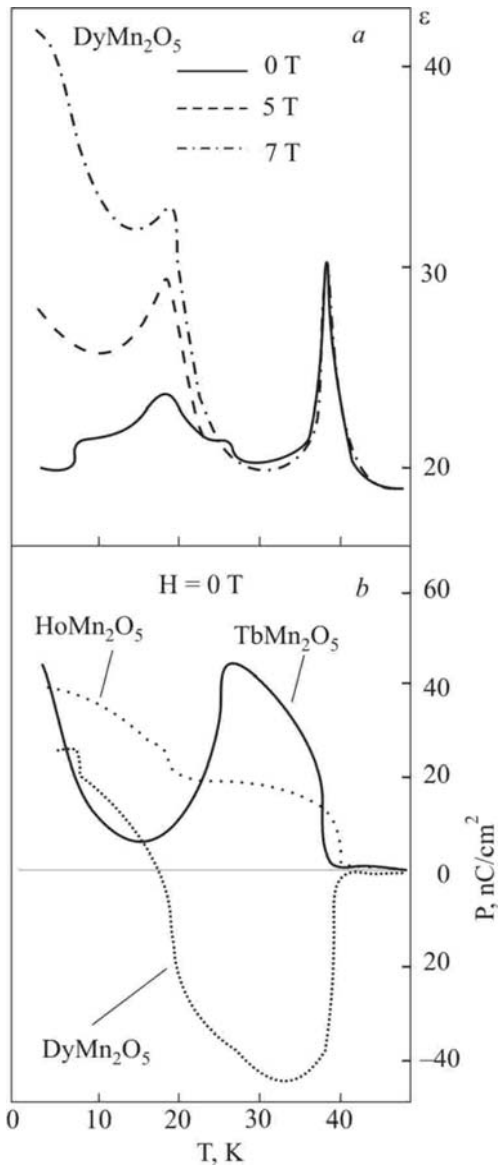

in a dielectric constant $\varepsilon$ with the magnetic field of $RMn_2O_5$ ($R = Ho, Dy, Tb$) were detected [190]. The most striking feature is the large $H$-dependence of $\varepsilon_y$ in $DyMn_2O_5$ below $T_N \sim 40$ K, when $H$ was applied along the $x$-axis (Fig. 20 $a$). If $H$ is zero, then $\varepsilon_y = \varepsilon$ exhibits a sharp peak at $T_c \approx 39$ K and shows the steplike anomalies at $T'_N \approx 27$ K and at the temperature of the spin ordering of $Dy$ near 8 K. The magnitude of $\varepsilon$ drastically increases with the increasing of $H$ and exhibits a maximum change of $\sim 109\%$ at 3 K by the application of 7 T. The spontaneous polarization ($P$) of all three studied crystals (Fig. 20 $b$) increases just below $T_c$ and shows a peculiar anomaly near $T'_N$. The absolute value of $P$ suddenly increases for $HoMn_2O_5$ or decreases for $(Tb, Dy)Mn_2O_5$ below $T'_N$. This complicated behavior was explained in the model of ferrielectricity, in which the net $P$ was composed of more than one component. It was proposed to compose the net $P$ of two components: one component appearing below $T_c$ and another below $T'_N$. These components are parallel to each other in

**Fig. 20.** *Adapted from* [190]. (a) T dependence of the dielectric constant $\varepsilon_y$ of $DyMn_2O_5$ in various $H_x$. (b) Spontaneous polarization $P_y$ of three crystals measured in zero H.

$HoMn_2O_5$ but antiparallel in $(Tb,Dy)RMn_2O_5$ [190, 191]. Neutron diffraction measurements [192] for $DyMn_2O_5$ are consistent with this model, provided that the FE sublattices are associated with different coexisting magnetic phases. The presence of multiple ordering wave vectors (along the $x$- and $z$-axes) observed in [192] means the presence of the competing ground states. A magnetic frustration, the presence of many low-energy magnetic states are the reasons of strong ME effects in $RMn_2O_5$. The appearance of the electric polarization and ME effects in $RMn_2O_5$ are connected mainly with the $Mn$ subsystem and magnetoelastic interaction [192]. Neutron diffraction measurements on $TbMn_2O_5$ [193] revealed the correlations between the magnetic and structural anomalies. Sign inversion of the spontaneous polarization $P$ was connected with the commensurare (CM) – incommensurate (ICM) magnetic transition. Geometrically frustrated magnetic structure in $TbMn_2O_5$ is stabilized by the antiferroelectric displacement of the $Mn$ ions, an example of the magnetic Jahn-Teller effect [193].

The lattice effects connected with the size of the rare-earth ion $R$ were studied by using the heat capacity measurements for single crystals $RMn_2O_5$ ( $R = Sm, Eu, Gd, Tb, Dy,$ and $Y$ ). Monotonic increasing of the transition temperatures with the decreasing of the size of $R$ correlates with the changes in structural parameters [194].

Compounds $RMn_2O_5$ have a complex orthorhombic structure with the space symmetry group $Pbam$. Manganese ions, in contrast to maganite $RMnO_3$, may have a different valency: $Mn^{3+}$ or $Mn^{4+}$. Competitive exchange interactions between the $Mn^{3+} - Mn^{3+}$, $Mn^{4+} - Mn^{4+}$ and $Mn^{3+} - Mn^{4+}$ rise to strong magnetic degeneration and numerous states with a small difference in energy. For example, in $YMn_2O_5$ the AF ordering at $T_N =45$ K is incommensurate along two directions $x$ and $z$ (2D-ICM). Electric polarization in this state is absent (Fig 21) [195]. At $T = T_D = 40.8$ K, the incommensuration along the $z$ axis disappears and electric polarization along the $y$-axis rises. The magnetic phase (1D-ICM) takes place in the narrow temperature interval $[T_{CM}, T_D]$ ( $T_{CM} = 40.0$ K). At $T_{CM}$, the ICM-CM ("lock-in") transition occurs and then the reverse transition CM-ICM ("lock-out") is observed. The last magnetic transition induces the second FE transition at $T_{c2} = 19\,\mathrm{K}$. A sequence of magnetic and FE temperature transitions reveals clear spontaneous ME effects.

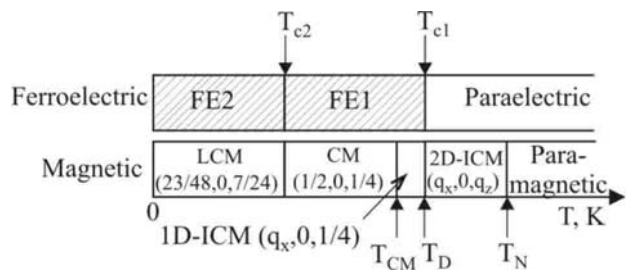

**Fig. 21**. The ferroelectric and magnetic phase transitions in $YMn_2O_5$. FE1 and FE2 represent ferroelectric phases below $T_{C1}$ and $T_{C2}$, respectively [195].

The frustrated magnetic structure of the compounds $RMn_2O_5$ is also very sensitive to the **external magnetic and electric fields**.

**The magnetic** field H ~ 1 T induces transition between the FE and paraelectric phases in $DyMn_2O_5$ at low temperature when the $Dy$ subsystem is ordered [196]. The reversal phase transition from the FE to paraelectric state induced by the magnetic field $H \sim 10$ T was detected in $HoMn_2O_5$ [197]. The direction of electric polarization in $TbMn_2O_5$ was highly reproducible, reversed by by magnetic field $H \sim 2$ T at low temperature, and the application of magnetic fields left a permanent imprint in the polarization [191].

In strong magnetic fields the novel ME effects were added to a cascade of ME effects [198 -200]. A strong magnetic field, besides the spin flop and spin-flip transitions, also induces other magnetic orientation transitions due to the existence of the states with small difference in energies in the frustrated $RMn_2O_5$. Magnetic phase transitions become apparent in dielectric anomalies. The spin- flop, spin- flip, lock-in and reorientation of spins are accompanied by the jump of electric polarization. In $ErMn_2O_5$ electric polarization changes its sign under the spin reorientation from the $(x,z)$ into $(x,y)$ plane [201]. The investigations of magnetic, magnetoelastic and ME properties of the manganates family $RMn_2O_5$ ($R = Eu, Y, Gd, Er$) in strong pulsed magnetic fields up to 250 $kOe$ [200] showed a nonmonotonic behavior of electric polarization and a change of its sign in the strong magnetic field. These anomalies are correlated with the changes of magnetic structure. Because of nonmagnetic ground states of the ions $Eu^{3+}$ and $Y^{3+}$, the ME effects in $(Eu,Y)Mn_2O_5$ take place only due to spins of $Mn^{3+}$. The influence of the rare-earth ions on the ME properties in other compounds is significant at low temperature when the spins of these ions are ordered. Then the value of the ME effects enhances, and new phase transitions appear since the magnetic field of the rare-earth ions changes the spin state of manganese subsystem [198].

Phenomenological treatment of the observed ME effects in $RMn_2O_5$ was suggested [200, 202]. This model, based on the consideration of two AF states, $\vec{A}$ and $\vec{G}$, was developed in the exchange approximation for homogeneous magnetic structure. There was shown the presence of the ME exchange energy term

$$F_{ME}^{ex} = \alpha_1 P_y (\vec{A}^2 - \vec{G}^2). \qquad (4.18)$$

A crystal has domains of two types, $\vec{A}$ and $\vec{G}$, with opposite directions of electric polarization $P_y$ in the $\vec{A}$ and $\vec{G}$ domains. This state is possible if $T < T^* < T_N$, where $T^*$ is the temperature of the specific spin reorientation observed experimentally. However, there is a relativistic anisotropic ME term

$$F_{ME}^{rel} = \alpha_2 P_y [\vec{G} \times \vec{A}] \qquad (4.19)$$

which exists in the temperature interval $T^* < T < T_N$ if manganese spins lie in the $(x, y)$ plane. The ME terms (4.18) and (4.19) exist together with the inhomogeneous ME energy which can also induce electric polarization. In any case, the FE transition is of magnetic origin. If spins lie in the $(x, y)$ plane just below $T_N$, then the electric polarization $P_y$ can appear due to relativistic energy (4.19) or some inhomogeneous ME interaction from the ICM structure. In the case of orientation of manganese spins in the $(x, z)$ plane at $T_N$, the electric polarization can rise only due to ICM structure. The exchange term (4.18) appears only at $T < T^*$. In the model considered the AF structure of $RMn_2O_5$ is noncollinear exchange structure of the type "exchange cross". Just below $T_N$ the AF ordering pairs $\vec{S}_1, \vec{S}_2$ and $\vec{S}_3, \vec{S}_4$ are perpendicular each other. Below $T^*$ the intrinsic spin flop takes place when the AF pair $\vec{S}_1, \vec{S}_2$ orients along the pair $\vec{S}_3, \vec{S}_4$. The $\vec{G}$-domain state occurs when $S_3 \sim S_1$, $S_4 \sim S_2$. In the case when $S_3 \sim -S_1, S_4 \sim -S_2$ the $\vec{A}$- domains appear. Near the temperature $T^*$ the crystal displays sharp ME properties in the magnetic field [200].

**The influence of the electric field on magnetic structure** is not much studied. Last time resonant magnetic soft $x$-ray diffraction experiments were done in the ferroelectromagnet $ErMn_2O_5$ [203]. The consequence of the phase transitions in this compound is similar to that in $YMn_2O_5$ (Fig.21). The electric polarization appears under magnetic phase transition from 2D-ICM into 1D- ICM state and leaves in the CM phase. The electric field $E \sim 13 kV / cm$ promotes the transition from ICM to CM phase changing the value and direction of magnetic moments.

**The ME dynamics** in $RMn_2O_5$ ($R = Gd, Eu$) was studied by using optical measurements in the frequency and temperature ranges 20- 300 $GHz$ and 5- 50 K, respectively, [176, 204]. These compounds possess the coinciding FE and magnetic ordering temperatures (40 K and 30 K accordingly for $R = Eu$ and $Gd$). The magnetic resonance spectra of $GdMn_2O_5$ was found to be characteristic for the homogeneous long-range magnetic order for both the AF $Mn$-subsystem and the magnetic, possessing a large FM moment, $Gd$ − subsystem. The gap in the magnetic spectrum decreasing with the increasing of temperature was $130\,GHz$ near $T_c$. Absorption lines were excited also by the microwave electric field oriented with respect to crystal axis $x, y$ and $z$. These lines with $\omega \sim 130\ GHz$ showed a sharp temperature dependence with maximum near $T_c$ and it was mainly of electrical origin. Remind that the frequency $\sim 130\ GHz$ is close to the magnetic resonance frequency observed near $T_c$. Absorption lines of electrical nature were significantly changed by the magnetic field $H \sim 1$ T at low temperature. All observed excitations disappeared in the paraelectric –paramagnetic phase at $T > T_c \approx 30$ K. Hence one may conclude that the observed AF-FE excitations are of mixed electric and magnetic nature, i.e., they are ferroelectromagnons (in a modern terminology "electromagnons") predicted in ferroectromagnets back in 1969 [77]. Thus, the evidence of **electromagnons** was **at first** found by Golovenchits and Sanina [176] in the ferroelectromagnet $GdMn_2O_5$ with $T_m = T_c$. In $EuMn_2O_5$, where the ion $Eu^{3+}$ is nonmagnetic (ground state $^7F_0$ ) , electromagnons were not detected. Therefore, it was concluded that the magnetic rare-earth ions (like $Gd$ ) play a significant role in the observation of this ME phenomenon by enhancing the ME effect.

The role of the rare-earth f-electrons in the observed electromagnons was recently investigated in the single crystals $RMn_2O_5$ ( $R = Y, Tb$) [205]. In these compounds the ground states of $Y^{3+}$ and $Tb^{3+}$ are nonmagnetic and magnetic, respectively. Electromagnons with the frequency $\omega \sim 7 - 10\ cm^{-1}$ in the FE- ICM phase were detected by spectral measurements in alternating electric field directed along the spontaneous polarization ($\vec{e} \parallel y$). Electromagnons with higher frequency $\omega \sim 20\ cm^{-1}$ were observed in the FE-CM phase, but the frequency peaks were weaker. In the paraelectric - paramagnetic phase electromagnons were absent. The spectra of $YMn_2O_5$ and $TbMn_2O_5$ were very similar to those which prove the $Mn$ nature of electromagnons in these compounds. Additional far-infrared and neutron measurements are necessary to clear up the role of a rare-earth magnetic subsystem in the dynamic ME coupling. Note that strong coupling of magnetic and electro-dipole excitations in the considered compounds [205] was found only in $\vec{e} \parallel y \parallel \vec{P}_s$ orientation that is (as well as the results [188, 189]) in contrary with the theory [182].

*4.1.3  Other helical compounds*

Ferroelectricity driven by magnetic ordering was discovered not only in manganites and manganates. In centrocymmetrical $Ni_3V_2O_8$ (NVO) in noncollinear ICM phase (spins lie in the $(x,y)$ plane) there was detected the electric polarization along the y-axis ($P_y$) existing at low temperature in the narrow interval $3.9 < T < 6.3$ K where the noncollinear ICM state takes place. Below 3.9 K the canted AF phase without FE exists. The ferroelectricity in NVO was explained by the phenomenological Landau-like model [206] similar that in $TbMnO_3$ [165].

Recently the FE ordering was discovered in the spiral magnetic state of the quantum quasi-one-dimensional (1D) $S = 1/2$ magnet $LiCu_2O_2$ [207]. This crystal contains an equal number of $Cu^{1+}$ and $Cu^{2+}$ ions, only the latter of which carries spin $S = 1/2$. The ions $Cu^{2+}$ form chains running along the $y$-axis. This compound is one of few 1D strong frustrated magnets where the competing FM and AF interactions may result in the magnetic structure with quantum properties. In classical magnets such a competition usually results in the ICM ground state. In the quantum system a strong frustration can destroy the long-range ordering producing only local spin correlations, for example, spin-liquid phase. In a conventional model for the spin $S = 1/2$ chain with competing nearest-neighbor (NN) $J_1$ (FM) and next-nearest-neighbor (NNN) $J_2$ (AF) interactions, the spin-liquid state can be realized for , $0.24 < \alpha < 0.5$ where $\alpha = J_2 / J_1$. A competition between the ICM structure and CM quantum properties in $LiCu_2O_2$ was noticed in the neutron diffraction experiment [208] and spins were supposed to lie in the $(x,y)$ plane. However, the observation of the electric polarization $P_z$ along the $z$-axis [207] and investigation of the behavior of electric polarization in external fields led to the conclusion about the disposition of spins in the $(y,z)$ plane. The behavior of electric polarization in $H = 0$ can be understood in a simple model of the spiral-magnetic FE [233]. The value of the parameter $\alpha$ $R^{3+}, R^{3+}, Mn^{4+}$ was obtained as $\alpha \approx 0.50 - 0.65$. Similar as in $TbMnO_3$, the magnetic field $H_y$ decreased $P_z$ and increased $P_x$. But it was unclear why the magnetic field $H_z$ directed along the spontaneous polarization increased its value. It was supposed that the magnetic structure of $LiCu_2O_2$ was more complex with nonzero $x$-component of spins. The last experiment with the polarized neutrons [209] confirmed the model [207] and indicated the agreement with the spin-current theory [233]. It means that the Dzyaloshinskii-Moriya mechanism is applicable even to the $e_g$-electron quantum spin-system. But the received value of the parameter $\alpha = 0.09 \sim 0.20$ is in discrepancy with the results [207, 173]. For the thorough understanding there would be needed further analysis of the magnetic structure and its quantum dynamics.

The second example of the quasi-one –dimensional ferroelectromagnets with quantum spin chains ($S = 1/2$) is cuprate $LiCuVO_4$ which has an orthorhombic inverse spinel structure. The orthorhombic distortion results from a cooperative Jahn-Teller effect of the $Cu^{2+}$ ions at the octahedral sites. The FE properties in $LiCuVO_4$ were discovered at low temperature simultaneously with the appearance of helical magnetic order at $T_m = T_c \sim 2.5$ K [210]. In the ordering state spins are modulated along the $y$-axis and lie in the $(x, y)$ plane with helix (spiral) vector $\vec{e}$ (it is perpendicular to the spin plane) along the $z$ axis. The spiral vector can easily be switched by the magnetic field [211]. Spontaneous electric polarization is parallel to the $x(a)$-axis. The ferroelectromagnet $LiCuVO_4$ is convenient for checking theoretical models as its magnetic structure is a simple one-dimensional chain with the spins $S = 1/2$. The studying of dielectric properties of $LiCuVO_4$ as functions of temperature, magnetic-field strength, and field direction was recently presented [212]. The results are in a good consistence with phenomenological predictions [208] (Fig. 22). According to the spin-current theory [233] and the paper [213], the vector of electric polarization is proportional to $[\vec{e} \times \vec{k}]$. The modulation vector $\vec{k}$ is directed along the $y$-axis. At zero the magnetic field spins lie in the $(x, y)$ plane, electric polarization $\vec{P} \parallel x$. At a critical field $H_1 \approx 2.5$ T, the spiral vector $\vec{e}$ is turned into the direction of the magnetic field. If a magnetic field is applied along the $z$-axis, the vector $\vec{e}$ remains parallel to $z$, and $\vec{P} \parallel x$ is maintained up to $H_2 \sim 7.3$ T, where a collinear structure is established. Paraelectric state takes place if the external magnetic field is directed along the $y$-axis since then electric polarization $\vec{P} \sim [\vec{e} \times \vec{k}] = 0$. In the case of $\vec{H} \parallel \vec{e} \parallel x$, the electric polarization is along the $z$-axis. Thus, the experimental measurements (see also the referencies in [212]) are in agreement with the theoretical model [233, 213].

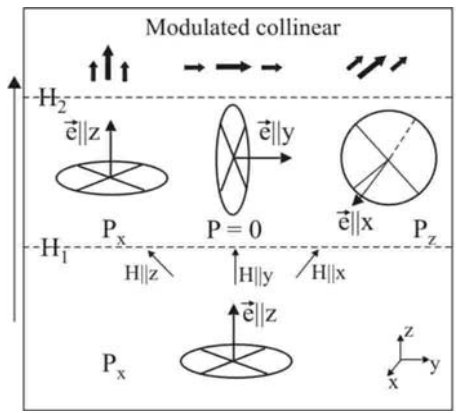

**Fig. 22**. *Adapted from* {212}. Schematic sketch of the spin configurations as a function of the external magnetic field in the ferroelectromagnet $LiCuVO_4$.

A new ferroelectromagnet $MnWO_4$ was recently discovered [214]. This crystal has the wolframite structure and one type of magnetic ions $Mn^{2+}$ with spin $S = 5/2$. Spins are aligned in the zigzag chains along the $c$-axis. Although this structure may be regarded as a one-dimensional Heisenberg spin chain, the $MnWO_4$ undergoes two modulated phases below $T_N \sim 13.5\,$K and the phase transition of the first kind from the noncollinear ICM to CM phase at $T \sim 7.6$ K. The electric polarization $P_y$ appears at $T \sim 12.7$ K. A reorientation electric polarization from the $y$ direction to the $x$ direction was observed when a magnetic field above 10 T was applied along the $b$-axis. This result shows that the reorientation of electric polarization can be induced by a magnetic field in a simple system without the rare-earth $4f$ moments. Note that such a reorientation (the so-called "polarization flop") was first observed in the $TbMnO_3$. It was also detected at temperatures above the ordering temperature of terbium when the rare-earth magnetic subsystem is paramagnetic [26].

The FE transition was discovered in a ferrimagnetic spinel oxide of $CoCr_2O_4$ upon the transition to the conical spin order below 26 K [215]. In a cubic spinel $CoCr_2O_4$ the ferrimagnetic (FIM) transition occurs at $T_m = 93$ K with the magnetic moment along the $z$-axis. With further lowering of the temperature the compound undergoes the transition to conical spin state with the incommensurate propagation vector of $[kk0]$ ($k \sim 0.63$) at $T \approx 26$ K and electric polarization directed along the $[\bar{1}10]$ (Fig. 23). The observed magnitude of the electric polarization ($\sim 2\mu C/m^2$) is 2-3 orders of magnitude smaller than those of $RMnO_3$ and $RMn_2O_5$. The reversal of the spontaneous magnetization by a small magnetic field ($\sim 0.1$ T) induced the reversal of the spontaneous polarization, indicating a strong clamping of the FM and FE domain walls in this cubic compound with weak magnetic anisotropy.

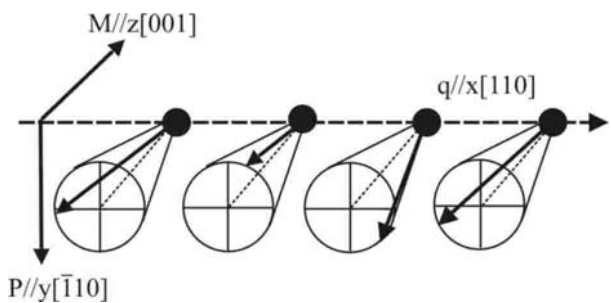

**Fig. 23.** *Adapted from* [215]. Relation among the net magnetization $\vec{M}$, the spiral vector $\vec{k}$, and induced polarization $\vec{P}$ in $CoCr_2O_4$.

In all mentioned above helical compounds the direction of electric polarization was in the agreement with the prediction made by Katsura *et al.* [233]: $\vec{P}$ lies in the spin plane and is perpendicular to the modulation vector $\vec{k}$. However, this so-called "spin-current" mechanism is not valid in $RbFe(MoO_4)_2$ which is the AF on a triangular lattice [216]. Neutron measurements showed the existence of the ICM structure along the $z$-axis below $T_N = 3.8$ K. Electric polarization in this ICM state, $P_z \sim 5\mu C/m^2$, was observed along the same $z$-axis as a modulation vector. The observed ME properties of $RbFe(MoO_4)_2$ [216] were explained by the symmetry-based Landau theory. It means that the model [233, 213] for FE in spiral magnets is not general (see Sec. 4.1.4).

**Semiconductor- ferroelectromagnet**

Recently the single crystals $Tb_{0.95}Bi_{0.5}MnO_3$ were grown and studied in a wide range of temperatures [217]. A small substitution of terbium by metallic bismuth led to a colossal change in dielectric constant ($\varepsilon \sim 10^4 - 10^5$) and ferromagnetic ordering of manganese spins. In spite of the isovalency of the substitution of $Tb^{3+}$ by $Bi^{3+}$ the charge bearers and $Mn^{3+}$ and $Mn^{4+}$ ions appeared.

As a result, the $Tb_{0.95}Bi_{0.5}MnO_3$ in contrary to $TbMnO_3$ is a semiconductor. The $A$ ($Tb^{3+}$) and $B$ ($Mn^{3+}$) ions in the perovskite structure $ABO_3$ lie in the alternating layers. The size of the alloy $Bi^{3+}$ ions is significantly larger than the size of $Tb^{3+}$ that results in the local lattice distortion. This distortion creates advantageous conditions for the displacement on the nearest neighbouring layers to the bismuth layer of the $Mn^{4+}$ ions with the size smaller than $Mn^{3+}$. Thus, the manganese ions with different valency and redundant electron appear ($Mn^{3+} \leftrightarrow Mn^{4+} + e$). Electronic exchange between $Mn^{3+}$ and $Mn^{4+}$ (double exchange [218]) leads to ferromagnetism. Hence, different phases occur in the crystal. The similar result was received when the compounds $RMn_2O_5$ ($R = Gd, Eu$) were alloyed by $Ce^{4+}$ [219]. The primary structure of $RMn_2O_5$ is more complex than that of $TbMnO_3$. The $R^{3+}, Mn^{4+}$ and the Jahn-Teller ions $Mn^{3+}$ lie in different layers and have different local surrounding. The ions $Ce^{4+}$ substitute the rare-earth ions in the investigated single crystals $Gd_{0.75}Ce_{0.25}Mn_2O_5$ and $Eu_{0.8}Ce_{0.2}Mn_2O_5$. The alloying of the $R^{3+}$ by metallic $Ce^{4+}$ leads to the appearance of a free electron ($R^{3+} = Ce^{4+} + e$) which moves in the manganese subsystem. The appearance of the electron in the neighboring layer with ions $Mn^{4+}$ results in the change of manganese valency ($Mn^{4+} + e = Mn^{3+}$). In all compounds $Tb_{0.95}Bi_{0.5}MnO_3$, $Gd_{0.75}Ce_{0.25}Mn_2O_5$ and $Eu_{0.8}Ce_{0.2}Mn_2O_5$ the thin quasi-2D layers appear containing alloying ions, ions $Mn^{3+}, Mn^{4+}$ and electrons. At low temperatures ($T < 100$ K) the bulk of the crystal volume is dielectric; the layers with charges occupy a small phase volume. Jump electron conductivity rises with the increasing of temperature and leads to the formation of periodic alternative layers containing ions $Mn^{3+}, Mn^{4+}$ and primary layers with manganese of definite valency. Lattice distortions appear at the boundaries of the layers. A periodic space distribution of the charge in the whole volume of the crystal at temperatures $T > 180$ K induces the appearance of the ferroelectricity with dielectric permittivity $\varepsilon \sim 10^4 - 10^5$. Moreover, a double exchange between the ions $Mn^{3+} - Mn^{4+}$ leads to the ferromagnetic orientation of spins of these ions. A magnetic field induces the novel phase transitions in the considered compounds. The proposed model [219] is confirmed by the measurements of frequency dependence of dielectric constant and conductivity and by being compared with the conventional theories. The investigations [219] opened inspiring perspectives for using semiconductor- ferroelectromagnets in the sciences, techniques, and practices.

### 4.1.4 Theories of magnetic induced FE in spiral magnets

A colossal ME effect discovered in the spiral magnets $RMnO_3, RMn_2O_5$ and other compounds was observed in the noncollinaer (frustrated) magnetic structures. A microscopic mechanism of magnetic induction of electric polarization in noncollinear magnets was investigated by Katsura *et al.* [233] in the cluster model with two transition metal ions $M1, M2$ with an oxygen atom $O$ between them (Fig. 24). In the octahedral ligand field the d-states of the transition metal $M$ are split, and the orbitals $d_{xy}, d_{yz}$ and $d_{zx}$ have lower energy. A magnetic moment of the j-th site is described by the unit vector $\vec{e}_j$ in Fig. 24. The inversion symmetry exists at the middle point of the two magnetic ions, i.e., there is no the Dzyzloshinskii-Moriya (DM) interaction, and the generic noncollinear magnetic ordering is assumed to be realized by the competing exchange interactions. It is supposed that the oxygen $p$-orbitals are empty. Because of the existence of the oxygen atom, there are hopping processes between the $M$ site and the $O$ site, $V$ is the transfer integral, $\Delta$ is the energy difference between the $p-$ and $d-$ states. Using the second-order perturbation theory for the vector of the electric polarization $\vec{P} = \langle e\vec{r} \rangle$, the following expression was obtained:

$$\vec{P} \cong -\frac{eV}{3\Delta} I \frac{\vec{e}_{12} \times (\vec{e}_1 \times \vec{e}_2)}{|\cos\frac{\vartheta_{12}}{2}|} \quad , \qquad (4.20)$$

where $e$ is the electron charge, $\vec{e}_{12}$ is the unit vector parallel to the direction of the bond from site $M1$ to site $M2$, and $\vartheta_{12}$ is the angle between the two vectors $\vec{e}_1$ and $\vec{e}_2$, i.e., $\vec{e}_1 \cdot \vec{e}_2 = \cos \vartheta_{12}$.

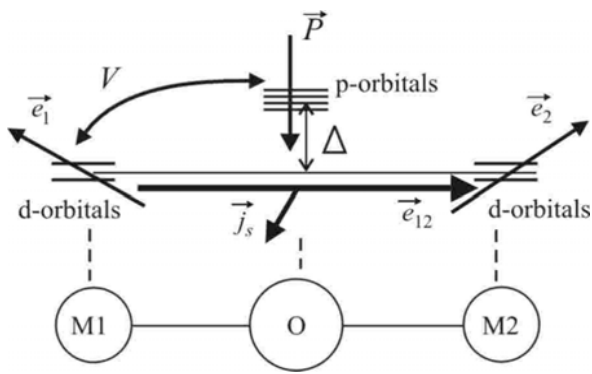

**Fig. 24**. *Adapted from* [233]. The cluster model by Katsura et al.[233]. Spin directions $\vec{e}_1, \vec{e}_2$ ; $\vec{P}$, electric polarization; $\vec{e}_{12}$ is unit vector connecting M1 and M2; $\vec{j}_s \propto \vec{e}_1 \times \vec{e}_2$ is spin current vector.

The integral $I$ is the following matrix element:

$$I = \int d_{yz}^{(j)}(\vec{r}) y p_z(\vec{r}) d^3\vec{r}, \qquad (j = 1,2). \tag{4.21}$$

The estimation of the integral $I$ in the paper [233] gives the value $I \sim 1\text{Å}$.

The noncollinear spin directions $\vec{e}_1$ and $\vec{e}_2$, due to the spin orbital interaction, induce the spin-current $\vec{j}_s \sim \vec{e}_1 \times \vec{e}_2$ between $M1$ and $M2$ (see Fig. 24). The direction of the electric polarization $\vec{P}$ is given by

$$\vec{P} \propto \vec{e}_{12} \times \vec{j}_s \quad. \tag{4.22}$$

The supposed microscopic mechanism is valid in centrosymmetrical magnets with exchange nature of spiral structure without the DM interaction. The last interaction occurs in noncentrosymmetrical magnets and leads to the existence of the long-wave relativistic spiral structure.

Phenomenological treatment of the magnetic origin of the electric polarization in spiral magnets was presented by Mostovoy [213]. For the cubic crystal with one magnetic vector $\vec{M}(\vec{r})$ the magnetically induced electric polarization is

$$\vec{P} \propto [\vec{M}(\nabla\vec{M}) - (\vec{M}\nabla)\vec{M}]. \tag{4.23}$$

If we introduce the vector modulation $\vec{k}$ and the vector chirality $\vec{C}$, which is perpendicular to the spin plane, i.e. $\vec{C} \sim \vec{j}_s$, then the relation (4.23) may be rewritten as

$$\vec{P} \sim \vec{C} \times \vec{k} \quad. \tag{4.24}$$

The relation (4.24) is in agreement with the observed direction of electric polarization in $CoCr_2O_4$, $MnWO_4$, $LiCu_2O_2$, and $LiCuVO_4$, but it is in contradiction with the corresponding data for $RbFe(MoO_4)_2$ [216]. Furthermore, the simple model [213] can not describe the reorientation of the electric polarization by magnetic field in $TbMnO_3$. This simple model is based on the assumption that the spin state can be described by a single magnetization vector $\vec{M}$ and ignores the presence of other magnetic states in the AF with a number of spins, for example, $G, C$ states in $TbMnO_3$ [167]. It also ignores the symmetry of the magnetic unit cell that resulted in incorrect conclusions drawn by Mostovoy [220- 222]. The importance of the symmetry of the magnetic unit cell, i.e. the permutation of magnetic sublattices under the symmetry operations in AF, was still pointed by Turov [32]. Recent detailed microscopic analysis of the ME interactions in cuprates $LiCu_2O_2$, and $LiCuVO_4$ showed that the in-chain spin current does not produce an electric polarization [223]. It means that the spin-current theory [233] can not explain the appearance of the electric polarization in these compounds, and the FE in the cuprates originates from an out-of-chain stuff. Thus, although the observed direction of the electric polarization $\vec{P}$ in the cuprates (if this result is undoubted) is in accordance with the rule (4.24), the understanding of its magnetic origin needs further microscopic analysis.

In summary, the popular model [213] does not represent a universal phenomenological description of the FE induced by the ICM order.

The model of Kitsura *et al*. [233] takes into account only electron orbital states and ignores the lattice displacements also leading to the appearance of electric polarization. The microscopic mechanism taking into account the oxygen displacements in $RMnO_3$ was recently proposed [224]. This mechanism is based on the antisymmetric DM type magnetoelastic coupling. It is worth noting that the Dzyaloshinskii-Moriya (DM) interaction energy is proportional to the relativistic term $\vec{M} rot \vec{M}$ and exists only in noncentrosymmetric crystals [225, 226]. However, the agreement of the theoretical result [224] with the experimental one requires the value of the DM parameter to be 2 orders of magnitude larger than the reasonable microscopic estimations [224].

A detailed construction of the Landau expansion, where the symmetry of the magnetic unit cell in the ICM ferroelectromagnets $TbMnO_3$, $Ni_3V_2O_8$, $MnWO_4$, $TbMn_2O_5$, $YMn_2O_5$, and $RbFe(MO_4)_2$ was also taken into account, was presented by Harris [222]. This work convincingly demonstrated that the spin-current model in its simplest form is not applied in the case of $RbFe(MO_4)_2$, where the electric polarization lies along the modulation vector $\vec{k}$. The Landau theory was used to give a qualitative description of the behavior of various susceptibilities near the phase transition.

No conventional application of the theory of irreducible corepresentations employed antiunitary operators was represented for the description of the FE of magnetic origin [227]. It has shown that FE can appear even in the collinear structure, that helical structure is not always polar, and in some cases symmetry allows parallel orientation of $\vec{P}$ and $\vec{k}$ .

Theoretical models considered above were designed for conventional insulators. For the insulators with a partially filled band and strong electron-electron interaction (Mott's insulators), an additional mechanism of contribution into electric polarization was supposed in the case of a strong spin-orbital interaction [228]. It was assumed that the magnetic ordering of the localized spins generates the same magnetic ordering for the electrons in the partially filled band. The electric current inducing magnetic ordering in insulator is compensated by the electric current induced from the strong spin-orbital coupling. This mechanism creates a typical in $RMnO_3$ value of electric polarization $P \sim 0.1 \mu C / cm^2$ if the spin-orbital energy is of order $0.1\ eV$ . The strong spin orbital energy means that it is much larger than the Coulomb energy. This condition is unrealistic for the $3d$ -oxides [223].

Last time *ab initio* computations using the local spin-density approximation (LSDA) for the calculations of band structures are very popular. In the conventional nonmagnetic FE the value of electric polarization computed by the modern *ab initio* band structure methods agrees well with the ones observed experimentally. In ferroelectromagnets, however, such computations usually predict significantly larger values [229]. This discrepancy in $HoMn_2O_5$ disappeared when the Coulomb interaction between the manganese $3d$ electrons was calculated by using *ab initio* computations [230]. The ionic part from the lattice displacements and electronic part from the valence electrons in the electric polarization have opposite signs and almost canceled each other. Thus, the electron-electron interactions drive a decimation of the resulting net polarization.

Unphysical overestimation of the spin-induced electric polarization in theoretical calculations was noted recently by Moskvin *et al.* [223]. It was also marked that Katsura *et al.* [233] proceeded from unrealistic scenario assuming that some effective Zeeman field align noncollinearly the spins of the $3d$ electrons. To justify the authors' [233] approach, the energy of the Zeeman splitting should be enormously large, at least of order of several $eV$ . Thus, Katsura *et al.* [233] started with an unrealistic for the $3d$ -oxides strong spin-orbital coupling and fully neglected the low-symmetry crystal field and orbital quenching effect. In addition, the authors overestimated the value of the integral $I$ (4.21), which defines a maximal value of the electric-dipole moments. Their estimation of $I$ erroneously used the $3d$ and $2p$ functions localized at the same site. In fact, this integral is estimated to be $I \approx R_{dp} S_{dp\pi}$ , where $R_{dp}$ is the cation-anion separation and $S_{dp\pi}$ is the $dp\pi$ overlap integral. Thus, the electric polarization induced by the spin current is 1-2 orders of magnitude smaller than the authors' estimations [233]. Moskvin *et al.* [223] also criticized the using of basic starting points of the current versions of LSDA for the estimation of the spin-dependent electric polarization in ferroelectromagnets. They emphasize two weak points of the so-called *first-principles calculations*. First, these calculations suppose that the spin configuration induces breaking of spatial symmetry what is quite the opposite to the conventional picture when the charge and orbital anisotropies induce the spin anisotropy. Second, the LSDA approach neglects quantum fluctuations.

A systematic microscopic theory of the spin-dependent electric polarization, which implies the derivation of effective spin operators for nonrelativistic and relativistic contributions to the electric polarization of the generic three-site two-hole cluster such as $Cu_1 - O - Cu_2$ and does not imply any fictious Zeeman fields to align the spins, was presented by Moskvin *et al.* [223]. The authors used the conventional [231, 232] approaches taking into account the quenching of orbital moments by low-symmetry crystal field, strong intra-atomic correlations, the $dp$-transfer effects, and rather small spin-orbital coupling. For an effective dipole operator $\hat{\vec{P}}$ there were obtained the expressions for the nonrelativistic exchange interaction

$$\hat{\vec{P}}_{ex} = \vec{\Pi}(\vec{s}_1 \cdot \vec{s}_2)$$

(4.25)

and the exchange-relativistic DM interaction

$$\hat{\vec{P}} = \vec{\Pi}[\vec{s}_1 \times \vec{s}_2] \ .$$

(4.26)

The expressions (4.25) and (4.26) take into account both local (covalent) and nonlocal (overlap) effects. The estimations of the contributions of the exchange (4.25) and DM (4.26) mechanisms in the electric polarization of the $3d$ oxides showed that the redistribution of the local on-site charge density due to $pd$ covalency and exchange coupling, i.e. exchange mechanism, was predominant. In the frames of a mean-field approximation, the simple expression for electric polarization arising due to the exchange mechanism is the following:

$$\vec{P} = \sum_n \vec{\Pi}_n (\vec{s} \cdot \vec{S}_n) \ ,$$

(4.27)

where $\vec{S}_n$ is a spin from the surroundings of $\vec{s}$, and

$$\vec{\Pi}_n = 2I_{gu}(\vec{R}_n)\langle g \mid e\vec{r} \mid u\rangle(\varepsilon_u - \varepsilon_g)^{-1} \ .$$

(4.28)

Here the particle states with energies $\varepsilon_g$, $\varepsilon_u$ have a definite spatial parity, even (g) or odd (u), respectively; $I_{gu}$ is the exchange energy between the states of different parity. It is seen from (4.28) that in the molecular field approximation the nonzero electric dipole moment shows up only for spin-noncentrosymmetric surrounding, that is, if the condition $\langle \vec{S}(\vec{R}_n) \rangle = \langle \vec{S}(-\vec{R}_n) \rangle$ is broken.

In summary, the paper [223] cited above contains a correct criticism of strong overestimation of the computing results for electric polarization and demonstrates a traditional way of accurate microscopic calculations.

The role of the bond-bending in the spin-current model [233] was recently analyzed [234]. Microscopic considerations showed the possibility of the enhancement of electric polarization due to orbital hybridization. A simple relation between the electric polarization and the helical wave vector $\vec{k}$ was revealed, $\vec{P}(\vec{k}) = -\vec{P}(-\vec{k})$. It should be noted that this relation is in the expressions (4.7) for the electric polarization $P_{z,x} \sim k_y$ in $TbMnO_3$ [167]. If the helical vector $\vec{k}$ changes its sign, then the change $A_z \to -A_z$ in (4.5) and the change of the clockwise spin helicity into the counterclockwise one take place (Fig. 18, [173]).

**The theory of the collective ME modes in a ferroelectromagnet with helical FM structure** was developed for the case of a simple FM spiral with helical vector $\vec{C} \parallel y$ (spins lie in the ($x,z$) plane) and wave modulation vector $\vec{k} \parallel z$ [182]. The authors (the same as of the paper [233]) use their spin-current model in which electric polarization in the ground state is along the $x$-axis (4.24). This consideration is focused on the pnonon modes $u_y$ polarized transversely to the spontaneous polarization $P_x$ and neglects the longitudinal lattice excitations $u_x$. However, exactly these excitations with the atomic displacements $\vec{u} \parallel \vec{P}_s$ revealed the coupling with spin excitations in the compounds $(Eu,Y)MnO_3$ [188, 189] and $RMn_2O_5$ [205] and, moreover, these longitudinal (with respect to the spontaneous polarization) ME excitations correspond to the phason mode at the FE phase transition in the manganite $TbMnO_3$ [178].

Our brief review of the recent theoretical investigations (certainly, not of all of them) of the spin-induced electric polarization in spiral magnets showed that although some important aspects of the FE emergence (magnetic frustrations, orbital hybridization, spin-orbital and exchange couplings, local magnetic symmetry of the unit cell) were revealed, there has not been clear understanding of microscopic mechanism leading to the observed values of electric polarization. A number of publications about ferroelectromagnets have appeared last time testifying to active development of this field of science.

*4. 2  Spin-induced ferroelectricity in the collinear magnetic structures*

A spiral magnetic order is not the only possible rout towards spin-induced ferroelectricity (recall $Ni-I$ boracite [11]). In the $RMn_2O_5$, for example, beside the spiral nature of the FE, the mechanism of the nearly-collinear acentric magnetic order with broken inversion symmetry was proposed [235]. In this mechanism, the FE results from the exchange striction caused by the symmetric superexchange coupling. Another promising example is the so-called $E-$type magnetic order that has been observed in the perovskite manganites [236] and nickelates [237].

Ferroelectricity in the magnetic $E$-phase of the orthorhombic perovskite $HoMnO_3$ was predicted in [238]. In this compound the $Mn$ atoms with parallel spins form zigzag chains in the $(x,y)$-plane, with the chain link equal to the nearest-neighbor $Mn-Mn$ distance. The neighboring zigzag chains in the $y$-direction have antiparallel spins. The $(x,y)$-planes are stacked antiferromagnetically along the $z$-direction. The $HoMnO_3$ has an ICM AF structure at temperature $T < T_N \sim 42-47$ K. The magnetic transition from the ICM to CM state with $k_y = \frac{1}{2}$ occurs at temperature $T_c = 26-29.6$ K . Fig. 25$a$ and 25$b$ show the simple sine-wave ICM structure at $T_c < T < T_N$ and the CM E- phase $(++--)$ below $T_c$, respectively. The E-phase admits the existence of electric polarization along the $x$-direction. There are two types of spin orientation in the unit cell: $E_1(++--)$ and $E_2(+--+)$ with the polarization along the $x$-direction and different signs for $E_1$ and $E_2$. The ME energy in this phase is the following:

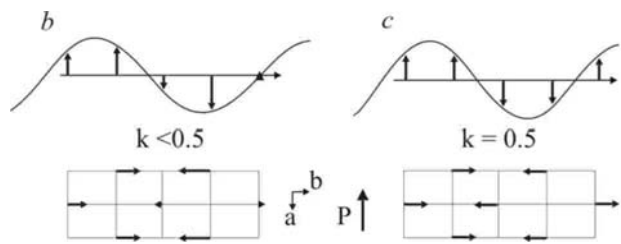

**Fig. 25**. (a). The simple sine-wave magnetic structure of $HoMnO_3$ for $T_c < T < T_N$; (b)The $E$-phase magnetic structure below $T_c$ [238].

$$F_{ME} \propto (E_1^2 - E_2^2)P_x,$$

i.e., $P_x \propto (E_1^2 - E_2^2)$, and in the domains $(\pm E_1)$ and $(\pm E_2)$ the electric polarization has different signs. In the considered collinear E-structure, the anisotropic DM interaction is absent and the enhancement by up to 2 orders of the induced electric polarization with respect to that in the spiral magnets [238] is expected. The polycrystalline $HoMnO_3$ has been experimentally studied to test this theoretical prediction, but the induced polarization turned out to be too small to support the proposed theory [239].

A spin-induced FE in the collinear magnetic state was recently discovered in the Ising chain magnet $Ca_3Co_{2-x}Mn_xO_6$ ( $x = 0.96$ ) [240], where there was used a simple Ising chain model with the competing nearest-neighbor FM ( $I_F$ ) and next-nearest-neighbor AF ( $I_{AF}$ ) interactions. For $| I_{AF} / I_F |> 1/2$ , the magnetic ground state is of the up-up-down-down ($\uparrow\uparrow\downarrow\downarrow$) type. If the charges of magnetic ions alternate along the chain, this magnetic ordering breaks the inversion symmetry on magnetic site and can induce the electric polarization via exchange striction. This mechanism is shown in Fig.26. The exchange striction associated with the symmetric superexchange interaction shortens the bonds between the parallel spins, while stretching those connecting antiparallel spins. As a result, the electric polarization $P$ is induced in the direction of the chain. As shown in Fig. 26, there are two ways to combine the $\uparrow\uparrow\downarrow\downarrow$ order with the ionic charge order, giving rise to the opposite electric polarization vectors. To prove this idea experimentally there was chosen a compound $Ca_3Co_2O_6$ because it is the Ising chain magnet in which about a half of $Co$ ions can be replaced by the $Mn$ ions. Neutron diffraction measurements were performed on the $Ca_3Co_{2-x}Mn_xO_6$ ( $x = 0.95$ ) and showed that $Co^{2+}$ and $Mn^{4+}$ ions alternating along the chains exhibit a $\uparrow\uparrow\downarrow\downarrow$ spin order. The appearance of the electric polarization coincides with the onset of this magnetic order at $T = 16.5$ K. The electric polarization $P_z$ is oriented along the chain and its maximum value is weak, of order $10^{-2} \mu C / cm^2$. Unlike in the spiral ferroelectromagnets, where the antisymmetric DM interaction is active, in the $Ca_3(Co, Mn)_2O_6$ the symmetry breaking occurs via the exchange striction associated with symmetric superexchange. In spite of the exchange nature of the ME coupling in this compound, the observed magnitude of the electric polarization is smaller than that in the $TbMnO_3$ where electric polarization has an exchange-relativistic origin.

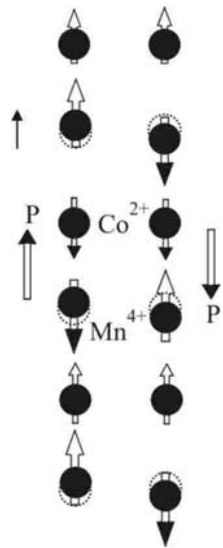

**Fig. 26.** *Adapted from* [240]. Ising chains with the up-up-down-down spin order and alternating ionic order, in which electric polarization is induced through symmetric exchange striction. Two possible magnetic configurations leading to the opposite polarizations are shown. The atomic positions in the undistorted chains are shown with dashed circles.

**Conclusions**

After a long time the observations of the colossal ME phenomena appeared again. According to the expectations [12], the first colossal ME effects were observed in the ferroelectromagnet with strong ME coupling due to the clear FE and magnetic properties in $YMnO_3$ [271] (see Sec.5), and in the ferroelectromagnet $TbMnO_3$ [26] with the coinciding temperatures of the magnetic and electric transitions. The electric polarization induced by spiral magnetic order (i.e. by space magnetic frustration) occurred to be very small, 2-3 orders smaller than the electric polarization of a classic FE $BaTiO_3$, since it appears owing to a weak inhomogeneous exchange-relativistic interaction. But it is the weakness of the electric polarization that allows to observe a colossal (10%) change of dielectric constant in not so large (some Teslas) magnetic field. After this discovery the similar and more colossal changes of $\varepsilon$ in the magnetic field were detected in other compounds. For example, this change can achieve up to 500% in the $DyMnO_3$. A magnetic origin of the electric polarization makes it possible to observe the significant change of $P$ under spin reorientation. Conversely, the electric control of the spin helicity in $TbMnO_3$ and magnetic structure in $TbMn_2O_5$ becomes possible due to a small spin inhomogenity in these compounds. From that time the word "control" is often used in the titles of publications about ME effects.

The first evidence of electromagnons [176] predicted merely 40 years ago [89] marked the beginning of an active investigation of the high-frequency properties in ferroelectromagnets. Significant progress was made in the theoretical considerations of the magnetic induced FE. Phenomenological treatments and microscopic models were proposed for the explanation of the FE origin and ME behavior of spiral ferroelectromagnets in the external constant and alternating fields.

One should be particularly marked, namely, the experimental realization of the enhancement of the ME effects due to the electron contribution in electric polarization at room temperature in the semiconductors -ferroelectromagnets with high dielectric permittivity $\varepsilon \sim 10^4 - 10^5$ [219].

## 5 Ferroelectromagnets with proper ferroelectric and magnetic transitions

### 5.1 Perovskites

The first discovered and best studied ferroelectromagnets have the perovskite structure. Most of the compounds have either $ABO_3$ or $A_2BB'O_6$ as the general chemical formula, and the variety of the existing compounds is greatly increased by chemical substitutions of the type $AB'_{1-x}B''_xO_3$.

Usually a unit cell of the perovskite does not possess the ideal cubic symmetry, it is slightly deformed as, for example, in the first ferroelectromagnet $PbFe_{1/2}Nb_{1/2}O_3$.

The bismuth ferrite $BiFeO_3$, which has a rhombohedrally-distorted perovskite structure, is also one of the first ferroelectromagnets.

### 5.1.1 $BiFeO_3$, bismuth ferrite

Bismuth ferrite is, perhaps, the most extensively studied ferroelectromagnet. The $BiFeO_3$ (space group $R3c$) is FE ($T_c = 1083$ K) and AF ($T_N = 643$ K). The electric polarization $\vec{P}$ is oriented along the rhombohedral axis $[111]_c$ (Fig. 27) and appears due to the displacement of $Bi$ and O ions. The participation of $Fe$ in the FE distortion is forbidden by the "$d^0$" rule discussed before in Sec.3. The calculations [241] have shown the predominant role of the so-called "lone pairs" (two valence electrons) from $Bi$ which are responsible for the FE ordering. Spins are coupled ferromagnetically in the plane $(111)_c$ and are AF ordered between these planes. The magnetic exchange structure (i.e., the mutual directions of magnetic moments in the crystal) is determined by the code $(I^-, 3^+_z, 2^+_x)$ according to the Turov's nomenclature [242] in the hexagonal setting. Here $I$ is the space-inversion element; $3_z$, the threefold axis, and $2_x$, the twofold axis, are the group generators; and $\pm$ are the indices of these elements that specify their parity about the transpositions of magnetic sublattices (i.e., "+" indicates that the symmetry element transposes the ions within the same magnetic sublattice of AF, while "-" indicates that the sublattice is transposed with the opposite spin direction). The exchange structure of $BiFeO_3$ is shown in Fig. 28 [243]. It is seen that the inversion $I$ permutes the magnetic sublattices $\vec{M}_1$ and $\vec{M}_2$, i.e., $I$ is an odd operator $I^-$. As a result, the AF vector $\vec{L} = \vec{M}_1 - \vec{M}_2$ changes its sign under the inversion operation, $\hat{I}\vec{L} = -\vec{L}$.

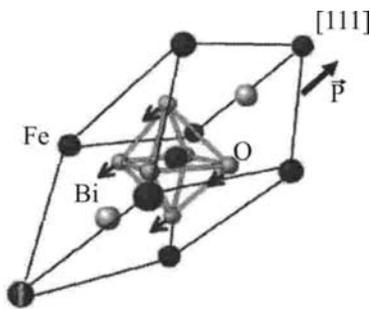

**Fig. 27.** *Adapted from* [19]. Rhombohedral unit cell of $BiFeO_3$; spontaneous polarization $\vec{P}$ is along the cubic axis [111].

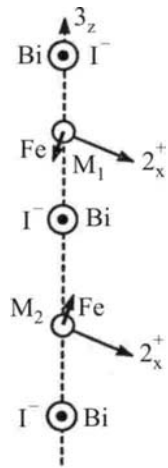

**Fig. 28.** The exchange magnetic structure of $BiFeO_3$ [243].

**Spontaneous ME effects** in the bismuth ferrite were described in Landau's theory by the terms which are invariable with respect to the elements of a "parent" centrosymmetrical group $R\bar{3}c$ [243, 244]. This symmetry allows the following lowest order ME invariants linear on electric polarization:

$$F_{ME}^{(1)} = \gamma_1 \vec{P}[\vec{L}(\nabla \cdot \vec{L}) + \vec{L} \times (\nabla \times \vec{L})], \qquad (5.1\ a)$$

$$F_{ME}^{(2)} = \gamma_2 \vec{P} \cdot (\vec{L} \times \vec{M}). \qquad (5.1\ b)$$

The first term (5.1a) is responsible for the spin-cycloid structure with a period of 62 nm in $BiFeO_3$ [245, 246]. The $F_{ME}^{(1)}$ is a relativistic Lifshitz-like invariant named by analogy to the Lifshitz invariant $\vec{L}rot\vec{L}$. The inhomogeneous magnetic structure in $BiFeO_3$ appears due to the ME interaction (5.1 $a$) and its appearence is possible only when the spontaneous polarization takes place. In fact, the spin cycloid structure appears in the FE state below $T_N < T_c$.

The second ME term (5.1 $b$) shows the possibility of a weak ferromagnetism (WFM) and LMEE in the AF –FE state below $T_N$. However, neither WFM nor LMEE was observed in a bulk $BiFeO_3$ because of its inhomogeneous magnetic structure leading to zero values of the net $M_s$ and $\alpha_{ik}$. The necessary condition to manifest WFM and LMEE in the bismuth ferrite is the suppression of the spin cycloid that may be done, for example, in epitaxial films and by ionic substituents (see below). It should be noted that in contrast to the conventional WFM induced by the Dzyaloshinskii-Moriya term $L_i M_k$, the WFM in the bismuth ferrite is possible only in the presence of the spontaneous polarization, $\vec{M}_s \sim (P_z L_y, -P_z L_x, 0)$ ( in hexagonal axes where $z$ - axis is $[111]_c$ ) [243, 244].

The magnetic symmetry of $BiFeO_3$ also admits the existence of the toroidal moment determined as an antisymmetric part of the linear ME tensor $T_i = \varepsilon_{ikl} \alpha_{kl}$. For the bismuth ferrite, $T_i \sim L_i$ [247]. However, the net value of the toroidal moment is zero in the inhomogeneous magnetic state.

The analysis of the ME coupling in the bismuth ferrite by LSDA calculation method shows the possibility of WFM if the spiral structure is suppressed. But these calculations have predicted that WFM is determined by the rotations of the oxygen octahedral rather than the FE polarization [248].

Data about the spontaneous polarization value are different: from 6 -9 $\mu C / cm^2$ in bulk crystals to ~100 $\mu C / cm^2$ in films [249]. A large FE polarization of $90 - 100 \mu C / cm^2$ was also theoretically predicted [249]. The discrepancy of the data requires further experimental and theoretical investigations.

Due to the ME nature of the spin cycloid and possible WFM these properties of $BiFeO_3$ can be controlled by the magnetic and electric fields.

**Induced ME effects in the bismuth ferrite** were observed in the static electric and magnetic fields.

The phase transition of first kind from the ICM cycloid state into the homogeneous magnetic structure was revealed in the strong magnetic field of 180 $kOe$ [243]. This magnetic phase transition was accompanied by the electric polarization jump. Electron spin resonance measurements in high magnetic field confirmed this phase transition [250]. Spectral data showed the resonant mode and significant hysteresis effect at $H \approx 18$ T. The Landau-Ginzburg phenomenological theory was developed theoretically to model the electron spin resonance spectra in the homogeneous spin state by taking into account the DM –like interaction (5.1 $b$) [250]. The agreement of the theory and the high-field electron spin resonance mode demonstrated that the induced phase was the homogeneous AF state.

The existence of the electric field can change the critical value of the magnetic field under transition between the ICM and CM phases. Electric field – magnetic field phase diagram was presented for the case $\vec{H} \parallel \vec{E} \parallel \vec{P}_s$ according to the results of calculations [244]. It was shown that the metastable states associated with the electric polarization hysteresis of the butterfly-type can exist in the range of weak electric fields.

The ME coupling in the bulk $BiFeO_3$ was very recently effectively demonstrated by using neutron scattering measurements [251]: the spin reorientation induced by electric field was observed at room temperature. The mutual orientations of magnetic moments, electric polarization and modulation vector $\vec{k}$ in $BiFeO_3$, created by the DM interaction (5.1 $a$), are the same as in the spiral ferroelectromagnets [233], i.e., $\vec{P}_s$ and $\vec{k}$ lie in the spin plane. But in contrary to the spiral magnets with improper FE order, where spins determine the value and direction of $\vec{P}_s$, in $BiFeO_3$ with the proper FE order the electric polarization $\vec{P}_s$ itself coordinates spin orientation according to the ME coupling (5.1 $a$). Neutron measurements [251] at room temperature revealed two ME domains with the same modulation vector $\vec{k} = [1,0,-1]$ and different orientations of spin plane and $\vec{P}_s$ (Fig. 29). A rhombohedral symmetry of the bismuth ferrite single crystal allows the orientation of $\vec{P}_s$ along the directions [111] and [1-11]. The ME coupling leads to a different orientation of the helicity vector in the FE domains. Under poling in the electric field $\vec{E} = [0,1,0]$ the switching of electric polarization between FE domains induces the rotation of the spin plane, i.e. the reorientation of spins. It should be noted that at first the AF domain switching induced by electric field at room temperature was demonstrated in the $BiFeO_3$ thin films (see below) [175].

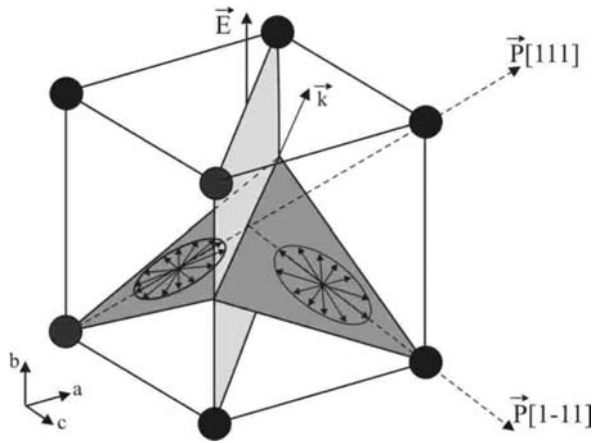

**Fig. 29.** *Adapted from* [251]. Scheme of the planes of spin rotations and helix vector $\vec{k}$ for the two FE domains.

The magnon and phonon spectra in the $BiFeO_3$ was calculated recently in the framework of the phenomenological Landau theory [252]. The static cycloid structure in the bismuth ferrite is produced by a Lifshitz-like term (5.1 $a$), i.e., it is a long-wave relativistic spiral. Its wavelength is $\lambda = 2\pi / q = 62 nm$, where $q = \gamma_1 P_s / c$, and $c$ is the coefficient of the inhomogeneous AF energy $c \sum_i (\nabla L_i)^2$. Conventional phenomenological techniques for the consideration of the ground states and excitations in the magnetic relativistic and exchange spiral structures can be found in the monograph [179]. The response of phonons and magnons on the alternating electric field in the $BiFeO_3$ was analyzed and revealed the ME coupling of spin waves and phonons only for the light polarized along the direction perpendicular to the cycloid plane [252].

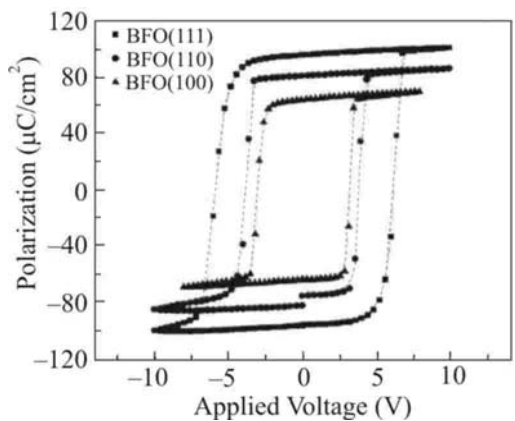

**Fig. 30.** FE polarization loops measured on epitaxial $BiFeO_3$ films with different crystallographic orientations [253].

The weak FM and LMEE are allowed by symmetry but not observed in the bulk $BiFeO_3$ because of the ICM structure and are revealed in the bismuth ferrite epitaxial thin-layers.

**Epitaxial thin films of** $BiFeO_3$ grown on $(001)_c$ $SrTiO_3$ revealed an enhancement of electric polarization and related properties by more than an order of magnitude in comparison with the bulk $BiFeO_3$ (see Fig. 30) [253]. The magnitude of the electric polarization significantly depended on the crystallographic orientation of the film. So, the polarizations $P$ of $\sim 55$ $\mu C/cm^2$ for $(001)_c$ films, $\sim 80$ $\mu C/cm^2$ for $(110)_c$ films, and $\sim 100$ $\mu C/cm^2$ for $(111)_c$ films were reported [254]. Magnetic measurements showed a weak, saturated magnetic moment of $8-10\,e.m.u.cm^{-3}$, which was consistent with the predicted magnitude [248]. Besides, the (001)-oriented $BiFeO_3$ thin layers were reported to have dramatically higher ME coefficients than the bulk crystal. The value of the ME coefficient of the thin layers was $\sim 3.5\,V/Oe\,cm$. Such a large value of ME susceptibility was known only in one compound - $TbPO_4$. The reason of the enhancement, probably, is the in-plane compressive stress which leads to a tetragonal distortion of the perovskite lattice [253]. Analysis has shown that the in-plane epitaxial constraint in $(111)_c$ $BiFeO_3$ thin films is sufficient enough to break the cycloidal spin order [255, 256]. Neutron diffraction confirmed the absence of any bulk-like cycloid modulation in the $BiFeO_3$ thin films [257]. Thus, there is the possibility to connect the enhanced ME properties in the $BiFeO_3$ films with the CM state where the homogeneous ME coupling $(5.1\,b)$ is more significant than the ME interaction in the ICM state [258].

The first observation of the AF domain switching by electric field in the films at room temperature was reported by Zhao *et al.* [175]. The FE domain structure in the 600-nm-thick $BiFeO_3$ films was controlled by piezoelectric force microscopy (PFM). The AF domain structure was studied before and after electrical poling by using photoemission electron microscopy (PEEM). The FE polarization in the $BiFeO_3$ can have eight possible orientations, corresponding to positive and negative orientations along the four cube diagonals, and the direction of polarization can be switched by 180º, 109º and 71º. The PEEM measurements showed that every FE domain coincides with only one AF domain. Switching of the polarization by either 109º or 71º changes the FE domain state as well as an easy magnetization plane perpendicular to the polarization changes. The demonstration of the room-temperature ME coupling in the $BiFeO_3$ is not only interesting from a fundamental standpoint, but it presents a great potential for controlling magnetism with an applied electric field.

*5. 1. 2  $BiMnO_3$, bismuth manganite*

The $BiMnO_3$ is a comparatively rare type of ferroelectromagnets with FM ordering ($T_m \sim 100$ K) [12]. If its FM order was repeatedly confirmed, a type of the electrical-dipole order below $T_c \sim 750 - 770$ K in the bulk crystal is not finally known. The $BiMnO_3$ has monoclinic (noncentrosymmetric) structure, and the possibility of the existing of the FE order was predicted by LSDA calculations [259]. The FE hysteresis loops for polycrystalline thin films and bulk powders of $BiMnO_3$ were reported in the FM state of impure samples [260]. The change in the dielectric constant induced by magnetic field in polycrystalline bismuth manganite was measured near the magnetic order temperature [261]. The relation (2.4) between the change of dielectric constant with magnetic field and the magnetization $\Delta\varepsilon / \varepsilon(0) = \chi_0 \gamma M^2$ ($\chi_0$ is dielectric susceptibility in paramagnetic phase) was examined and showed a good agreement with the law $\Delta\varepsilon \sim M^2$ (Fig.30) [12]. Our estimations of the ME constant $\gamma$ in the $BiMnO_3$ from the results [261] have given the value $|\gamma| \sim 10^{-6} \, dyne^{-1}cm^2$ confirmed the exchange nature of this ME effect (see Sec.2).

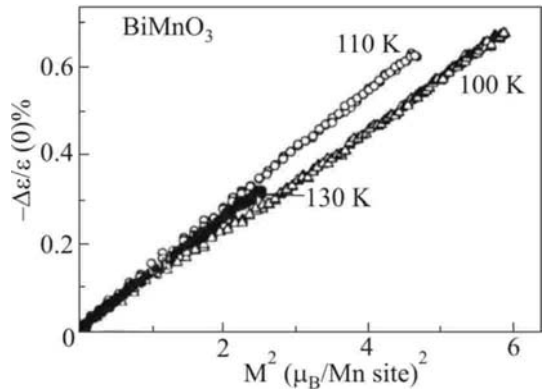

**Fig. 31.** Magnitude of field-induced change of dielectric constant $-\Delta\varepsilon / \varepsilon(0)$ as a function of square of magnetization $M^2$ at temperatures in the vicinity of $T_m$ [261].

The evidence for the FE order in the $BiMnO_3$ epitaxial films was received from optical second-harmonic generation (SHG) measurements [262]. There are 12 possible FE domains in the bismuth manganite film which can be changed by electric field. The SHG signal correlates with these changes in the domain structure. The giant enhancement of SHG by 3 to 4 orders of magnitude from a 900 nm incident light was observed at room temperature under the electric field of ~707 V/mm. It was supposed that possible mechanisms of the observed enhancement may be the domain contributions and (or) electron transfer between $3d$ and $2p$ orbitals. The last mechanism is typical for the electronic FE [147].

*5. 1. 3  New perovskite ferroelectromagnets*

The synthesis of new ferroelectromagnets by ionic substitution started in 1962, when the studies of ferroelectromagnets only began. However, the enhanced ME properties were not reported until recently.

In search of the $BiFeO_3$-based materials with enhanced ME properties [263], attention has been drawn to solid solutions of the $BiFeO_3 - xPbTiO_3$ ceramic substituted with $La$ on the $Bi$ sites. For the $Bi_{0.9-x}Tb_xLa_{0.1}FeO_3$ an anomalously high polarization and a small spontaneous magnetization were observed at room temperature. For example, this compound with $x = 0.075$ is the FE − FM (or FIM), where $T_m = 513$ K, $T_c = 1079$ K, $P_s \sim 0.5 \mu C / cm^2$, and a high dielectric constant $\varepsilon \sim 460$. Strong ME coupling was revealed in the magnetic field when the field $H = 1$ $T$ has changed electric polarization on the value $\Delta P_s = P_s$.

The converting of the conventional FE $PbTiO_3$ with high polarization into ferroelectromagnet was realized by partly substituting $Fe$ into $Ti$ site [264]. The new ferroelectromagnets $Pb(Fe_{1-x}Ti_x)O_3$ exhibited the FE and FM ordering at room temperature. So, for $x = 1/2$ the values $T_m = 543$ K, $T_c = 690$ K were obtained. The results demonstrated very good dielectric properties: the spontaneous polarization $P_s \sim 10$ $\mu C / cm^2, \varepsilon \sim 200$. The ME interaction displayed itself in a break in the temperature dependence of dielectric susceptibility at $T_m$ (such a break in $\varepsilon$ was predicted back in 1962 [67, 68]), in the 12% enhancement in magnetization after poling the sample by means of electric field $10 kV / cm$. This new material carries more importance since the basic compound $PbTiO_3$ has been thoroughly studied for non-volatile memories and other applications. The substitution of the $3d$-ion into the $Ti$ site in $PbTiO_3$ was found to be an effective method to receive new ferroelectromagnets. New FE-FM $Pb(Mn_xTi_{1-x})O_3$ was recently synthesized in this way [265]

In the $Bi$ − containing perovskite $Bi_2FeCrO_6$, the FE order was predicted by using first-principles density functional theory [266]. The prognostic values of the electric polarization $P_s \sim 80 \mu C / cm^2$ and magnetization of $\sim 160$ $emu / cm^3$ far exceeded the properties of any known ferroelectromagnet. To the best of our knowledge, this prognosis has not been experimentally confirmed yet. However, a colossal nonlinear optical ME effect (changes in the SHG) has been reported very recently [267]. The compound $Bi_2FeTiO_6$ is also a possible FE − FM with large magnetization $M \sim 4$ $\mu_B$ per unit cell and electric polarization $P_s \sim 27 \mu C / cm^2$ predicted theoretically [268].

A new family of ferroelectromagnets, the rare-earth chromites $LnCrO_3$ ($Ln = Ho, Er, Yb, Lu, Y$), recently was discovered. These compounds exhibit WFM at low temperature ($113 - 140$ K) and the FE ordering with $T_c$ in 472 -516 K range [269].

The synthesis of new ferroelectromagnets, especially of the perovskite type, is developing so quickly and successfully that it is impossible to include all newly appeared compounds. The number of real and potential ferroelectromagnets has been constantly increasing since 2008, when, by the *Web of Knowledge* [20], there were carried out about 100 investigations of ferroelectromagnets (multiferroics).

## 5. 2  Hexagonal rare-earth manganites

Hexagonal rare-earth manganites with the general formula $RMnO_3$ ( $R = Y, Ho, Er, Tm, Yb, Lu$, or $Sc, In$ ) are FE and AF (or WFM) with $T_c = 570 - 990$ K and $T_N = 70 - 130$ K [12]. Their crystal structure is noncentrosymmetric in contrast to the centrosymmetric perovskite structure (Fig. 32). The trigonal bipyramids, which are connected by their vertices, form layers perpendicular to the six-fold axis. The *Mn* atoms in hexagonal manganites lie inside the pyramids in the surrounding of 5 oxygen ions, while the manganese atoms in orthorhombic perovskites occupy the octahedral $B$-positions and are surrounded by 6 oxygen ions (Fig.1). The rare-earth ions in hexagonal manganites have rather small radius in comparison with orthorhombic perovskites. The difference in crystal structure determines different electric, magnetic and ME properties of these manganites. Magnetic frustration appeared due to the competition of the exchange interactions in hexagonal manganites results a long-range order manifested itself geometrically in the noncollinear triangular AF spin orientation in the plane perpendicular to the sixfold axis (Fig. 11) [126, 270]. The neutron study of magnetic structure in the hexagonal manganites is a rather difficult task since it is not easy to grow a single crystal of sufficient size to measure the neutron diffraction. Therefore, the novel method by Pisarev *et al* [124,126] of nonlinear spectroscopy (SHG) (see Sec. 3.2.2) became a key method for studying the ME coupling in hexagonal manganites.

### 5.2.1 Colossal ME coupling between AF and FE domains

A coupling between the electric and AF domains was assumed earlier [101]. The first direct experimental proof of a strong ME interaction between the FE and AF domain structure in $RMnO_3$ was received by Fiebig *et al.* [271].

The investigations of the hexagonal manganites by SHG revealed that for ferroelectromagnets the nonlinear SHG susceptibility in (3.1) has to be replaced by a multiple-order parameter expansion

$$\hat{\chi} = \hat{\chi}(0) + \hat{\chi}(P_z) + \hat{\chi}(L) + \hat{\chi}(P_z L) + ..., \tag{5.2}$$

where $\hat{\chi}(P_z L)$ refers to four domains with different combinations of sign $(\pm)$ $P_z$ and $L$ in the domain. The sets of nonzero tensor elements $\chi_{ikl}$ in $RMnO_3$ system are different for different contributions in (5.2) so that the appropriate choice of polarizations of the light fields participating in the nonlinear optical process allows studying the respective forms of ordering as well as their interaction separately. The FE and AF domain structures revealed by the SH light in $YMnO_3$ [271, 272] are shown in Fig. 32 $a$ and Fig. 32 $b$, respectively. The dark and bright regions correspond to the domains with opposite orientation of the order parameter $P_z$ or $L$. The FE domain structure was visualized by interference of the SH signal wave from $\chi_{zyy}(P_z)$ with a planar reference light field. In a similar way, the AF domain structure was visualized by interference of the SH waves from $\chi_{yyy}(P_z L)$ and $\chi_{zyy}(P_z)$. Depending on the orientation of $L$, the interference is constructive or destructive, leading to a different brightness for the opposite AF domains. The comparison of Fig. 32 $a$ and Fig. 32 $b$ shows that any reversal of the FE polarization is accompanied by a simultaneous reversal of the AF order parameter.

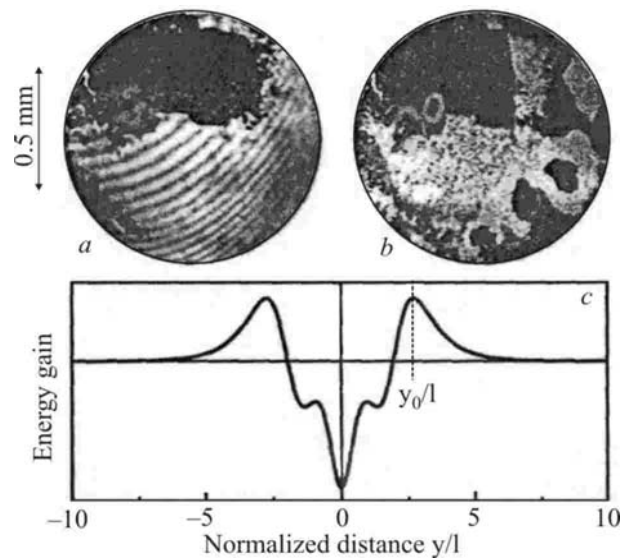

**Fig. 32.** FE(a) and AF (b) domains in $YMnO_3$ [272];(c) Energy gain through interacting FE and AF domain walls at distance $y$ [273].

Consequently, two types of the AF domain walls are found in Fig. 32 $b$: "clamped" walls occurring at any location of the FE domain walls in Fig. 32 $a$, and additional "free" walls within the FE domain. The clamping cannot originate in a bulk coupling between $P_z$ and $L$ because such a linear coupling is forbidden. However, a piezomagnetic interaction between the FE and AF domain walls can explain the observed effect [273]. The piezomagnetic contribution to the free energy as a function of the distance between the AF and FE domain walls is shown in Fig. $32c$. It is seen that the largest energy gain is achieved by centrically overlapping the AF and FE domain walls whereas walls at distance $y_0$ even repel each other.

A novel magnetic phase was detected in the $HoMnO_3$ below the Neel temperature $T_N = 76$ K [274]. A sharp $Mn$-spin reorientation transition took place at $T_{SR} \approx 33$ K accompanied by a sharp peak in the dielectric constant. The anomaly appeared both as a function of temperature and as a function of magnetic field without hysteresis. The anomaly in $\varepsilon$ at $T = T_{SR}$ was much larger than at $T_N$, it depended on the magnetic field and disappeared with its increasing. The investigation by SHG method of AF domain structure at the spin-reorientation transition [275] showed the formation of spin-rotation domains supplementing spin-reversal of the $Mn^{3+}$ spins. The low symmetry in the domain walls, apparently, allows additional linear ME coupling between the wall magnetization and electric polarization (inhomogeneous ME effect) which modifies the dielectric constant.

The microscopic model of the clamping of AF domain wall at FE domain boundary in $YMnO_3$ was also proposed [276]. The clamping could be understood to originate in the antisymmetric exchange energy and spin anisotropy energy which depend on the sign of electric polarization. This term can appear due to low symmetry within the domain wall which is described in terms of soliton theory.

A strong coupling of the FE and AF domains observed in hexagonal manganites appears due to magnetic contribution into electric polarization. According to the well-known "$d^0$" rule, the $Mn$ ions do not participate in the FE displacement directly. Their contribution in $P$ is possible via covalent bond between the $Mn$ ($3d$) and $O(2p)$ bands [146]. This mechanism was supposed in the theoretical considerations of SHG spectra and dispersion of the dielectric constant $\varepsilon(\omega)$ by first principal calculations [277] and exciton theory [278]. Experimental investigation of optical spectra in the hexagonal $RMnO_3$[279] detected the intensive electro-dipole absorption which results from the transition between the $Mn$ ($3d$) and $O(2p)$ bands. Some theoretical predictions are in a qualitative agreement with experiment but a strong difference between them was noticed in the value and the frequency dependences of dielectric constant calculated from first principles [277].

However, the role of manganese ions in the observed colossal ME effect is not yet as clear as it may seem. So, the structural study [280] revealed that the electric dipole moment in $YMnO_3$ is mainly formed not by the $Mn - O$, but by $Y - O$ pairs (see also [281]).

In contrary to perovskite manganites with modulated magnetic structure, the role of the rare-earth magnetic subsystem in hexagonal manganites is more important. A long-range magnetic order of the rare-earth ions is established at low temperature $T'$ which is lower than the Neel temperature for the $Mn$ ions. In the $HoMnO_3$, the recent x-ray resonant magnetic scattering [282] confirmed the AF ordering of $Ho^{3+}$ spins along the $z$-axis below $T' = 40$ K [126] and the reorientation of these spins into $(x, y)$ plane below 4,5 K [300].

In hexagonal manganites there are two long-range magnetic subsystems interacting actively with each other. Magnetic phase transitions in hexagonal manganites occurred via the $3d - 4f$ exchange interaction induce anomalies in dielectric properties, for example, the break or sharp peak in the temperature dependence of dielectric constant [274].

The interaction between the $Mn$ and rare-earth spin subsystems in $ErMnO_3$ was revealed by the Faraday rotation and SHG [283]. The magnetization of the $Er$ lattice along the $z$-axis is perpendicular to the direction of the $Mn$ spins and triggers the $AF - to - FM$ phase transition of the $Mn$ lattice due to strong $R(4f) - Mn(3d)$ exchange.

**Electric control of ferromagnetic ordering** was reported in the hexagonal $HoMnO_3$ [174]. The external electric field $E_0 = 10^5 V/cm$ noticeably changed magnetic state. The SHG and Faraday rotation measurements showed that the reaction of the $Ho^{3+}$ magnetic sublattice to electric field was more fundamental than that of $Mn^{3+}$. Below $T_N$ the electric field induced the phase transition from the paramagnetic or AF state (below $T' = 40$ K) into the FM state. This effect is the first observation of the creation of magnetic order by electric field (in the $Ni - I$ boracite the electric field only rotates magnetization [11]). Such changes in magnetic structure, namely in $Ho^{3+}$ subsystem, were checked by the Faraday rotation measurements in the zero electric field and in the case $E \neq 0$. The FM $Ho^{3+}$ ordering was activated or deactivated, and FM component was controlled by the sign of the electric field. The driving mechanisms for the ME control are the microscopic ME interactions originating from the interplay of $Ho^{3+} - Mn^{3+}$ exchange interaction and FE distortion. The described electric control of the FM ordering was monitored by magnetooptical techniques in $HoMnO_3$ [174].

### 5.2.2 Other displays of the magnetoelectric coupling

The break in the temperature dependence of a static dielectric constant at the Neel temperature was observed in the single crystals $RMnO_3$ ($R = Y, Lu$) [284]. This effect is general for ferroelectromagnets of any symmetry because of its exchange nature [12]. The change in dielectric constant is proportional to the magnetic correlation function, $\Delta\varepsilon \sim \langle \vec{S}_i \vec{S}_j \rangle$, i.e., $\Delta\varepsilon$ is proportional to the square of the order parameter (see Sec.2.3.1).

In the case of the AF transition, the $\Delta\varepsilon \sim L^2$, where $\vec{L}$ is an AF vector. The ME energy inducing this change in $\varepsilon$ is $F_{ME} \sim P^2 L^2$ (Sec.2).

The measurements of frequency and temperature dependencies of optical conductivity and $\varepsilon$ in $LuMnO_3$ [285] showed the change in phonon frequencies at $T_N \sim 90$ K. The value of this change in the lowest phonon mode with $\omega = 270 cm^{-1}$ is of order 1%. It is noteworthy to mention that firstly the change of the phonon mode at $T_N$ (the break) was observed in the $BaCoF_4$ in 1975 [57] (Fig.6). The value of the frequency change $\Delta\omega/\omega$ was also of order of $10^{-2}$ [57, 58].

Interesting results were recently received on the electromagnon spectrum in hexagonal ferroelectromagnets in the Holstein-Primakoff approach [286]. The ME interaction energy was taken in the form $H_{ME} \sim (u_i - u_j)S_i \cdot S_j$, where $u_i$ is the displacement of atom in the $i$-site. The triangular AF ground state and the ME coupling significantly modified magnon spectrum. The existence of the roton-like minimums in spectrum at the wave vector $(2\pi,0)$ and its symmetry-related points in the Brillouin zone were predicted.

## Conclusions

Although ferroelectromagnets with the proper ($T_c \neq T_m$) electric and magnetic ordering were mainly discovered about 50 years ago, the ME phenomena in them have been actively studied. Some of these compounds (for example, $BiFeO_3$, $BiMnO_3$) have rather simple crystal and magnetic structures and are convenient for theoretical and experimental investigations.

Many a long year before the "second ME renascence", the ME effects were investigated mainly in ferroelectromagnets with proper FE and magnetic ordering and these effects were weak. There was formed an opinion that the ME interaction can not display itself clearly in the ferroelectromagnets of this type. However, the development of the new exact optical measuring technique [124, 125] made it possibile to detect strong intrinsic coupling electric and magnetic phenomena in crystals. A colossal ME interaction between the FE and AF domain walls was observed in the hexagonal $RMnO_3$ compounds. The interaction of the domain walls due to inhomogeneous ME effect and local LMEE in the magnetic walls led to a strong coupling of the AF and FE domains, and a contribution to the LMEE induced in the AF domain walls was identified [287]. Gigantic ME bulk effects, where magnetic phase control was exerted by applied electric and magnetic fields, was revealed.

### 5. 3 Electronic ferroelectromagnets

In modern theory of solids the crystal lattice and electrons are the two different (though interacting) subsystems. In the convenient FE theory, the electric polarization is connected mainly with the lattice displacements, though the electron participation in electric polarization may be significant even in an adiabatic approximation. It takes place, for example, in the classical perovskites $BaTiO_3$ and $PbTiO_3$. At FE transition the displacement of an ion, namely its nuclear and electrons, in the insulator, where the electrons are localized, occurs as a whole. This simple classic picture is good for the ionic crystal, but it is distracted by a covalent bond in the magnetic crystal, where the exchange interaction causes the magnetic order. Covalent bonds mix the electron orbitals and diminish the electron localization. The frustration of the exchange bonds between electrons, the presence of magnetic ions of different valency and metallic admixture lead to the redistribution of charge (electronic) density in crystal and form the so-called charge ordering (CO). Electronic CO in some cases can be accompanied by the appearance of the electric polarization of electron nature.

Electronic mechanism of the FE in the perovskite manganites with CO was shown in the $R_{1-x}Ca_xMnO_3$ [281,288]. For example, the half-doped manganite $Pr_{0.5}Ca_{0.5}MnO_3$ is known to be a cite-centered CO with the alternation of magnetic ions with different valence, $Mn^{3+}$ and $Mn^{4+}$ ions (Fig. 33 $a$ ). There is also the possibility to form a bond-centered ordering or a bond-centered charge density wave: all sites are equivalent, but there appears an alternation of short and long bonds, short bonds having higher electron density (Fig. 33 $b$ ). The last model was proposed for $Pr_{1-x}Ca_xMnO_3$ with $x \sim 0.4$ as a result of structural study [289]. The state with electron localization on a pair of $Mn$ ions, or on the respective $Mn - Mn$ bonds, was called the Zener polaron [289]. Theoretical study [288] predicted the existence of an intermediate state with coexising site- and bond- centered states for the $x$ values between 0.4 and 0.5 (Fig. 33 $c$): the state in Fig. 33 $a$ gradually transforms into a bond-centered state in Fig. 33 $b$ under decreasing $x$. There is an important difference between the states in Figs. 33 $(a,b)$ and 33 $c$: the states in Fig. 33 $a,b$ have no electric dipole moment, whereas the state in Fig. 33 $c$ has well-formed dimmers, and each dimmer has a dipole moment due to the charge difference on the sides. Subsequently, a total nonzero dipole moment of each $MnO_2$- plane, oriented in $[1\bar{1}0]$ direction, appears (Fig. 33 $c$).

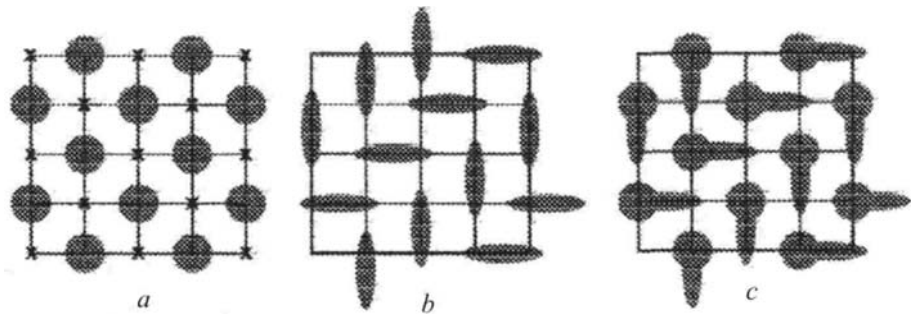

**Fig. 33. (a)** Site-centered charge ordering in half-doped manganites like $Pr_{0.5}Ca_{0.5}MnO_3$. (b) bond-centered ordering, and (c) combined ordering, giving FE [288].

Very recently in the $Pr_{0.6}Ca_{0.4}MnO_3$ and $Nd_{0.6}Ca_{0.4}MnO_3$ the evidence of the existence of spontaneous electric polarization in CO and AF states at 4.2 K was received [290] that is the confirmation of the theory [288]. The change in electric polarization under suppression of CO by the magnetic field and the transition from insulator AF into metallic FM state was observed. The sign of electric polarization induced by the magnetic field depended on the sign of the electric field applied under previous cooling that is the evidence of spontaneous polarization.

The very recent study of the electric field gradient in AF $(Pr,Ca)MnO_3$ ($T_N \sim 150$ K) revealed the electro-dipole transitions at temperatures between CO and AF ordering temperatures [291]. The FE transition was not limited below $x = 0.5$ and spreaded over the entire CO region for these compounds ($0.35 \leq x \leq 0.85$).

Electronic FE was detected [292] in the mixed valence compound $LuFe_2O_4$ from a family of the $RFe_2O_4$, where R are the rare-earth elements from $Dy$ to $Lu$ and $Y$. The $LuFe_2O_4$ is FIM with $T_m = 250$ K. The crystal structure consists of the alternate stacking of triangular lattices of the rare-earth elements, iron and oxygen. An equal amount of $Fe^{2+}$ and $Fe^{3+}$ coexists at the same site in the triangular lattice. The coulombic preference for pairing of $Fe^{2+}$ and $Fe^{3+}$ causes the degeneracy in the lowest energy for the charge configuration in the triangular lattice, similarly to that in the triangular AF Ising spins. Therefore, the $RFe_2O_4$ is considered to be a charge-frustrated system of triangular lattices. The postulated charge structure allows the presence of local electric polarization, since the centres of $Fe^{2+}$ and $Fe^{3+}$ do not coincide in the unit cell. This indicates the possibility of the FE ordering via only electron density modulation. The later $x$-ray scattering experiment which is sensitive for detecting the charge distribution confirmed the appearance of electric polarization at the charge ordering temperature $T_{CO} = 330$ K [293]. The temperature dependence of the observed electric polarization is shown in Fig.34. The direction of the electric polarization depends on the direction of the cooling electric field, which indicates that $LuFe_2O_4$ possesses macroscopic electric polarization. The value of a spontaneous polarization below $T_m$ is of the same order as in $BaTiO_3$ $P_s \sim 30 \ \mu C/cm^2$. The shoulder of the electric polarization at the magnetic transition temperature $T_m = 250 \, \text{K}$ shows the ME coupling of magnetization with electric polarization. One can estimate the change in $P_s$ at magnetic ordering in Fig. 34. This change is $\Delta P/P \sim 10^{-1}$. **It is a colossal ME effect!** Remind that the spontaneous ME effect in the insulator ferroelectromagnets (the change in phonon frequency at magnetic transition due to exchange interaction) is of one order smaller, $\Delta \omega / \omega \sim \Delta P/P \sim 10^{-2}$ [57, 285]. The value of a dielectric constant at the FE order temperature is large, $\varepsilon \sim 5000$. Observed frequency dispersion of $\varepsilon$ has a common feature with the order-disorder type of FE materials, where the motion of the FE domain boundaries gives rise to dispersion. Thus, the first colossal ME effect in the electronic ferroelectromagnets was discovered in $LuFe_2O_4$, where the FE ordering arises from the electron correlations acting on a frustrated geometry.

Very recently **a colossal ME optical effect** was observed in the ferroelectromagnet with magnetically driven FE $TbMn_2O_5$ [311]. Normalized SHG yields were up to seven orders of magnitude larger than in ferroelectromagnets with proper FE and magnetic transitions. This giant coupling was supposed to be determined by electronic nature of the spontaneous polarization.

The magnetite $Fe_3O_4$ is the oldest known magnet. The Fe ordering in it was detected below the Verwey transition at $T_v \sim 120 \, \text{K}$ in the end of twentieth century [121]. The $Fe_3O_4$ is a FIM ($T_m = 851 \, \text{K}$) with the cubic inverse spinel structure and mixed iron valency: the tetrahedral (A) sites are occupied by $Fe^{3+}$, whereas the twice as abundant octahedral B sites are randomly occupied by $Fe^{3+}$ and $Fe^{2+}$. The Verwey (MI) transition between metallic and insulating phases is accompanied by a structural phase transition from the cubic inverse spinel to a distorted structure. One usually connects the appearance of the FE properties in the magnetite with the charge ordering below $T_v$ [294, 295]. However, in the band calculations [296, 297] it was conjectured that the ordering of $t_{2g}$ orbital (orbital ordering) rather than CO itself defines the state of the magnetite in insulating phase. Very recently a direct observation of $t_{2g}$ orbital ordering by resonant $x$-ray diffraction study confirmed the theoretical predictions [296, 297] of the importance of orbital ordering in the coexistence of orbital and charge order in $Fe_3O_4$ below $T_v$ [298]. The microscopic picture of FE origin in the magnetite is not completely known today and is actively discussed.

The nickelate perovskites $RNiO_3$ ($R$ − rare earth) are, probably, also electronic ferroelectromagnets. The change in the crystal symmetry at the MI transition and the coexistence of charge ordering with the spin ordering of the up-up-down-down-type ($E$-type) was observed in the insulating phase of $RNiO_3$ [154, 237].

# 6 Conclusions

## 6.1 Summary

The purpose of this book is to view a way which the sciences of ferroelectromagnetic compounds passed the last fifty years. It is seen that the hopes for control of magnetic (electric) properties of crystal by electric (magnetic) field began to be realized due to the discovery of a strong ME coupling. It should be noted that the so-called colossal ME effects were observed mainly in ferroelectromagnets as it was expected. The technical progress in the preparing and studying of oxides materials, to which most of ferroelectromagnets belong, and the development and using of the novel measurement methods such as optical SHG method [124], the non-reciprocal linear dichroism and birefringence [299] paved the way to the discovery of a novel colossal ME effects. Most ferroelectromagnets are not magnetoelectrics, i.e. they have no LMEE, but they show large spontaneous and nonlinear ME effects. Spontaneous ME effects took place near the phase transition temperatures when a new ordering appears (the cases of $BaMnF_4, YMnO_3, etc.$). These observed effects were usually weak and they displayed themselves as a break in temperature dependences of a dielectric constant, phonon frequency and electric polarization at the ordering temperature $T_m < T_c$. A promising exception is the gigantic change of electric polarization of order of 10% below $T_m$ in the electronic ferroelectromagnet $LuFe_2O_4$ [293]. However, it should be checked whether the influence of non negligible conductivity in these materials exists.

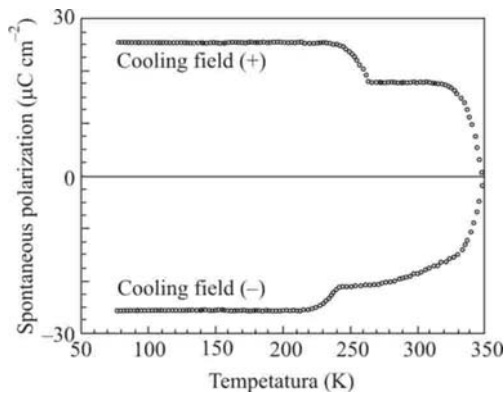

**Fig. 34.** Temperature dependence of the electric polarization of $LuFe_2O_4$ [293].

The number of the observed colossal induced (field) ME effects is larger than of spontaneous ones. Magnetic control of the electric polarization (the change of the value and direction by magnetic field) was detected in many ferroelectromagnets beginning from $TbMnO_3$. Electric control (the change of magnetic order by electric field) has been demonstrated in the $Ni-I$ boracite, $HoMnO_3, TbMnO_3, BiFeO_3, ErMn_2O_5$ and $YMnO_3$.

Generally speaking, the ME effect appears under any change of the order in space (static ME effects) and in time (dynamic ME effects). A static ME effect is a result of the space frustration: magnetic spiral inhomogeneity in the spin-driven FE, geometrical spin frustration, and charge frustration in electronic ferroelectromagnets. What is peculiar, the magnetic inhomogeneity of the domain walls induces a local electric polarization and LMEE that, apparently, play an important role in a colossal ME coupling of the AF and FE domains in the hexagonal manganites [270, 300, 287]. Thus, the complex magnetic and domain structures in ferroelectromagnets promote the ME phenomena. A dynamic ME effect is a result of an intrinsic unity of the electric and magnetic fields displayed in electromagnetic wave. The existence of the coupled spin-phonon excitations (electromagnons) in ferroelectromagnets is one of demonstrations of a dynamical ME effect in medium. Microscopic mechanisms of the ME effect are the Heisenberg exchange, the DM exchange-relativistic interaction, the spin-orbital and anisotropic energy, the electric- and magnetostriction interactions.

During the last 50 years the number of the single-phase ferroelectromagnets was filled up. In additional to the compounds pointed in the Table of the review [12], the following novel families and compounds appeared: orthorhombic $RMnO_3$, $RMn_2O_5$, $Ni_3V_2O_8$, $RbFe(MoO_4)_2$, $MnWO_4$, $LiCu_2O_2$, $LiCuVO_4$, $CoCr_2O_4$, $LuFe_2O_4$, $LnCrO_3$ ($Ln = Ho, ErYb, Lu, Y$), $Fe_3O_4$, nickelates $RNiO_3$, FM- relaxor FE spinel $CdCr_2S_4$ [301], pyroxenes with general formula $AMSi_2O_6$ ($A =$ mono- or divalent metal, $M =$ di- or trivalent metal) are magnetically driven FE [302]. It is quite possible that the rare-earth ferroborate $GdFe_3(BO_3)_4$ also is a ferroelectromagnet. Strong correlations between its AF ordering ($T_N = 37$ K) and dielectric properties were detected [303]. The supposition about a noncollinear antiferroelectric order below the structural phase transition at $T_c = 156$ K in $GdFe_3(BO_3)_4$ allowed to explain the ME properties observed experimentally in this compound [304]. Besides, a number of solid solutions, mainly of perovskite structure, were received.

*6. 2 Outlook*

Ferroelectromagnets are exceptionally important objects for studying of coupling electric and magnetic phenomena in nature. Aside from its fundamental importance, the mutual ME control of electric and magnetic properties open quite a big new field of applications. Ferroelectromagnets are promising materials for electric (magnetic) field controlled magnetic (electric) resonance devices, ME memory systems, sensors, actuators, transducers, tunable microwave devices etc. Large change of dielectric susceptibility in the magnetic field may be used for construction of the device with magnetic tunable capacity. In the electronic ferroelectromagnets with low activation the energy of the electron motion because of weaker coupling of polarization switching with the lattice distortion, may enable the creation of a fatigue-free solid charge capacitor [293]. Single-phase ferroelectromagnets could also find applications, for example, in spintronics [305]. For the realization of these hopes it is necessary to have good single insulators with the magnetic and FE orderings at room temperature and a strong ME coupling.

However, most ferroelectromagnets have a weak ME interaction and the low temperatures of magnetic ordering that is unsuitable for practical applications. Strong ME properties can be expected in the electronic ferroelectromagnets where the magnetic and electric (charge) orderings are created by one and the same object – electron. The absence of intermediate chains in the ME interaction has to increase the value of the ME effect significantly. Thus, the electronic ferroelectromagnets seem to be the most perspective for applications. It is possible to diminish the conductivity of these compounds using thin films and ferrimagnets which are more likely to be insulating than ferromagnets [301]. Electronic compounds showed also a high dielectric response and a possibility of magnetic order at the temperature close to room temperature [219].

The enhancement of the magnetic and FE properties in the thin epitaxial films of $BiFeO_3$ [253] was not confirmed later [301]. However, the thin-film systems, which are often insulators, are popular in studying and applications [306]. A chemical doping [264, 219] may appear to be a more successful way of receiving ferroelectromagnets with the strong ME interaction at room temperature.

There is also an alternative route to fabricate artificially the composite FE and magnetic materials with high transition temperatures in the form of laminates, epitaxial multilayers, nanopillar composites etc. The ME coupling between different materials is indirect; it occurs when there is a strain and an intimate contact. The multi-phase systems are not the subject of this review. More detail information about ME composites can be found in the publications [19, 31, 307].

There is a great interest to the ME phenomena, especially last years. Materials of the conferences MEIPIC-4 [308], MEIPIC-5 [309] and the reviews [19, 200, 281, 301, 310] detected general results and perspective directions for further investigations of ferroelectromagnets. Theoretical suggestions, synthesis and experimental studying move forward very quickly and one may expect that mainly ferroelectromagnets of helical type will be discovered. The premise of further progress in the discovery of ferroelectromagnets with strong ME coupling at room temperature is a deep understanding of the ME microscopic mechanisms in the single-phase ferroelectromagnet. However, at present such understanding is at the initial stage. The calculations from "first principles" by LSDA method are often not irreproachable and too optimistic [232]. But studying of the ME effects continues very intensively.

The prediction, discovery and study of ferroelectromagnets started in Russia have attracted attention of scientists from Europe, Japan, USA, India and China. One may say that the ME "boom" has spread among numerous scientific laboratories all over the world. It gives us every hope that ferroelectromagnets will find practical application in the nearest future.

References

1. Smolenskii G. A. and Agranovskaya A.I. *Zh.. Eksp. Teor. Fiziki,* **28**, 1491 (1958).

2. Smolenskii G. A. , Agranovskaya A. I., Popov S. N., and Isupov V. A. *Zh. Eksp. Teor. Fiziki*, **28,** 2152 (1958).

3. Smolenskii G. A. and Ioffe V. A. *Communications de Colloque International de Magnetism de Grenoble* (France), 2-6 Jullet, Communication No.1 (1958).

4. Smolenskii G. A., Isupov V. A., Agranovskaya A. I., and Krainik N. N. *Fiz. Tverd. Tela,* **2**, 2982 (1960).

5. Smolenskii G. A., Isupov V. A., Krainik N. N., and Agranovskaya A. I. *Izv. Akad. Nauk SSSR, Ser. Fiz.,.* **25**, 1333 (1961).

6. Bokov V. A., Myl'nikova I. E., and Smolenskii G. A. *Zh. Eksp. Teor. Fiziki ,* **42**, 643 (1962).

7. Venevtsev Yu. N., Zhdanov G. S., and Solovjev S. P. *Kristallographiya*, **5**, 520 (1960).

8. Smolenskii G. A., Yudin V.M., Sher E. S., and Stolypin Yu. E. *Zh. Eksp. Teor. Fiziki,* **43**, 877 (1962).

9. Kiselev S.V., Ozerov R. P., and Zhdanov G. S. *Dokl. Akad. Nauk SSSR*, **145**, 1255 (1962).

10. Bertaut F., Forrat F., and Fang P. *Compt .Rend .Aadc. Sci.,* **256**, 1958 (1963).

11. Ascher E., Rieder H., Schmid H., and Stössel H. *J. Appl. Phys.,* **37**, 1404 (1966).

12. Smolenskii G. A. and Chupis I. E. *Sov. Phys .Usp.,.* **25**, 475 (1982); in *Problems in Solid-state Physics* (Mir Publishers Moscow, 1984, p.81).

13. Venevtsev Yu. N., Gagulin V. V., and Lyubimov V. N. *Ferroelectromagnetics*, Nauka, Moscow, 1982 (in Russian).

14. Curie P. *J. Physique,* **3**, 393 (1894).

15.  Landau L. D. and Lifshits E. M. *Electrodynamics of Continuous Media* (Moscow 1957 in Russian;  Oxford: Pergamon 1960).

16.  Dzyaloshinskii I. E. *Sov. Phys. JETP,* **10**, 628 (1959).

17.  Astrov D. N. *Sov. Phys. JETP* **11**, 708 (1960); **13**, 729 (1961).

18.  Nedlin  G. M. F*iz. Tverd. Tela,* **4,** 3568 (1962).

19.  Fiebig M. *J. Phys. D: Appl. Phys.* **38,** R123 (2005).

20.  *Web of Knowledge* http://isi10.isiknowledge.com/

21.  O'Dell T. H. *The Electrodynamics of Magneto-Electric Media* (Amsterdam: North-Holland) (1970).

22.  Freeman A. J. and Schmid H. (ed) 1975 *Magnetoelectric Interaction Phenomena in Crystals Proc.  MEIPIC-1 (Seattle, USA, 21 -24 May 1973)* (London: Gordon and Breach).

23.  Baturov L. N., Al'shin B. I., and Yarmukhamedov Yu. N. *Fiz. Tverd. Tela,* **20**, 2254 (1978).

24.  Schmid H., Janner A., Grimmer H., Rivera J. P., and Ye Z. G. (ed) 1994 *Proc. MEIPIC-2 (Ascona,  Switzerland, 13-18 September 1993) Ferroelectrics* , **161-162**.

25.  Bichurin M. (ed) 1997 *Proc. MEIPIC-3 (Novgorod, Russia, 16-20 September 1996) Ferroelectrics,* **204.**

26.  Kimura T., Goto T., Shintanl H., Ishizaka K., Arima T., and Tokura Y. *Nature (London),* **426,** 55  (2003).

27.  Aizu K. *Phys. Rev.* B, **2,** 754 (1970).

28.  Zel'dovich Ya. B. Zh. Eksp. Teor. Fiz, **6**, 1184 (1958).

29.  Gorbatsevich A. A. and Kopaev Yu. V. *Ferroelectrics,* **161**, 321 (1994).

30.  Schmid H. *Ferroelectrics,* **162**, 317 (1994).

31.  Bichurin M. I., Petrov V. M., Filippov D. A., and Srinivasan G. *Magnetoelectric effect in composites*  Novgorod, 2005 (in Russian).

32.  Turov E. A. *Physical properties of the magnetic ordering crystals* (Moscow, 1963, Izd. Akad. Nauk   SSSR); Turov E.A., Kolchanov A. V., Men'shenin V. V., Mirsaen I. F., and Nikolaev V. V.  *Symmetry  and physical properties of antiferromagnets* (Moscow, 2001, Fizmatlit) (in Russian).

33.  Bulaevskii L. N. and Fain V. M. *Pis'ma v Zh. Eksp. Teor. Fiziki,* **8,** 268 (1968).

34.  Balkarei Yu. I. and Nikitov V. A. *Fiz. Tverd. Tela,* **17**, 2089 (1975).

35.  Chupis  I. E., *Proc. of Conf. on Low Temperature Physics,* Kharkov, 1980, Part II, p.173 (in Russian).

36.  G. T. and Ferrari J. M. *Phys. Rev. B. Solid State,* **15**, 290 (1977).

37.  Schmid H., in: *Magnetoelectric Interaction Phenomena in Crystals,* London, New Mitsek A. I. and Smolenskii G. A. *Fiz. Tverd. Tela,* **4**, 3581 (1962).

38.  Rado G. T. *Phys. Rev. Lett.* **6**, 609 (1961).

39.  Englman R., Yatom H. *Phys. Rev* **187**, 793;  **188**, 803 (1969)

40.  Kovalev O. V. *Fiz. Tverd. Tela,* **14**, 307 (1972).

41.  Gufan Yu. M. *Pis'ma Zh. Eksp. Teor. Fiziki,* **8**, 271 (1968).

42.  Rivera J. P., Schmid H., Moret J. M., and Bill H. *Intern. J. Magnet.,* **6**, 211 (1974).

43.  Minetaka H., Kay K., and Jinzo K. *J. Phys. Soc. Japan,* **39**, 1625 (1975).

44.  Baturov L. N., Al'shin B. I., and Zorin R. V., in: *Abstracts of Papers to the All-Union Conf. on Magnetic Phenomena Physics,* Kharkov, 1979, p. 165.

45.  Rado York, Paris, 1975,  p. 121.

46.  Astrov D. N., Al'shin B. I., Zorin R. V., and Drobyshev L. A. *Zh. Eksp. Teor. Fiz.,* **55**, 2122 (1968).

47.  Al'shin B. I., Astrov D. N., and Baturov L. N. *Pis'ma v Zh. Eksp. Teor. Fiz.,* **22**, 444 (1975).

48.  Baturov L. N.,  Al'shin B. I., and Astrov D. N. *Fiz. Tverd. Tela,* **19**, 916 (1977).

49.  Baturov L. N., Zorin R. V., Al'shin B. I., and Bugakov V. I. *Fiz. Tverd. Tela,* **23**, 908 (1981).

50.  Krainik N.N., Khuchua N. P., and Zhdanova V. V., in: *Proc. of Intern. Meeting of Ferroelectrics,* Prague, 1966, V. 1,  .p 377.

51.  Roginskaya  Yu. E., Tomashpol'skii Yu. Ya., Venevtsev Yu. N., Petrov V. M., and Zhdanov G. S. *Zh. Eksp. Teor. Fiz.,* **50**, 69 (1966).

52.  Drozhdin S.N., Bochkov B. G., Gavrilova N. D., Popova T. V., Koptsik V. A., and Novik V. K.  *Kristallografiya,*  **20**, 854 (1975).

53.  Samara G. A. and  Richards P. M., *Phys. Rev. Ser B,* **14**, 5073 (1976).

54.  Samara G. A. and  Scott I. F. *Solid State Comm.* **21**, 167 (1977).

55.  Fox D. L., Tilley D. R., Scott I. F., and Guggenheim H. I. *Phys. Rev. Ser. B,* **21**, 2926 (1980).

56.  Cox D. E., Shapiro S. M., Cowley R. A., Eibschütz I. M., and Guggenheim M.I. *Phys. Rev. Ser. B,* **49**, 5754 (1979).

57.  Popkov Yu. A., Petrov S. V., and Mokhir A. P. *Fiz. Nizkikh  Temp.,* **1**, 189 (1975).

58.  Chupis I. E.  *Fiz. Nizkikh  Temp.,* **9**, 998 (1983).

59.  Koptsik V. A. *Kristallografiya,* **5**, 932 (1960).

60.  Neronova N. N. and Belov N. V. *Dokl. Akad. Nauk USSR,* **129**, 556 (1959*); Kristallografiya*, **4,**  807 (1959).

61. Shuvalov L. A. and Belov N. V. *Kristalografiya,* **7**, 192 (1962).

62. Lyubimov V. N. *Fiz. Tverd. Tela,* **5**, 951 (1963); *Kristallografiya,* **8**, 699 (1963).

63. Kovalev O. V. *Fiz. Tverd. Tela,* **14**, 1961 (1972); *Kristallografiya,* **18**, 221 (1973).

64. Brunskill  J. H. and Schmid H. *Ferroelectrics,* **36**, 395 (1981).

65. Ismailzade I.H. and Yakupov  R. Y. *Phys. Stat. Sol., Ser.(a),* **32**, K161 (1975).

66. Chupis I. E. *Fiz. Nizkikh Temp.,* **2**, 762 (1976).

67. Smolenskii G. A. *Fiz. Tverd. Tela,*  **4**, 1095 (1962).

68. Nedlin G.M.  *Izv. Akad. Nauk USSR, Ser. Fiz.,* **29**, 890 (1965).

69. Newnham R. E., Kramer J. J., Shulze W. A., and  Cross L. E. *J. Appl. Phys,.* **49**, 6088 (1978).

70. Bar'yakhtar V.G., L'vov V.A., and  Yablonskii D. A. *Pis'na v Zh. Eksp. Teor. Fiz.,* **37**, 565 (1983).

71. Barr'yakhtar V.G., Stefanovskii E. P., and  Yablonskii D. A. *Pis'ma v Zh. Eksp. Teor. Fiz.,* **42**, 258 (1985).

72. Chupis I. E. , in: *Solid State Physics,* Vladivostok, 1972 (in Russian); *Proc. of  1$^{st}$ Int. Conf. on  Magnetism MCM-73,* Moscow, 1973, **1**, 1, Nauka, Moskow, 1974, p.354.

73. Chupis I. E. and Plyushko N. Ya. *Fiz. Tverd. Tela,* **14**, 3444 (1972).

74. Chupis I. E., in: *Some Questions in the Physics of Thin Ferromagnetic Films,* Vladivostok, 1974, p. 3.

75. Chupis I. E. *Fiz. Nizkikh Temp.,* **1**, 183 (1975).

76. Chupis I. E. *Fiz. Nizkikh Temp.,* **7**, 203 (1981).

77. Bar'yakhtar V. G. and Chupis I. E. *Fiz. Tverd. Tela,*  **11**, 3242 (1969).

78. Bar'yakhtar V. G. and Chupis I. E. , in: *Magnetoelectric Interaction Phenomena in Crystals,* London-  New York –Paris, 1975, p. 57.

79. Bar'yakhtar V. G. and Chupis I. E. *Ukr. Fiz. Zh.,* **17**, 652 (1972).

80. Bar'yakhtar V. G. and Chupis I. E. *Intern. Journ. Magnet.,* **5**, 337 (1974).

81. Chupis I.E. *Fiz. Nizkikh Temp.,* **2**, 622 (1976).

82. Akhiezer A. I. and Akhiezer I. A. *Zh. Eksp. Teor. Fiz.,* **39**, 1009 (1970).

83. Davydov L. N. and Spol'nik Z. A.  *Ukr. Fiz. Zh.,* **18**, 1368 (1973).

84. Chupis I. E. and Aleksandrova N. Ya. *Fiz. Tverd. Tela,* **21**, 3166 (1979)

85. Chupis I. E. *Fiz. Nizkikh Temp.,* **6**, 771 (1980).

86. Plyushko N. Ya. and Chupis I. E. *Ukr. Fiz. Zh.,* **19**, 826 (1974).

87. Savchenko M. A. and Khabakhpashev M. A. *Fiz. Tverd. Tela,* **18**, 2699 (1976).

88. Akhiezer I. A. and Davydov L. N. *Fiz. Tverd. Tela,* **12**, 3171 (1970).

89.  Bar'yakhtar V.G. and Chupis I. E. *Fiz. Tverd. Tela,* **10**, 3547 (1968).

90.  Asher E. *Phil. Mag.,* **17**, 149 (1968).

91.  Akhiezer I. A. and Davydov L. N. *Ukr. Fiz. Zh.* **15**, 1747 (1970); *Pis'ma v Zh. Eksp. Teor. Fiz.* **13**, 380 (1971).

92.  Chupis I. E. and Aleksandrova N. Ya. *Ukr. Fiz. Zh.,* **27**, 300 (1982).

93.  Chupis I. E. and Savchenko V. N. , in: *Some questions in the Physics of Thin Ferromagnetic Films* Vladivostok, 1974 (in Russian), p. 93.

94.  Akhiezer I. A. and Davydov L. N *Fiz. Tverd. Tela,* **13**, 1795 (1971).

95.  Nikitov V. A. *Trudy MFTI, Ser. Radiotekhnika i Elektronika,* № 9, 118 (1975).

96.  Bakay A. S. and Chupis I. E. *Fiz. Nizkikh Temp.,* **3**, 1153 (1977).

97.  Akhiezer A. I. , Bar'yakhtar V. G., and Peletminskii S. V. *Spin Waves,* Nauka, Moscow, 1967 (in Russian), Sec. 14.

98.  I. Van den Boomgaard and Born R. A. I. A. *J. Mater. Sci.,* **13**, 1538 (1978).

99.  Wood E. and Austin A. E. *Intern. J. Magnet.,* **5**, 303 (1974).

100. Kizhaev S.A. and Pisarev R. V. *Sov. Phys. Sol. St.,* **26**, 1012 (1984).

101. Huang Z. J., Cao Y., Sun Y. Y., Xue Y. Y., and Chu C. W. *Phys. Rev.* B **56**, 2623 (1997).

102. Sanina V. A., Sapozhnikova L. M., Golovenchits E. I., and Morozov N. V., *Soviet Phys.- Solid State,* **30**, 1736 (1988).

103. Golovenchits E. I., Morozov N. V., Sanina V. A., and Sapozhnikova L. M. Soviet Phys. Solid State, **34**, 56 (1992).

104. Ikeda A. and Kohn K. *Ferroelectrics*, **169**, 75 (1995).

105. Koyata Y., Nakamura H., Iwata N., Inomata A., and Kohn K. *J. Phys. Soc. Jpn.,* **65**, 1383 (1996).

106. Koyata Y. and Kohn K. *Ferroelectric*, **204**, 115 (1997).

107. Kohn K. *Ferroelectrics,* **162**, 1 (1994).

108. Kagomiya I., Iwata N., Uga M., Fujita T., and Kohn K. *J.of the Magn. Soc. of Jpn,* **22**, Suppl. S1, 55 (1998).

109. Nakamura H., Ishikawa M., Kohn K. *J. Phys. 1V France,* **7**, C1-365 (1997).

110. Murashov V.A., Rakov D. N., Bush A, A., Venevtsev Yu. N., Ionov V. M., Dubenko I. S., Titov Yu. V., Klimenko A. N., Sergeev V. S., Mrost S. E., and Prozorovskii A. E. , in: *Ferroelectromagnetic substances,* Nauka, Moscow, 1990 (in Russian), p.86.

111. Perekalina T. M. and Cherkezyan S. A. , in: *Ferroelectromagnetic substances,* Nauka, Moscow, 1990 (in Russian), p. 96.

112. Iida J. *J. Phys. Soc. Jpn.,* **62**, 1723 (1993).

113. Yamada Y., Kitsuda K., Noholo S., and Ikeda N. *Phys. Rev. B* **62**, 12167 (2000).

114. Ye Z.-G., Rivera J.-P., Schmid H., Haida M., and Kohn K. *Ferroelectrics,* **161**, 99 (1994).

115. Nakamura H. and Kohn K. *Ferroelectrics,* **204**, 107 (1997).

116. Iwata N., Uga M., and Kohn K. *Ferroelectrics,* **204**, 97 (1997).

117. Tehranci M.-M., Kubrakov N. F., and Zvezdin A. K. *Ferroelectrics,* **204**, 181 (1997).

118. Popov Yu. F., Kadomtseva A. M., Vorob'ev G. P., Sanina V. A., Tehranchi M.-M., and Zvezdin A. K., *J Magn. Magn. Mater.,* **188**, 239 (1998).

119. Popov Yu. F., Zvezdin A. K., Vorob'ev G. P., Kadomtseva A. M., Murashov V. A., and Rakov D. N. *JETP Lett.,* **57**, 69 (1993).

120. Rado G. T. and Ferrari J. M. *Phys. Rev.* B, **15**, 290 (1977).

121. Kato K. and Iida S. *J. Phys. Soc. Japan,* **50**, 2844 (1981); **51**, 1335 (1982); Miyamoto Y. and Chikazumi S. *J. Phys. Soc. Japan,* **57**, 2040 (1988); and the references cited there in: Khomskii D. I. *J. Magn. Mater.,* **306**, 1 (2006).

122. Rivera J.-P. and Schmid H. *Ferroelectrics,* **204**, 23 (1997).

123. Golovenchits E. I. and Sanina V. A. *Fiz. Tverd. Tela,* **28**, 713 (1986).

124. Pisarev R. V., Fiebig M., and Fröhlich D. *Ferroelectrics,* **204**, 1 (1997).

125. 125. Birss R. R., *Symmetry and Magnetism* (North-Holland, Amsterdam, 1966).

126. Fiebig M., Fröhlich D., Kohn K., Leute S., Lottermoser T., Pavlov V. V., and Pisarev R. V. *Phys. Rev. Lett.,* **84**, 5620 (2000).

127. Manzhos I. V. *Fiz. Nizk. Temp.,* **16**, 361 (1990)**.**

128. Manzhos I. V. and Chupis I. E. *Phys. Stat. Sol. (b),* **157**, K65 (1990).

129. Manzhos I.E. *Ferroelectrics,* **162**, 281 (1994).

130. Chupis I. E. and Manzhos I. V. *Ferroelectrics,* **204**, 189 (1997).

131. Soboleva T. K. *Ferroelectrics,* **162**, 287 (1994).

132. Schmid H. *Ferroelectrics,* **221**, 9 (1999).

133. Stephanovskii E. P. and Yablonskii D. A. *Fiz. Tv. Tela,* **28**, 1125 (1986).

134. Chupis I. E. *Ferroelectrics,* **161**, 287 (1994).

135. Toledano P., Schmid H., Clin M., and Rivera J. P. *Phys. Rev.* B, **32**, 6006 (1985).

136. Toledano P. *Ferroelectrics,* **161**, 257 (1994).

137. Vitebskii I. M. and Lavrinenko N. M. *Fiz. Nizk. Temp.,* **12**, 1193 (1986); Pashkevich Yu. G., Fedorov S. A., Eremenko A. V., and Sobolev V. L. *Ferroelectrics,* **162**, 237 (1994).

138. Belykh V. G., Vitebskii I. M., Sobolev V. L., and Soboleva T. K. *Fiz. Nizk. Temp.,* **14,** 992 (1988).

139. Clin M., Rivera J. P., and Schmid H., in: *Internal Meeting on Ferroelectrics,* Saarbrucken, Germany (1989).

140. Clin M., Rivera J. P., and Schmid H. *Ferroelectrics*, **79**, 173 (1988).

141. Chupis I.E. *Fiz. Nizk. Temp.*, **18**, 306 (1992).

142. Ivanov S. A., Kurlov V. N., Ponomarev B. K., and Red'kin B. S. *Izvest. Akad. Nauk, ser. Phys.,* **56**, 146 (1992).

143. Chupis I. E. *Ferroelectrics,* **162**, 375 (1994).

144. Troyanchuk I. O., Kasper N. V., Mantytskaya O. S., and Pastushonok S. N. *JETP,* **78**, 212 (1994).

145. Hill N. A. and Rabe K. M. *Phys. Rev.* B, **59**, 8759 (1994).

146. Hill N. A. *J. Phys. Chem. B,* **104**, 6694 (2000).

147. Portengen T., Östreich Th., and Sham L. J. *Phys. Rev.* B, **54**, 17452 (1996).

148. Venevtsev Yu. N. and Gagulin V.V. *Ferroelectrics,* **162**, 23 (1994).

149. Gagulin V. V., Korchagina S. K., Shevchuk Yu. A., Fadeeva N. V., and Bogatko V. V. *Ferroelectrics*, **204**, 345 (1997).

150. Ye Z.- G. and Schmid H. *Ferroelectrics*, **162**, 119 (1994).

151. Crottaz O., Schobinger-Papamantellos P., Suard E., Ritter C., Gentil S., Rivera J.-P., and Schmid H. *Ferroelectrics,* **204**, 45 (1997).

152. Lautenschläger G., Weitzed H., Vogt T., Hock R., Bohm A., Bonnet M., and Fuess H. *Phys. Rev.* B, **48**, 6087 (1993).

153. Garcia-Muñoz J. –L., Rodriguez-Carvajal I., and Lacorre P. *Phys. Rev.* B, **50**, 978 (1994).

154. Alonso J. A. , Garsia –Muñoz J. L., Fernández- Diaz M. T., Aranda M. A. G., Martinez-Lope M. J., and Casias M. T. *Phys. Rev. Lett.,* **82**, 3871 (1999).

155. Popov Yu. F., Kadomtseva A. M., Vorob'ev G. P., Timofeeva V. A., Tehranchi M.-M., and Zvezdin A. K. *Ferroelectrics,* **204**, 269 (1997).

156. Popov Yu. F., Zvezdin A. K., Kadomtseva A. M., Tehranchi M.-M., Vorob'ev G. P., Timofeeva V. A., and Ustinin D. M. *JETP,* **87,** 146 (1998).

157. Katsnel'son E. Z. and Bashkirov L. A., in: *Ferroelectromagnetic substances,* Nauka, Moscow, 1990, p.102 (in Russian).

158. Sirota N. N., Bashkirov L. A., Katsnelson E. Z., and Shapovalova E. V. *Phys. Stat. Sol. B.,* **100,** K39 (1980); *Phys. Stat. Sol.* A, **67**, K67 (1981).

159. Chupis I. E. and Mamaluy D. A. *JETP Lett.,* **68**, 922 (1998).

160. Chupis I. E. and Mamaluy D. A. *J. Phys.: Condens. Matter,* **12**, 1413 (2000).

161. Mamaluy D. A. and Chupis I. E. *Zh. Eksp. Teor. Fiz.,* **117**, 175 (2000).

162. Chupis I. E. *Low Temp. Phys.,* **27**, 1028 (2001).

163. Goto T., Kimura T., Lawes G., Ramirez A. P., and Tokura Y. *Phys. Rev. Lett.,* **92**, 257201 (2004).

164. Quezel S., Tcheou F., Rossat- Mignod J., Quezel G., and Roudaut E. *Physica* B, **86-88**, 916 (1977).

165. Kenzelmann M., Harris A. B., Jonas S., Broholm C., Schefer J., Kim S. B., Zhang C. L., Cheong S.-W., Vajk O. P., and Lynn J. W. *Phys. Rev. Lett.,* **95**, 087206 (2005).

166. Kimura T., Lawes G., Goto T., Tokura Y., and Ramirez A. P. *Phys. Rev.* B, **71**, 224425 (2005).

167. Chupis I. E. *Low Temp. Phys.,* **34**, 422 (2008).

168. Kajimoto R., Yoshizawa H., Shintani H., Kimura T., and Tokura Y. *Phys. Rev.* B, **70**, 012401 (2004).

169. Meier D., Aliouane N., Argyriou D. N., Mydosh J. A., and Lorenz T. *New Journal of Physics,* **9** (4), 100-100 (2007).

170. Chupis I. E. *Izv. Ros. Akad. Nauk Ser. Fiz.,* **71**, 1098 (2007).

171. Arima T., Goto T., Yamasaki Y., Miyasaka S., Ishii K., Tsubota M., Inami T., Murakami Y., and Tokura Y. *Phys. Rev.* B, **72**, 100102 (2005).

172. Aliouane N., Argyriou D. N., Stempfer J., Zegkinoglou I., Landsgesell, and Zimmermann M. V. *Phys. Rev.* B, **73**, 020102 (R) (2006).

173. Yamasaki Y., Sagayama H., Goto T., Matsuura M., Hirota K., Arima T., and Tokura Y. *Phys. Rev. Lett.,* **98**, 147204 (2007).

174. Lottermoser T., Lonkai T., Amann U., Hohlwein D., Ihringer J., and Fiebig M. *Nature* (London), **430**, 541 (2004).

175. Zhao T., Scholl A., Zavaliche F., Lee K., Barry M., Doran A., Cruz M. P., Chu Y. H., Ederer C., Spaldin N. A., Das R. R., Kim D. M., Baek S. H., Eom C. B., and Ramesh R. *Nat. Matt.,* **5**, 823 (2006).

176. Golovenchits E. and Sanina V., *in: Magnetoelectric Interaction Phenomena in Crystals* ed. Fiebig M., Eremenko V. V., and Chupis I. E. (Dordrecht:Kluwer), p.139 (2004).

177. Pimenov A., Mukhin A. A., Ivanov V. Yu., Travkin V. D., Balbashov A. M., and Loidl A. *Nature Phys.* **2**, 97 (2006).

178. Chupis I. E. *Low Temp. Phys.,* **33**, 715 (2007).

179. Izumov Yu. A. *Neutron diffraction on the long-periodical structures*, Moscow, Energoatomizdat, p. 98, 1987 (in Russian).

180. Chupis I. E. *Fiz. Nizk. Temp.*, **35**, 1101 (2009).

181. Senff D., Link K., Hradil K., Hiess A., Regnault L. P., Sidis Y., Aliouane N., Argyriou D. N., and Braden M. *Phys. Rev. Lett.,* **98**, 137206 (2007).

182. Katsura H., Balatsky A. V., and Nagaosa N. *Phys. Rev. Lett.,* **98**, 027203 (2007).

183. Goto T., Kimura T., Lawes G., Ramirez A. P., and Tokura Y. *Phys. Rev. Lett.,* **92**, 257201 (2004).

184. Arima T., Tokunaga A., Goto T., Kimura H., Noga Y., and Tokura Y. *Phys. Rev. Lett.,* **96**, 097202 (2006).

185. Ivanov V. Yu., Mukhin A. A., Travkin V. D., Prokhorov A.S., Kadomtseva A. M., Popov Yu. F., Vorob'ev G. P., Kamilov K. I., and Balbashov A. M. *JMMM ,* **300**, 130, (2006).

186. Hemberger J., Schrettle F., Pimenov A., Lunkenheimer P., Ivanov V. Yu., Mukhin A. A., Balbashov A. M., and Loidl A. *Phys. Rev. B*, **75**, 035118 (2007).

187. Ivanov V. Yu., Mukhin A. A., Travkin V. D., Prokhorov A. S., Glushkov V. V., and Balbashov A. M. *Proc. of Conf. on Low Temperature Physics* Rostov, 2006, v. 1, p. 48 (in Russian).

188. Aguilar R. Valdes, Suchkov A. B., Zhang C. L., Choi Y. I., Cheong S. –W., and Drew H. D. *Phys. Rev.* B, **76,** 060404 (R) (2007).

189. Pimenov A., Loidl A., Mukhin A. A., Travkin V. D., Ivanov V. Yu., and Balbashov A. M. *Phys. Rev. B*, **77**, 014438 (2008).

190. Hur N., Park S., Sharma P. A., Guha S., and Cheong S-W. *Phys. Rev. Lett.,* **93**, 107207 (2004).

191. Hur H., Park S., Sharma P. A., Ahn J., Guha S., and Cheong S- W. *Nature* (London), **429**, 392 (2004).

192. Ratcliff II . W., Kiryukhin V., Kenzelmann M., Lee S.-H., Erwin R., Schefer J., Hur N., Park S., and Cheong S- W. *Phys. Rev. B*, **72**, 060407 (R) (2005).

193. Chapon L. C., Blake G. R., Gutmann M. J., Park S., Hur N., Radaelli P. G., and Cheong S- W. *Phys. Rev. Lett.,* **93**, 177402 (2004).

194. Tachibana M., Akiyama K., Kawaji H., and Atake T. *Phys. Rev. B*, **72**, 224425 (2005).

195. Kobayashi S., Osawa T., Kimura H., Noda Y., Kagomiya I., and Kohn K. *J. Phys. Soc. Jpn.,* **73**, 1593 (2004).

196. Higashiyama D., Miyasaka S., Kida N., Arima T., and Tokura Y. *Phys. Rev.* B, **70**, 174405 (2004).

197. Higashayama D., Miyasaka S., and Tokura Y. *Phys. Rev.* B, **72**, 064421 (2005).

198. Popov Yu. F., Kadomtseva A. M., Vorob'ev G. P., Krotov S. S., Kamilov K. I., and Lukina M. M. *Fiz. Tverd. Tela,* **45**, 2051 (2003).

199. Popov Yu. F., Kadomtseva A. M., Krotov S. S., Vorob'ev G. P., Kamilov K. I., Lukina M. M., and Tegranchi M. M. *Zh. Eksp. Teor. Fiziki,* **123**, 1090 (2003).

200. Kadomtseva A. M., Krotov S. S., Popov Yu. F., and Vorob'ev G. P. *Fiz. Nizkikh Temp.,* **32**, 933 (2006).

201. Kobayashi S., Osawa T., Kimura H., Noda Y., Kagomiya I., and Kohn K. *J. Phys. Soc. Jpn.,* **73**, 1031 (2004).

202. Kadomtseva A. M., Krotov S. S., Popov Yu. F., Vorob'ev G. P., Lukina M. M. *JETP,* **100**, 305 (2005).

203. Bodenthin Y., Staub U., Garcia-Fernández M., Janoschek M., Schlappa J., Golovenchits E. I., Sanina V. A., and Lushnikov S. G. *Phys. Rev. Lett.,* **100**, 027201 (2008).

204. Golovenchits E. I., Sanina V. A. *JETP Lett.,* **78**, 88 (2003).

205. Sushkov A. B., Valdés Aguilar R., Park S., Cheong S-W., and Drew H. D. *Phys. Rev. Lett.* **98**, 027202 (2007).

206. Lawes G., Harris A. B., Kimura T., Rogado N., Cava R. J., Aharony A., Entin-Wohlman O., Yildirim T., Kenzelmann M., Broholm C., and Ramirez A. P. *Phys. Rev. Lett.,* **95**, 087205 (2005).

207. Park S., Choi Y. J., Zhang C. I., and Cheong S- W. *Phys. Rev. Lett.,* **98**, 057601 (2007).

208. Masuda T., Zheludev A., Bush A., Markina M., avd Vasiliev A. *Phys. Rev. Lett.,* **92**, 177201 (2004).

209. Seki S., Yamasaki Y., Soda M., Matsuura M., Hirota K., and Tokura Y. *Phys. Rev. Lett.,* **100**, 127201 (2008).

210. Naito Y., Sato K., Yasui Y., Kobayashi Y., and Sato M. *J. Phys. Soc. Jpn.,* **76**, 023708 (2007).

211. Buttgen N., Krug von Nidda H.-A., Svistov L. E., Prozorova L. A., Prokofiev A., and Aszmus W. *Phys. Rev.* B, **76,** 014440 (2007).

212. Schrettle F., Krohns S., Lunkenheimer P., Hemberger J., Buttgen N., Krug von Nidda H.-A., Prokofiev A. V., and Loidl A. *Phys. Rev.* B, **77**, 144101 (2008).

213. Mostovoy M. *Phys, Rev. Lett.,* **96**, 067601 (2006).

214. Taniguchi K., Abe N., Takenobu T., Iwasa Y., and Arima T. *Phys. Rev. Lett.,* **97**, 097203 (2006).

215. Yamasaki Y., Miyasaka S., Kaneko Y., He J.-P., Arima T., and Tokura Y. *Phys. Rev. Lett.,* **96**, 207204 (2006).

216. Kenzelmann M., Lawes G., Harris A. B., Gasparovic G., Broholm C., Ramirez A. P., Jorge G. A., Jaime M., Park S., Huang Q., Shapiro A. Ya., and Demianets L. A. *Phys. Rev. Lett.,* **98**, 267205 (2007).

217. Golovenchits E. I. and Sanina V. A. *Pis'ma v Zh. Eksp. Teor. Fiziki,* **81**, 630 (2005); **84**, 222 (2006).

218. Gor'kov L. P. *Uspekhi Fiz. Nauk,* **168**, 665 (1998).

219. Sanina V. A., Golovenchits E. I., and Zalesskii V. G. *Physics of the Solid State,* **50**, 913, 922 (2008).

220. Kenzelmann M. and Harris A. B. *Phys. Rev. Lett.,* **100**, 089701 (2008).

221. Mostovoy M. *Phys. Rev. Lett.,* **100**, 089702 (2008).

222. Harris A. B. *Phys. Rev.* B, **76**, 054447 (2007).

223. Moskvin A. S. and Drechsler S.-L. *Phys. Rev.* B, **78**, 024102 (2008).

224. Sergienko I. A. and Dagotto E. *Phys. Rev.* B, **73**, 094434 (2006).

225. Dzyaloshinskii I. *J. Phys. Chem. Solids,* **4**, 241 (1958).

226. Moriya T. *Phys. Rev.*, **120**, 91 (1960).

227. Radaelli P. G. and Chapon L. C. *Phys. Rev.* B, **76**, 054428 (2007).

228. Hu J. *Phys. Rev.Lett.,* **100**, 077202 (2008).

229. Wang C., Guo G.-C., and He L. *Phys. Rev. Lett.,* **99**, 177202 (2007); *Phys. Rev.* B, **77**, 134113 (2008).

230. Giovannetti G. and J. van den Brink *Phys. Rev. Lett.,* **100**, 227603 (2008).

231. Moskvin A. S. *JETP,* **104**, 911 (2007).

232. Moskvin A. S. *Phys. Rev.* B, **75**, 054505 (2007).

233. Katsura H., Nagaosa N., and Balatsky A. V. *Phys. Rev. Lett.,* **95**, 057205 (2005).

234. Hu C. D. *Phys. Rev.* B, **77**, 174418 (2008).

235. Chapon L. C., Radaelli P. G., Blake G. P., Park S., and Cheong S.-W. *Phys. Rev. Lett.,* **96**, 097601 (2006).

236. Zhou J.-S. and Goodenough J. B. *Phys. Rev. Lett.,* **96**, 247202 (2006).

237. Catalan G. *Phase Transitions,* **81**, 729 (2008).

238. Sergienko I. A., Sen C., and Dagotto E. *Phys. Rev. Lett.,* **97**, 227204 (2006).

239. Lorentz B. Wang Y., and Chu C. W. *Phys. Rev.* B, **76**, 104405 (2007).

240. 240. Choi Y. J. , Yi H. T., Lee S., Huamg Q., Kiryukhin V., and Cheong S.-W. *Phys. Rev. Lett.,* **100**, 047601 (2008).

241. Seshadri R. and Hill N. A. *Chem. Mater.,* **13**, 1892 (2001).

242. Turov E. A. *Phys. Usp.,* **37**, 303 (1994).

243. Kadomtseva A. M., Zvezdin A. K., Popov Yu. F., Pyatakov A. P., and Vorob'ev G. P. *JETP Lett.,* **79**, 571 (2004).

244. Kadomtseva A. M., Popov Yu. F., Pyatakov A. P., Vorob'ev G. P., Zvezdin A. K., and Viehland D. *Phase Trans.,* **79**, 1019 (2006).

245. Sosnowska I., Peterlin-Neumaier T., and Steichele E. *J. Phys.* C, **15**, 4835 (1982).

246. Sosnowska I. and Zvezdin A. K. *J. Magn. Magn. Mater.,* **140 -144**, 167 (1995).

247. Popov Yu. F., Kadomtseva A. M., Krotov S. S., Belov D. V., Vorob'ev G. P., Makhov P. N., and Zvezdin A. K. *Low Temp. Phys.,* **27**, 478 (2001).

248. Ederer C. and Spaldin N. A. *Phys. Rev.* B, **71**, 060401(R) (2005).

249. Neaton J. B., Ederer C., Waghmare U. V., Spaldin N. A., and Rabe K. M. *Phys. Rev.* B, **71**, 014113 (2005).

250. Ruette B., Zvyagin S., Pyatakov A. P., Bush A., Li J. F., Belotelov V. I., Zvezdin A. K., and Viehland D. *Phys. Rev.* B, **69**, 064114 (2004).

251. Lebeugle D., Colson D., Forget A., Viret M., Bataille A. M., and Gukasov A. *Phys. Rev. Lett.,* **100**, 227602 (2008).

252. Rogerio de Sousa and Moore J. E. *Phys. Rev.* B, **77**, 012406 (2008).

253. Wang J., Zheng H., Nagarajan V., Liu B., Ogale S. B., Viehland D., Venugopalan V., Schlom D. G., Wuttig M., Ramesh R., Neaton J. B., Waghmare U. V., Hill N. A., and Rabe K. M. *Science,* **299**, 1719 (2003).

254. Li J., Wang J., Wuttig M., Ramesh R., Wang N., Ruette B., Pyatakov A. P., Zvezdin A. K.,and Viehland D. *Appl. Phys. Lett.* **84**, 5261 (2005).

255. Bai F., Wang J., Wuttig M., Li J. F., Wang N., Pyatakov A. P., Zvezdin A. K., Cross L. E., Viehland D. *Appl. Phys. Lett.* **86**, 032511 (2005).

256. Jiang Q. and Qiu J. H. *J. Appl. Phys.* **99**, 103901 (2006).

257. Bea H., Bibes M., Petit S., Kreisel J., and Barthelemy A. *Phil. Mag. Lett.* **87,** 165 (2007).

258. Zvezdin A. K. *Bulletin of the Lebedev Physics Institute,* **4**, 3 (2004).

259. Seshardi R. and Hill N. A. *Chem. Mater.,* **28**, 2892 (2001).

260. A. Moreira dos Santos, Parashar S., Raju A. R., Zhao Y. S., Cheetham A. K., and Rao C. N. R. *Solid State Commun.,* **122**, 49 (2002).

261. Kimura T., Kawamoto S., Yamada I., Azuma M., Takano M., and Tokura Y. *Phys. Rev.* B, **67**, 180401 (2003).

262. Sharan A., Lettien J., Jia Y., Tian W., Pan X., Schlom D. G., and Gopalan V. *Phys. Rev.* B, **69**, 214109 (2004).

263. Palkar V. R., Darshan, Kundaliua C., Malik S. K., and Bhattacharya S. *Phys. Rev.* B, **69**, 212102 (2004).

264. Palkar V. R. and Malik S. K. *Solid State Comm.,* **134**,783 (2005).

265. Stoupin S., Chattopadhyay S., Bolin T., and Segre C. U. *Solid State Comm.,* **144**, 46 (2007).

266. Baettig P. and Spaldin N. *Appl. Phys. Lett.,* **86**, 012505 (2005).

267. Ju S. and Guo G-Y. *Appl. Phys. Lett.,* **92**, 202504 (2008).

268. Feng H –J. and Liu F –M. *Phys. Lett* A, **372**, 1904 (2008).

269. Sahu J. R., Serrao C. R., Ray N., Waghmare U.V., Rao C. N. R. *J. Mater. Chemistry,* **17**, №1, ВР42 (2007).

270. Sugie H., Iwata N., and Kohn K. *J. Phys. Soc. Jpn.,* **71**, 1558 (2002).

271. Fiebig M., Lottermoser Th., Fröhlich D., and Pisarev R. V. *Nature,* **419**, 818 (2002).

272. Fiebig M., in: *Magnetoelectric Interaction Phenomena in Crystals* eds. Fiebig M., Eremenko V. V., and Chupis I. E. (Dordrecht:Kluwer), **164,** 163 (2004).

273. Goltsev A. V., Pisarev R. V., Lottermoser Th., and Fiebig M. *Phys. Rev. Lett.,* **90**, 177204 (2003).

274. Lorenz B., Livinchuk A. P., Gospodinov M. M., and Chu C. W. *Phys. Rev. Lett.,* **92**, 087204 (2004).

275. Lottermoser T. and Fiebig M. *Phys. Rev.* B, **70**, 220407 (R) (2004).

276. Hanamura E., Tanabe Y. *J. Phys. Soc. Jpn,.* **72**, 2959 (2003); *in: Magnetoelectric Phenomena in Crystals* eds. Fiebig M., Eremenko V. V., and Chupis I. E. (Dordrect: Kluwer), **164**, 151 (2004).

277. Qian M., Dong J., and Xing D. Y. *Phys. Rev.* B, **63**, 155101 (2001).

278. Iizuka-Sakano T., Hanamura E., and Tanabe Y. *J. Phys. Condens. Matter,* **13**, 3031 (2001).

279. Kalashnikov A. M. and Pisarev R. V. *Pis'ma v Zh. Eksp. Teor. Fiz.,* **78**, 175 (2003).

280. B.van Aken, Palstra T. T. M., Filippeti A., and Spaldin N. A. *Nature Mater.,* **3**, 164 (2004).

281. Khomskii D. I. *J. Magn. Magn. Mater.,* **306**, 1 (2006).

282. Nandi S., Kreyssig A., Tan L., Kim J. W., Yan J. Q., Lang J. C., Haskel D., McQueeney R. J., and Golgman A. I. *Phys. Rev. Lett.,* **100**, 217201 (2008).

283. Fiebig M., Degenhardt C., and Pisarev R. V. *Phys. Rev. Lett.,* **88**, 027203 (2002).

284. Katsufuji T., Mori S., Masaki M., Moritomo Y., Yamamoto N., and Takagi H. *Phys. Rev.* B, **64**, 104419 (2001).

285. Souchkov A. B., Simpson J. R., Quijada M., Ishibashi H., Hur N., Ahn J. S., Cheong S. W., Millis A. J., and Drew H. D. *Phys. Rev. Lett.,* **91**, 027203 (2003).

286. Kim J. H. and Han J. H. *Phys. Rev.* B, **76**, 054431 (2007).

287. Fiebig M., Lottermoser T., Lonkai T., Goltsev A., and Pisarev R. V. *J. Magn. Magn. Mater.* **290**, 883 (2005).

288. Efremov D. V., J. van den Brink, and Khomskii D. I. *Nature Mater.,* **3**, 853 (2004).

289. Daoud-Aladine A., Rodriguez-Carvijal J., and Pinsard-Gaudart L. *Phys. Rev. Lett.,* **89**, 97205 (2002).

290. Kadomtseva A. M., Popov YU. F., Vorob'ev G. P., Kamilov K. I., Ivanov V. Yu., Mukhin A. A., and Balbashov A. M. *Zh. Eks. Teor. Fiziki,* **133**, 156 (2008) (in Russian).

291. Lopes A. M. L., Araujo J. P., Amaral V. S., Correia J. G., Tomioka Y., and Tokura Y. *Phys. Rev. Lett.,* **100**, 155702 (2008).

292. Ikeda N. et al. *J. Phys. Soc. Jpn.,* **69**, 1526 (2000).

293. Ikeda N., Ohsumi H., Ohwada K., Ishii K., Inami T., Kakurai K., Murakami Y., Yoshii K., Mori S., Horibe Y., and Kito H. *Nature,* **436**, 1136 (2005).

294. Subias G., Garsia J., Blasco J., Proietti M. G., Renevier H., and Sanchez M. C. *Phys. Rev. Lett.,* **93**, 156408 (2004).

295. 295. Nazarenko E., Lorenzo J. E., Joly Y., Hodeau J. L., Mannix D., and Marin C. *Phys. Rev. Lett.,* **97**, 056403 (2006).

296. Leonov I., Yaresko A. N., Antonov V. N., Korotin M. A., and Anisimov V. I. *Phys. Rev. Lett.,* **93**, 146404 (2004).

297. Jeng H.-T., Guo G. Y., and Huang D. J. *Phys. Rev. Lett.,* **93**, 156403 (2004).

298. 298. Schlappa J., Schubler-Langeheine C., Chang C. F., Ott H., Tanaka A., Hu Z.., Havekort M. W., Schierle E., Weschke E., Kaindl G., and Tjeng L. H. *Phys. Rev. Lett.,* **100**, 026406 (2008).

299. Krichevtsov B. B. *Pis'ma v Zh. Eksp.Teor. Fiziki,* **74**, 177 (2001) (in Russian).

300. Radaelli P.G., Chapon L.C., Daoud-Aladine A., Vecchini C., Brown P.J., Chatterji T., Park S., and Cheong S-W. *Phys. Rev. Lett.,* **101**, 067205 (2008).

301. 301. Erenstein W., Mathur N. D., and J. F.Scott, *Nature,* **442**, 759 (2006).

302. Jodlauk S., Becker P., Mudosh J. A., Khomskii D. I., Lorenz T., Streltsov S. V., Hezel D. C., and Bohaty L. *J. Phys. Condens, Matter.,* **19**, 432201 (2007)

303. Pankrats A. I., Petrakovskii G. A., Bezmaternykh L. N., and Baikov O. A. *Zh.Eksp. Teor. Fiziki,* **246,** 887 (2004) (in Russian).

304. Zvezdin A. K., Krotov S. S., Kadomtseva A. M., Vorob'ev G. P., Popov Yu. F., Pyatakov A. P., Bezmaternykh L. N., and Popova E. N. *Pis'ma v Zh. Eksp. Teor. Fiziki,* **81**, 335 (2005) (in Russian).

305. Bea H., Gajek M., Bibes M., and Bartholemy A. *J. Phys. Condens. Matter.,* **20**, 434221 (2008).

306. Marti X., Sanchez F., Skumryev V., Laukhin V., Perrater C., Garcia- Cuenca M. V., Varela M., and Fontcuberta J. *J. Phys. Condens. Matter.,* **20**, 434220 (2008).

307. Nan C. W., Bichurin M. I., Dong S. X., Viehland D., and Srinivasan G. *J. Appl. Phys.,* **103**, 031101 (2008).

308. Bichurin M (ed) 2002 *Proc. MEIPIC-4(Novgorod, Russia, 16-19 Octi\ober 2001) Ferroelectrics,* **279- 280**.

309. Fiebig M., Eremenko V. V., and Chupis I. E. (ed) 2004 *Magnetoelectric Interaction Phenomena in Crystals (*Dordrecht: Kluwer) *Proc. MEIPIC-5 (Sudak, Ukraine, 21-24 September 2003).*

310. Cheong S-W. and Mostovoy M. *Nature Mater.,* **6**, 13 (2007).

311. Lottermoser Th., Meier D., Pisarev R. V., and Fiebig M. *Phys. Rev.* B, **80**, 100101 (R) (2009).

**Abstract**

Ferroelectromagnets — compounds with ferroelectric and magnetic orderings — were discovered about fifty years ago. They are of great interest due to the possibility of mutual control of electric (magnetic) properties by magnetic (electric) field. At present such possibility of magnetoelectric control has become reality which stimulated revival of scientific and practical interest in ferroelectromagnets. The progress in experimental and theoretical investigations of magnetoelectric phenomena in ferroelectromagnets for the last years, especially the colossal magnetoelectric effects observed recently, are considered. The problems and future prospects of enhancing magnetoelectric effects in ferroelectromagnets are being discussed.

The book is intended for researchers in ferroelectricity and magnetism and for students of physical specialities.

## List of chemical formulae

$Pb(Fe_{1/2}Nb_{1/2})O_3$      the first ferroelectromagnet

$BiFeO_3$      bismuth ferrite

$BiMnO_3$      bismuth manganite

$BaMnF_4$

$BaCoF_4$

$Li(Fe_{1/2}Ta_{1/2})O_2F$

$Fe_3O_4$      magnetite

$Cr_2BeO_4$

$Cr_2O_3$      chrome oxide

$BaTiO_3$

$Ni_3B_7O_{13}I$      nickel-iodine boracite

$TbMnO_3$      terbium manganite

$DyMnO_3$      dysprosium manganite

$RMnO_3$      rare-earth manganites

$RMn_2O_5$      rare-earth manganates

$LuFe_2O_4$

$RFe_2O_4$

$YCrO_3$      yttrium chromite

$RNiO_3$      rare-earth nickelates

$RbFe(MoO_4)_2$

$CoCr_2O_4$

$LiCu_2O_2$

$LiCuVO_4$

$Ni_3V_2O_8$

$CdCr_2S_4$

**List of abbreviations**

| | |
|---|---|
| ME | magnetoelectric |
| FE | ferroelectric |
| AF | antifeffomagnetic |
| FM | ferromagnetic |
| AWFM | antifferomagnet with weak ferromagnetism |
| FIM | ferrimagnetic |
| LMEE | linear magnetoelectric effect |
| FEDW | ferroelectric domain wall |
| DM | Dzyaloshinskii- Moriya |
| IME | inhomogeneous magnetoelectric effect |
| MI | metal-insulator (transition) |
| spin-flop | phase transition, accompanied by spins overturn |
| spin-flip | phase transition, accompanied by spins flipping |
| CO | charge ordering |
| CM | commensurate magnetic (structure) |
| ICM | incommensurate magnetic (structure) |
| SHG | second harmonic generation |

**List of Symbols**

| | |
|---|---|
| $F$ | free energy |
| $\vec{E}$ | electric field strength |
| $\vec{H}$ | magnetic field strength |
| $\vec{M}_j$ | magnetic moment of $j$ − sublattice |
| $\vec{M} = \vec{M}_1 + \vec{M}_2$ | magnetization |
| $\vec{L} = \vec{M}_1 - \vec{M}_2$ | antiferromagnetic vector |
| $\vec{P}$ | electric polarization vector |
| $\vec{T}$ | toroidal moment |
| $T_c$ | temperature of ferroelectric ordering |
| $T_m$ | temperature of ferro(ferri)magnetic ordering |
| $T_N$ | temperature of antiferromagnetic ordering (Neel temperature) |
| $\chi_{ik}^E$ | tensor of dielectric susceptibility |
| $\chi_{ik}^M$ | tensor of magnetic susceptibility |
| $\chi_{ik}^{ME}$ | tensor of magnetoelectric susceptibility |
| $\alpha_{ik}$ | tensor of a linear magnetoelectric effect |
| $u_E$ | electrostriction |
| $u_M$ | magnetostriction |
| R | rare-earth element |
| $\omega_s$ | spin frequency |
| $\omega_p$ | optical phonon frequency |
| 1D | one-dimensions |
| $\dot{A}$ | $\partial A / \partial t$ |
| $u_{ik}$ | tensor of elastic displacements |
| $V$ | sample volume |